\documentclass[aps,prb,twocolumn,showpacs,superscriptaddress,groupedaddress]{revtex4-1}
\usepackage{graphicx}
\usepackage{amsmath}
\usepackage{amssymb,color}
\usepackage[utf8]{inputenc}
\usepackage{braket}
\input{epsf}
\newcommand{\DDa}{\Delta_{\alpha}}

\newcommand{\kk}{\textbf{k}}
\newcommand{\DDho}{\Delta_{h_1}}
\newcommand{\DDht}{\Delta_{h_2}}
\newcommand{\DDe}{\Delta_{e}}
\newcommand{\lh}{\lambda_{hh}}
\newcommand{\lhe}{\lambda_{eh}}
\newcommand{\DD}{\Delta}
\newcommand{\nn}{\nonumber}

\newcommand{\EFe}{E_{F_e}}

\newcommand{\vt}{\vartheta}
\begin{document}

\title{\boldmath{$s+is$} Superconductivity with incipient bands: doping dependence and STM signatures}

\author{Jakob~B\"oker$^1$}
\author{Pavel~A. Volkov$^1$}
\author{Konstantin~B. Efetov$^{1,2}$}
\author{Ilya~Eremin$^1$}
\affiliation{1-Institut f\"ur Theoretische Physik III, Ruhr-Universit\"at Bochum, D-44801 Bochum, Germany}
\affiliation{2-National University of Science and Technology “MISiS” - 119049 Moscow, Russian Federation}

\begin{abstract}
Motivated by the recent observations of small Fermi energies and comparatively large superconducting gaps, present also on bands not crossing the Fermi energy  ({\it incipient} bands) in iron-based superconductors, we analyze the doping evolution of superconductivity in a four-band model across the Lifshitz transition including BCS-BEC crossover effects on the shallow bands. Similar to the BCS case, we find that with hole doping the phase difference between superconducting order parameters of the hole bands change from $0$ to $\pi$ through an intermediate $s+is$ state, breaking time-reversal symmetry (TRS). The transition, however, occurs in the region where electron bands are incipient and chemical potential renormalization in the superconducting state leads to a significant broadening of the $s+is$ region. We further present the qualitative features of the $s+is$ state that can be observed in scanning tunneling microscopy (STM) experiments, also taking incipient bands into account.
\end{abstract}
\date{\today}

\maketitle

\section{Introduction}

The discovery of unconventional and high-$T_c$ superconductivity in Fe pnictides and chalcogenides opened up several new directions in the study of nonphononic mechanisms of Cooper pairing in multiband correlated electron systems\cite{Kamihara.2008,Johnston.2010,Paglione.2010,Hirschfeld.2011,Hosono.2015}. One of the interesting issues that has attracted considerable attention is the observation of finite superconducting gaps on bands not crossing the Fermi level\cite{Miao.2015,Okazaki.2014}. Furthermore, photoemission and quantum oscillation experiments have shown that in many Fe-based superconductors either electron or hole pockets are smaller than previously thought and the corresponding bands barely cross the Fermi level or are located fully below or above it\cite{Charnukha.2015,Terashima.2014,Watson.2015}. Interestingly, a recent quasiparticle interference (QPI) study of FeSe\cite{Kasahara.2014} has revealed that the ratio of the superconducting gap to the Fermi energy on the shallow electron band is very large $\Delta/E_{F_e}\sim1$ with similar ratios $\Delta/E_F\approx 0.6$ being reported for $\text{FeSe}_x\text{Te}_{1-x}$\cite{Lubashevsky.2012, Okazaki.2014, Rinott.2017}. This has been followed up by observation of strong superconducting fluctuations in FeSe far above $T_c$\cite{Kasahara.2016,Naidyuk.2016,Sinchenko.2016}.

The large values of $\Delta/E_F$ observed initially in FeSe have been interpreted\cite{Kasahara.2014} as a signature of a crossover from conventional Cooper pairing (BCS) to Bose-Einstein condensation (BEC) regime\cite{Randeria.2014}. In the latter case, Cooper pairs are formed well above superconducting $T_c$ and should manifest themselves as unusually strong superconducting fluctuations.

A great deal of current understanding of the BCS-BEC crossover physics comes from the remarkable experiments on systems of ultracold fermionic atoms\cite{Regal.2004,Bartenstein.2004,Ketterle.2008,Zwerger.2012}. These systems offer the great advantage of an experimentally tunable interparticle attraction between constituents via Feshbach resonances, allowing us to perform experiments throughout the crossover region. Of particular relevance are experiments on quasi two-dimensional (quasi-2D) systems \cite{Sommer.2012,Makhalov.2014,Ries.2015}. More recent experiments report the observation of a Berezinskii-Kosterlitz-Thouless (BKT) transition to the superfluid state\cite{Murthy.2015} as well as the presence of pairing far above the critical temperature\cite{Murthy.2017}.

Similar ideas have been recently applied theoretically to study the potential BCS-BEC crossover in multiband systems with small Fermi energies\cite{Guidini.2014,Chubukov.2016}. One of the most important conclusions of these studies is that for small Fermi energies the chemical potential of the system is strongly renormalized in the superconducting state even in situations when the temperature of the pair formation roughly agrees with the superconducting transition. This renormalization of the chemical potential is especially important given the variety of the superconducting and magnetic states, which appear in the iron-based superconductors upon changing the control parameters such as disorder, pressure and doping.

Another interesting situation where related physics can appear is near a Lifshitz transition\cite{Lifshitz.1960}, where one of the bands continuously moves away from the Fermi level as a function of an external parameter (e.g. pressure, doping, external magnetic field\cite{Ptok.2017} or nanostructuring\cite{Bianconi.2001}). Theoretical proposals have been put forward that strong $T_c$ enhancement can be achieved close to Lifshitz transition in striped \cite{Perali.1996,Valetta.1997} and layered systems\cite{Innocenti.2010}. Superconductivity in two-band models close to a Lifshitz transition is also being studied\cite{Innocenti2011,Guidini.2014,Valentinis.2016}. Recent experiments on monolayer FeSe\cite{Shi.2017} do suggest enhancement of superconductivity by a Lifshitz transition.

A peculiar example of an iron-based superconductor with a Lifshitz transition is $\text{Ba}_{1-x}\text{K}_x\text{Fe}_2\text{As}_2$. Angle-resolved photoemission (ARPES)\cite{Sato.2009,Xu.2013} and thermopower\cite{Hodovanets.2014} measurements point toward the existence of such a transition in the overdoped compound with $x\sim0.7-0.9$. Intriguingly, in the same doping range, the structure of the superconducting gaps undergoes dramatic changes, seemingly inconsistent with a two-band description.

Multiple experiments such as ARPES\cite{Nakayama.2011,Ding.2008}, neutron scattering\cite{Christianson.2008} and thermal conductivity measurements\cite{Luo.2009} support a nodeless $s^{+-}$ gap structure around the optimal doping $x\approx0.4$ with the order parameter changing sign between the electron and the hole Fermi surface pockets. However, in the extremely overdoped case, the electron bands are located away from the Fermi energy with no signs of a superconducting gap observed up to now. On the other hand, superconducting gaps are present at the hole bands with the symmetry still under debate. Experiments support either strongly anisotropic $s$-wave pairing symmetry with accidental nodes where the order parameter changes sign between the two remaining hole pockets\cite{Sato.2009,Okazaki.2012,Watanabe.2014,Cho.2016} or $d$-wave-pairing symmetry with well-pronounced line nodes\cite{Tafti.2013,Reid.2012}. Moreover, in the intermediate doping region, frustration between the two superconducting channels has been theoretically predicted to result in a time-reversal symmetry-breaking $s+is$\cite{Stanev.2010,Carlstrom.2011,Hu.2012,Maiti.2013} state or $s+id$ state\cite{Thomale.2011,Platt.2009}. In these states, the phase difference $\phi$ between the order parameters at the two hole bands is not equal to a multiple of $\pi$ with the $\phi\leftrightarrow-\phi$ symmetry being spontaneously broken.

$s+is$ superconductors have been theoretically predicted to possess many unconventional properties.
Josephson critical current of a constriction junction between $s+is$ superconductors has been found to be anomalously suppressed\cite{Hu.2012}. As a result of simultaneous breaking of $U(1)$ and $Z_2$ vortex fractionalization and unusual vortex cluster states have been predicted\cite{Carlstrom.2011}. Additionally, excitations inside the vortex cores in the $s+is$ state have been found to lead to a new mechanism of vortex viscosity\cite{Silaev.2013}. Collective excitations of the phase differences between order parameters of different bands (Leggett modes) in the $s+is$ state have peculiar phase-density nature\cite{Carlstrom.2011} and have been predicted to soften at the $s+is$ critical points\cite{Maiti.2013,Lin.2012}.

The time-reversal symmetry-breaking in the $s+is$ state is most directly manifested in spontaneous currents around nonmagnetic impurities\cite{Maiti.2015} or quench-induced domain walls\cite{Garaud.2014}. The currents result in local magnetic fields in the superconducting phase and provide a signature of the $s+is$ state. This idea has been implemented in recent $\mu$Sr experiments on $\text{Ba}_{1-x}\text{K}_x\text{Fe}_2\text{As}_2$ aiming at detection of the $s+is$ state. While the first report\cite{Lotfi.2014} provided no evidence, recent results\cite{Grinenko.2017} are consistent with $s+is$ state at $x=0.73$, close to the region where the Lifshitz transition is considered to occur. Other possible experimental signatures of $s+is$ superconductivity have also been suggested \cite{Garaud.2016,Yerin.2015}. Crucially, the data presented in\cite{Grinenko.2017} encourages one to consider $s+is$ state in more detail with application to $\text{Ba}_{1-x}\text{K}_x\text{Fe}_2\text{As}_2$ and propose complementary techniques to study its properties.

In this paper, motivated by the recent observations of comparatively small Fermi energies and possible closeness of $s+is$ state to the Lifshitz transition, we analyze the doping evolution of superconductivity in a four-band model for the iron pnictides including the effects of BCS-BEC crossover physics. Similar to the BCS case, we find that with hole doping the phase difference between superconducting order parameters of the hole bands changes from $0$ to $\pi$ through an intermediate $s+is$ state breaking TRS. However, in contrast to the BCS treatment, we find that the region of the $s+is$ state is considerably expanded in the phase diagram due to additional renormalization of the chemical potential. In addition, the $s+is$ state is shown to extend to the region where the electron bands are already above the Fermi level, which agrees with recent experiments\cite{Grinenko.2017}. Finally, we consider possible signatures of the $s+is$ state in the STM quasiparticle interference patterns using a recent proposal\cite{Hirschfeld.2015}, extending the formalism also to the incipient band case. The paper is organized as follows: In Sec. \ref{sec_1} we present the model, the superconducting phase diagram with degenerate and non-degenerate hole bands is studied in Sec. \ref{sec_2}. Section \ref{sec_4} shows the qualitative effects of the $s+is$ state for quasiparticle interference. We present the conclusions in Sec. \ref{concl}. In Appendix A, we show that our solutions correspond to minima of the effective action by analyzing its second variation matrix. The details of the calculations of the corrections to the local density of states due to impurities for the $s+is$ state and incipient bands are presented in Appendix B.

\section{Model\label{sec_1}}

\begin{figure}[h]
 	\renewcommand{\baselinestretch}{1.0}
 	\includegraphics[angle=0,width=1\linewidth]{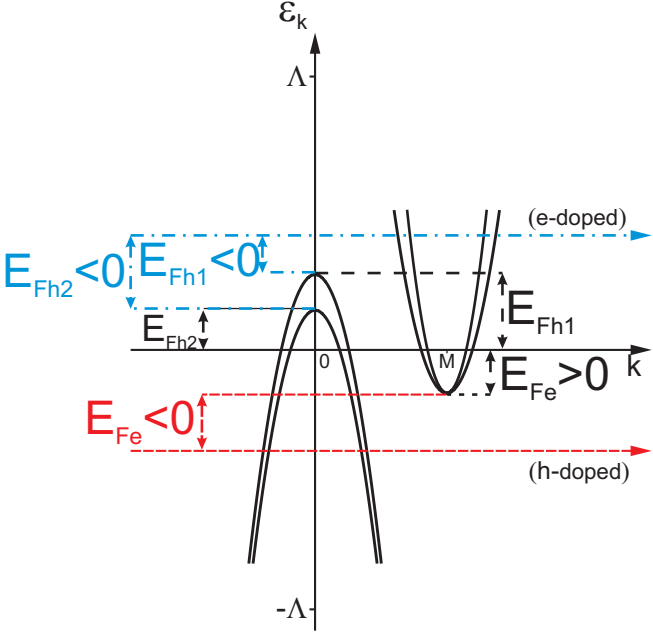}
 	\caption{Band structure of the considered 2D model. Two parabolic hole and electron bands are around the $\Gamma$ and $M$ points, respectively. $E_{F_{h_1}}$,$E_{F_{h_2}}$($E_{F_e}$) correspond to the Fermi energies of hole (electron) bands. The position of the $k$ axis marks the overall chemical potential of the system. Three cases are illustrated:
 \emph{Undoped or moderate doping} (black): both hole and electron bands cross the Fermi level. $E_{F_{h_i}}$ and $E_{F_e}$ are positive.  \emph{Hole-doped} (red dashed): hole bands cross the Fermi level while electron bands are incipient, $E_{F_e}$ is negative. \emph{Electron-doped} (blue dashed-dotted) both hole bands are incipient. $E_{F_{h_i}}$ negative. Energy difference $E_D=E_{F_{h_1}}+E_{F_e}$ is independent from doping. Interactions are assumed to be frequency- and momentum- independent up to an upper energy cutoff $\Lambda$.}
 	\label{Fig:Model}
\end{figure}

\begin{figure}[h]
	\renewcommand{\baselinestretch}{1.0}
	\includegraphics[angle=0,width=1\linewidth]{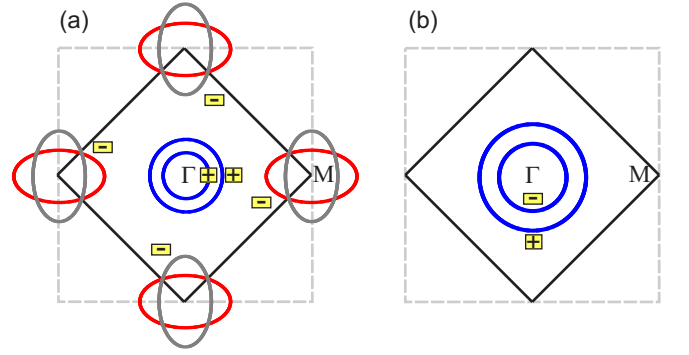}
	\caption{$k_x,k_y$-cut through the Fermi surface of the folded Brillouin-zone at (a) moderate doping and (b) large hole doping. In panel (b), Fermi surface consists of hole pockets only while electron bands are incipient. Plus and minus signs represent the pairing symmetry.}
	\label{Fig:FermiSurface}
\end{figure}

We consider a two-dimensional model with two hole bands centered at the $\Gamma$ point and two identical electron bands around the $M$ point of the Brillouin zone (see Fig.\ref{Fig:Model}), reproducing the qualitative features of $\text{Ba}_{1-x}\text{K}_x\text{Fe}_2\text{As}_2$ band structure. For simplicity, the masses of electrons and holes are taken to be isotropic and equal. We define the Fermi energies $E_{F_{h_1}}$, $E_{F_{h_2}}$ of hole bands as the difference between the band top energy and the chemical potential at zero temperature in the absence of interactions $\mu_0$ and $E_{F_{e_1}}=E_{F_{e_2}}$ for electron bands as the difference of $\mu_0$ and the band bottom energy. Thus if the Fermi energy of any band defined this way becomes negative, then the corresponding pocket of the Fermi surface vanishes and the band is incipient (see Fig. \ref{Fig:FermiSurface}).

We consider superconductivity to be driven by the repulsive inter-band interactions between the electron and the hole bands and between the two hole bands. Moreover, we assume $s$-wave symmetry throughout the phase diagram and ignore anisotropy within each band. The intraband interactions are taken to be repulsive but weaker than the interband ones, and as we are mostly interested in the hole-doped case we do not take the interaction between the electron bands into account. We assume the interactions to be frequency- and momentum independent up to a spin fluctuation cutoff energy $\Lambda$ assumed to exceed the Fermi energies of the bands\cite{Hirschfeld.2011}. This can be considered as a minimal approximation for the iron-based superconductors, where generally anisotropy can be relevant. However, as the $s+is$ state arises solely due to competition between two different $A_{1g}$-symmetric superconducting channels, momentum- and frequency- independent interactions are sufficient to examine its properties.

We study our model within a mean field approach where the chemical potential is found self-consistently as a function of the superconducting gap and temperature\cite{Leggett.1980}. This approach has been found to work well for single-band systems throughout in both the BCS and BEC cases at sufficiently low temperatures\cite{Zwerger.2012}. At elevated temperatures, pairing fluctuations start to play a very prominent role as the system goes over to the BEC regime\cite{Nozieres.1985}. On the other hand, for two-band systems there are indications that if one of the bands is strongly in BCS regime $E_F\gg |\Delta|$ then the behavior of system as a whole is closer to BCS type\cite{Chubukov.2016}. For multiband superconductors allowing for the $s+is$ state to occur, fluctuations have been shown to lead to new phases\cite{Bojesen.2013,Bojesen.2014} and re-entrant phase transitions\cite{Carlstrom.2015}. Moreover, softening of Leggett modes\cite{Maiti.2013,Lin.2012} suggests that close to the $s+is$ state boundaries fluctuations become progressively more important. In our work, we shall not address the issue of fluctuations further, concentrating on the effects of the chemical potential renormalization on the $s+is$ state.

Additionally, we shall not consider corrections to pairing interaction arising in the non-adiabatic limit $\Lambda\gg E_F$\cite{Gor'kov.1961}. These corrections result\cite{Chubukov.2016} in different pre-exponential factors; however, as the $s+is$ region occupies a relatively small part of the phase diagram, we assume that these corrections are independent of doping and can be effectively absorbed into a redefinition of the factors $E_{0e}$ and $E_{0h}$ in Eq. (\ref{E_0s}) without altering any of the qualitative results.

 The Hamiltonian of the system is then

\begin{equation}
H=\sum_{\kk, \alpha, \sigma}\xi^{\alpha}_{\kk} c^{\alpha+}_{\kk\sigma} c^{\alpha}_{\kk\sigma}\nn\\
+\sum_{\kk,\kk^{\prime},\alpha,\gamma}\left[U_{\alpha\gamma}c^{\alpha+}_{\kk\uparrow}c^{\alpha+}_{-\kk,\downarrow}c^{\gamma}_{-\kk'\downarrow}c^{\gamma}_{\kk',\uparrow} +H.c. \right]              \label{Eq:Hamiltonian},
\end{equation}
where $\{\alpha, \gamma\}\in\{h_1,h_2,e_1,e_2\}$ label the bands, and
\begin{align}
\xi^{h_i}_{\kk}&=-\epsilon_\textbf{k}+\mu_{h_i},\\
\xi^{e}_{\kk}&=\epsilon_\textbf{k}-\mu_{e_i}
\end{align}
are the bare energy dispersions for holes and electrons with $\epsilon_\textbf{k}=\frac{\kk^2}{2m}$, $\mu_{e_i} = E_{F_{e_i}}-(\mu_0-\mu)$, and $\mu_{h_i} = E_{F_{h_i}}+(\mu_0-\mu)$. We further set $U_{h_1h_1}=U_{h_2h_2}=U_h$, $U_{e_1e_1}=U_{e_2e_2}=U_e$,
$U_{h_1h_2}=U_{h_2h_1}=U_{hh}$, and $U_{h_ie_j}=U_{e_jh_i}=U_{he}$. The interaction term can be written then in a convenient matrix form:
\begin{equation}
\begin{gathered}
H_{\text{int}}=\sum_{\kk,\kk^{\prime}}\hat{b}^\dagger_\kk \hat{U} \hat{b}_{\kk'}
+\frac{U_e}{2}c^{e_-\dagger}c^{e_-},
\\
\hat{U}=\left(\begin{array}{ccc}
U_h & U_{hh} & U_{eh}\\
U_{hh} & U_h & U_{eh}\\
U_{eh}  & U_{eh} & U_e/2
\end{array} \right),
\end{gathered}
\label{Hint}
\end{equation}
where $\hat{b}_\kk = (c^{h_1}_{-\kk\downarrow}c^{h_1}_{\kk,\uparrow};\; c^{h_2}_{-\kk\downarrow}c^{h_2}_{\kk,\uparrow};\;c^{e_1}_{-\kk\downarrow}c^{e_1}_{\kk,\uparrow}+c^{e_2}_{-\kk\downarrow}c^{e_2}_{\kk,\uparrow})$ and $c^{e_-}=c^{e_1}_{-\kk\downarrow}c^{e_1}_{\kk,\uparrow}-c^{e_2}_{-\kk\downarrow}c^{e_2}_{\kk,\uparrow}$.
As the quantities $\mu_{\alpha}$ are not independent, we shall use the following relations:
\begin{align}
\mu_{h_1}+\mu_e = E_{F_{h_1}}+E_{F_e} \equiv E_D\\
\mu_{h_1}-\mu_{h_2} = E_{F_{h_1}}-E_{F_{h_2}}\equiv E_{h_1h_2}.
\end{align}
Hole doping corresponds to decreasing $E_{F_e}$ while keeping $E_{h_1h_2}$ and $E_D$ constant.

Introducing the order parameters $\DDho$, $\DDht$, $\DD_{e_1}$, and $\DD_{e_2}$, it is easy to see from  (\ref{Hint}) that there the $\DD_{e_1} = -\DD_{e_2}$ channel is decoupled from the others and has no attraction. Thus we can set $\DD_{e_1}=\DD_{e_2}=\DD_{e}$ and obtain the following set of self-consistent equations (a rigorous derivation can be obtained by Hubbard-Stratonovich (HS) decoupling of the field $(\DD_{e_1}+\DD_{e_2})/2$):
\begin{align}
\DDho&=-\frac{\vt_h}{2}\DDho L_{h_1}-\frac{\lambda_{hh}}{2}\DDht L_{h_2}-\DDe\lambda_{eh}L_e,\nn\\
\DDht&=-\frac{\vt_h}{2}\DDht L_{h_2}-\frac{\lambda_{hh}}{2}\DDho L_{h_1}-\DDe\lambda_{eh}L_e,\label{GapEquations}\\
\DDe&=-\frac{\vt_e}{2}\DD_e L_e-\frac{\lambda_{eh}}{2}\left(\DDho L_{h_1}+\DDht L_{h_2}\right),\nn
\end{align}

where we have introduced the dimensionless quantities

\begin{align}
\vt_{\alpha}&=N_0U_{\alpha},\quad
\lambda_{\alpha\gamma}=
N_{0}U_{\alpha\gamma},\nn\\
L_{\alpha}&=
\int_0^\Lambda d\epsilon_\textbf{k}\frac{\tanh\left(\frac{E^{\alpha}_{\textbf{k}}}{2T}\right)}{E^{\alpha}_{\textbf{k}}}\nn,
\end{align}
with $N_0=\frac{m S}{2\pi}$ being the 2D density of states where $S$ is the area of the 2D system and $E^{\alpha}_{\textbf{k}}=\sqrt{(\xi^{\alpha}_{\textbf{k}})^2+|\Delta_{\alpha}|^2}$ is the quasiparticle spectrum. Note that the system of equations (\ref{GapEquations}) is not a closed one as $\mu_\alpha$ (the chemical potential) depend on the total number of particles $N$. $N$ is given by the amount of electrons minus that of holes. As $E_{F_{\alpha}}$ are defined for normal state at $T=0$, we can express $N$ through $E_{F_{e}}$ and use it as a doping variable. In the superconducting state, we obtain

\begin{align}
N=&4E_{F_e}\Theta(E_{F_e})-2\left(E_{F_{h_1}}+E_{F_{h_2}}\right)\nn\\=-&\int_{0}^{\infty}d\epsilon_{\textbf{k}}\Bigg\{2\frac{\xi^e_{\textbf{k}}\tanh\left(\frac{E^e_{\textbf{k}}}{2T}\right)}{E^e_{\textbf{k}}}+\frac{\xi^{h_1}_{\textbf{k}}\tanh\left(\frac{E^{h_1}_{\textbf{k}}}{2T}\right)}{E^{h_1}_{\textbf{k}}}\nn\\
&+\frac{\xi^{h_2}_{\textbf{k}}\tanh\left(\frac{E^{h_2}_{\textbf{k}}}{2T}\right)}{E^{h_2}_{\textbf{k}}}\Bigg\}\label{Eq:NumberEquation}
\end{align}

\section{Doping Dependence of the \boldmath{$s+is$} state\label{sec_2}}
\begin{figure}[h]
	\renewcommand{\baselinestretch}{1.0}
	\includegraphics[angle=0,width=1\linewidth]{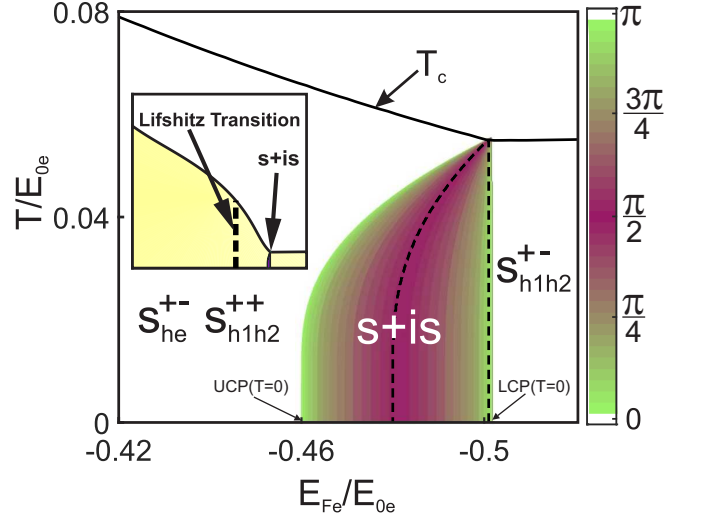}
	\caption{Phase diagram as function of temperature ($T$) and doping ($E_{F_e}$). Color encodes the phase difference $\phi$ between the order parameters of hole bands modulo $\pi/2$. Black dashed line marks the $s+is$ state boundaries if one ignores (\ref{Eq:NumberEquation}) and takes $\mu_i=E_{F_i}$. The parameters used are: $\lh/\lhe=0.95$, $\lhe=0.15$, $\vt_{\alpha}=0$, $E_{h_1h_2}=0$, $\Lambda=3000$, $E_D=110$, $E_{0e}=10.7$ and $E_{0h}=4.8\times10^{-3}$ in arbitrary energy units.}%
	\label{Fig:FinalPhasediagram}
\end{figure}

One of the main results of our analysis is that inclusion of the additional equation for the renormalization of the chemical potential [Eq.\ref{Eq:NumberEquation}] has a profound effect on the $s+is$ superconducting phase. In Fig.\ref{Fig:FinalPhasediagram}, we present a typical phase diagram obtained from numerical solutions of the self-consistency equations (\ref{GapEquations}) and (\ref{Eq:NumberEquation}) with an iterative procedure. We have also checked that the solutions obtained do correspond to minima of the free energy by analyzing the second variation matrix of the action, as discussed in detail in Appendix \ref{app}. Encoded in color is the phase difference $\phi$ between $\DD_{h_1}$ and $\DD_{h_2}$ modulo $\pi/2$ and the following scales are used as energy units:

    \begin{equation}
    \begin{gathered}
    E_{0e}:=2\Lambda\text{e}^{\left(-\frac{\lhe^2}{\lh}-\frac{\vt_e}{2}\right)^{-1}},
    \\
    E_{0h}:=2\Lambda\text{e}^{-2/(\lh-\vt_h)}.
    \end{gathered}
    \label{E_0s}
    \end{equation}

\begin{figure}[h]
	\includegraphics[width=\linewidth]{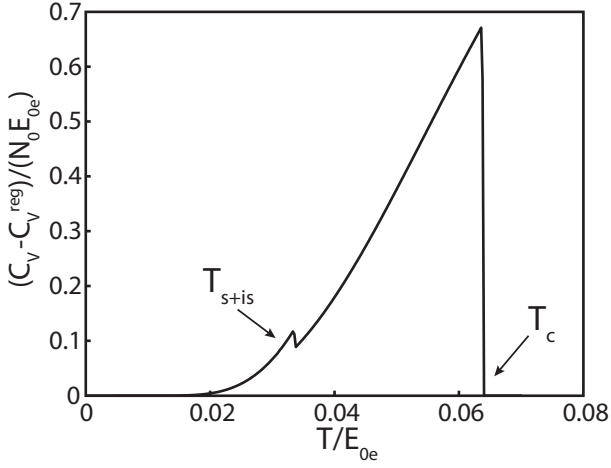}
	\caption{The discontinuous contribution to the specific heat $C_V-C_V^{reg}$ [Eq. \ref{Eq:Cv}] as a function of temperature for $E_{Fe}/E_{0e}= -0.467$. Marked are the critical temperatures for the transition into usual superconducting state $T_c$ and the $s+is$ state ($T_{s+is}$). The parameters of the model are taken to be the same as in Fig.\ref{Fig:FinalPhasediagram}.}
	\label{Fig:cv}
\end{figure}

One can see that there exists a region where $\phi$ is not a multiple of $\pi$, corresponding to the $s+is$ state. In this state the initial $Z_2$ symmetry $\phi\leftrightarrow -\phi$ of the system is broken and a second order phase transition is expected at the onset of $\phi\neq 0,\pi$. A clear thermodynamic signature of the $s+is$ state would then be the presence of a specific heat discontinuity {\it below} $T_c$. To verify this, we calculate $C_V=-T\left(\frac{d^2 \Omega}{d T^2}\right)_{\mu,V}$, approximating the grand canonical potential by the value of the action at the saddle point (\ref{app_gc_pot}). One obtains
\begin{equation}
 C_V =C_V^{reg}(T)+\sum_{\alpha}\frac{\partial |\Delta_\alpha|^2}{\partial T}\int_{-\mu_\alpha}^\infty
 \frac{N_0 d \xi}{4 T \cosh^2\frac{\sqrt{\xi^2+|\Delta_\alpha|^2}}{2T}},
\label{Eq:Cv}
\end{equation}
where $C_V^{reg}(T)$ is a continuous function of temperature. In Fig.\ref{Fig:cv}, we present $[C_V-C_V^{reg}](T)$ calculated for $E_{Fe}/E_{0e}= -0.467$.

One can see that the discontinuity at the $s+is$ transition is present but its value is much smaller than the one at $T_c$. The general reason for that can be deduced directly from the expression (\ref{Eq:Cv}). Below $T_c$ the value of the integral in the right-hand side. is suppressed by a factor $\sim e^{-|\Delta_\alpha(T)|/T}$ with respect to its value at $T_c$. Consequently, one can expect that for $T_{s+is}$ well below $T_c$ the discontinuity in the specific heat to be relatively small. Together with inevitable sample inhomogeneity\cite{Yeoh.2011,Song.2013}, which can significantly smear $T_{s+is}$ due to local variation of the doping level, this provides a possible explanation for the absence of features in $C_V$ at $T_{s+is}$ in the recent experimental data \cite{Grinenko.2017}.

For every temperature the $s+is$ region is bounded by a lower (right in Fig.~(\ref{Fig:FinalPhasediagram})) and an upper (left in Fig.~(\ref{Fig:FinalPhasediagram})) critical points (LCP, UCP) with respect to $E_{Fe}$, the region being largest at zero temperature and shrinking into a single point for $T= T_c$. The latter feature is actually due to $E_{h1h2}=0$ and in general case the $s+is$ region is bound from above by a temperature lower than $T_c$ as is shown in Sec.\ref{sec_3a}. Nevertheless, the onset temperatures of the $s+is$ state in Fig.\ref{Fig:FinalPhasediagram} are still well separated from $T_c$ for most dopings, consistent with the experimental observations of Ref. \onlinecite{Grinenko.2017}. Note also that the electron bands are incipient in the $s+is$ region; i.e.; it is located after the Lifshitz transition.

The black dashed line marks the phase boundary of the $s+is$ state if the chemical potential renormalization due to (\ref{Eq:NumberEquation}) is neglected, i.e., $\mu_{\alpha}=E_{F_{\alpha}}$. It is evident from Fig. \ref{Fig:FinalPhasediagram} that the renormalization of the chemical potential by the superconducting gap leads to a broadening of the $s+is$ region. Peculiarly, while the UCP shifts considerably when including (\ref{Eq:NumberEquation}), the LCP remains practically the same. However, for $T\rightarrow T_c$ LCP and UCP converge seemingly to the same point, regardless of the chemical potential renormalization being included or not. To understand these qualitative features, we study the positions of UCP and LCP analytically at $T=0$ and $T=T_c$.

\subsection{Analytical calculation for the \boldmath{$s+is$} order critical points}
Our calculation will follow a route, similar to that in Ref. \cite{Maiti.2013}, where LCP and UCP for variable $\lambda_{eh}$ have been found. Let us rewrite the system (\ref{GapEquations}) in the following form:
\begin{align}
\DDho+\DDht&=-\frac{\lh+\vt_h}{2}\left(\DDho L_{h_1}+\DDht L_{h_2}\right)-2\DDe\lhe L_e,\label{Eq:1}\\
\DDho-\DDht&=-\frac{\lh-\vt_h}{2}\left(\DDht L_{h_2}-\DDho L_{h_1}\right),\label{Eq:2}\\
\DDe&=-\frac{\vt_e}{2}\DD_eL_e-\frac{\lambda_{eh}}{2}\left(\DDho L_{h_1}+\DDht L_{h_2}. \right)\label{Eq:3},
\end{align}
We choose the global phase of the order parameters such that $\DDe$ is real. Considering then the imaginary and real parts of the system [\ref{Eq:1}-\ref{Eq:3}] separately yields
\begin{gather}
\sin(\phi_1)|\DDho|+\sin(\phi_2)|\DDht|=0,\label{Eq:reim2}
\\
L_{h_1}=L_{h_2}=\frac{2}{\lh-\vt_h},\label{Eq:reim1}
\\
L_e=\left(\frac{\lhe^2}{\lh}-\frac{\vt_e}{2}\right)^{-1},\label{Eq:reim4}
\\
|\DDho|\cos(\varphi_1)+|\DDht|\cos(\varphi_2)=-2\kappa\DD_e,\label{Eq:reim3}
\end{gather}
where we define the dimensionless constant $\kappa$
\begin{align}
\kappa:=\lhe\frac{\lh-\vt_h}{2\lhe^2-\vt_e\lh}.
\end{align}
Note that relations (\ref{Eq:reim1} and \ref{Eq:reim4}) hold for all temperatures.

\subsubsection{UCP and LCP at T=0}
At zero temperature, we find for (\ref{Eq:NumberEquation}) and $L_{\alpha}$
\begin{align}
N=&4E_{F_e}\Theta(E_{F_e})-2\left(E_{F_{h_1}}+E_{F_{h_2}}\right)\nn\\=&2\left(\mu_e+\sqrt{\mu_e^2+|\DDe|^2}\right)
-\mu_{h_1}-\sqrt{\mu_{h_1}^2+|\DDho|^2}\nn\\&-\mu_{h_2}
-\sqrt{\mu_{h_2}^2+|\DDht|^2},\label{Eq:NumberEquationT=0}\\
L_{\alpha}&=
\ln\left(\frac{2\Lambda}{\sqrt{\mu_{\alpha}^2+|\DDa|^2}-\mu_{\alpha}}\right),
\end{align}
and thus from (\ref{Eq:reim1}) and (\ref{Eq:reim4}) we get
\begin{equation}
\begin{gathered}
|\DDho| = \sqrt{E_{0h}^2+2 \mu_{h_1}E_{0h}},\\
|\DDht| = \sqrt{E_{0h}^2+2 (\mu_{h_1}-E_{h_1h_2})E_{0h}},\\
\mu_e = \frac{|\DDe|^2-E_{0e}^2}{2 E_{0e}}.
\end{gathered}
\label{DDh+mue}
\end{equation}
Using the relations above, we can also rewrite Eq.(\ref{Eq:NumberEquationT=0}) as
\begin{gather*}
N=4E_{F_e}\Theta(E_{F_e})-2\left(E_{F_{h_1}}+E_{F_{h_2}}\right)
\\
=2\frac{|\DDe|^2}{E_{0e}}
-2\mu_{h_1}-E_{0h}-2\mu_{h_2}-E_{0h},
\end{gather*}
leading to
\begin{equation}
E_{F_e} = 2 \mu_e+\frac{E_{0e}}{2}-\frac{E_{0h}}{2}.
\label{Eq:NumberEquation+}
\end{equation}

\subsubsection{Identical hole bands}
To study the effect of chemical potential renormalization let us first consider the hole bands to be identical ($E_{h_1h_2}=0$). It is evident from (\ref{DDh+mue}) that $|\DDho|=|\DDht|$ and from (\ref{Eq:reim2}) we find $\phi_{h_1}=-\phi_{h_2}\equiv\phi/2$. The system [\ref{Eq:1}-\ref{Eq:3}] has then three types of solutions:

\begin{itemize}
	\item[1.]  $s_{h1h2}^{+-}$: $\DDho=-\DDht$ ($\phi = \pi$);
	\item[2.]  $s+is$: $|\DD_{h_1}| = |\DD_{h_2}|$ $\phi\neq0,\pi$;
	\item[3.]  $s_{h1h2}^{++}$: $\DDho=\DDht$($\phi = 0$).
\end{itemize}

LCP and UCP are then found by matching the $s+is$ solution with $s_{h1h2}^{+-}$ and $s_{h1h2}^{++}$ one, respectively.

$LCP$: $s+is$ state coincides with $s_{h1h2}^{+-}$ when $\phi=\pi$. It follows then, that $\DD_e=0$ and from (\ref{DDh+mue}) and (\ref{Eq:NumberEquation+}) we get:
\begin{gather*}
    \mu_{e_{\text{min}}}\equiv-\frac{E_{0e}}{2},
    \\
    E_{F_e}^{min}\equiv-\frac{E_{0e}}{2}-\frac{E_{0h}}{2}.
\end{gather*}
Thus, LCP is shifted to larger hole dopings due to the chemical potential renormalization. However, for $E_{0h}\ll E_{0e}$ the correction to the LCP is insignificant, which is consistent with Fig.~\ref{Fig:FinalPhasediagram}. It is important to note that according to this result the electron bands are always incipient at the LCP.\newline

$UCP$:  The transition from $s+is$ to $s_{h1h2}^{++}$ takes place for $\phi=0$. From (\ref{Eq:reim3}) it follows that there is a sign change of the order parameter between hole and electron pockets. Inserting then the values of $|\Delta_h|,\;|\Delta_e|$ obtained from (\ref{Eq:reim1}), (\ref{Eq:reim4}), into (\ref{Eq:reim3}), we arrive at
 \begin{align}
	   1=\kappa\frac{\sqrt{E_{0e}^2+2\mu_eE_{0e}}}{\sqrt{E_{0h}^2+2(E_D-\mu_e)E_{0h}}},
\end{align}
where we used $\mu_h=E_D-\mu_e$. We can now find the corresponding $\mu_e$:
\begin{gather*}
 \mu_{e_{\text{max}}}\equiv\frac{E_{0h}^2+2E_DE_{0h}-\kappa^2E_{0e}^2}{2\kappa^2E_{0e}+2E_{0h}}.
 \\
 E_{F_e}^{\text{max}}\equiv \frac{
 	(1-\kappa^2)E_{0e}E_{0h}+E_{0h}^2+4E_DE_{0h}-\kappa^2 E_{0e}^2
 }
 {2\kappa^2E_{0e}+2E_{0h}}.
\end{gather*}

Collecting the results for $ \mu_e^{LCP}$ and $ \mu_e^{UCP}$ we find that the $s+is$ state is confined to the region:

\begin{align}
	-\frac{1}{2}E_{0e}<
	\mu_e< \frac{E_{0h}^2+2E_DE_{0h}-\kappa^2E_{0e}^2}{2\kappa^2E_{0e}+2E_{0h}}\label{Limits}.
\end{align}

\begin{figure}[h]
	\centering
	\includegraphics[width=0.45\textwidth]{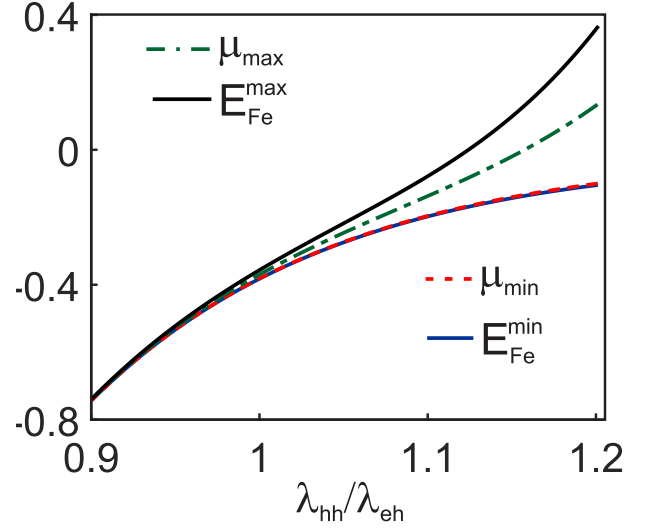}		
	\caption{The lower and the upper boundaries of the $s+is$ region for chemical potential $\mu_{e_{\text{min}}}$ and $\mu_{e_{\text{max}}}$ and doping $E_{F_e}^{max}$ and $E_{F_e}^{min}$ as functions of the ratio $\lh/\lhe$ taking $\vt_{\alpha}=0$.}	
	\label{Fig:Boundaries}
\end{figure}

Note that for $E_D\gg \kappa^2 E_{0e}^2/E_{0h}$, $E_{F_e}$ is positive at UCP, and $s+is$ state extends into the moderately doped region where electron bands cross the Fermi energy. In Fig.\ref{Fig:Boundaries}, we present the evolution of $\mu_{e_{\text{min}}}$, $\mu_{e_{\text{max}}}$, $E_{F_e}^{\text{min}}$, and $E_{F_e}^{\text{max}}$ for $\vt_{\alpha}=0$ as functions of the ratio $\lh/\lhe$.

One can see, that the $s+is$ region shrinks with decreasing $\lh$. As for small values of $\lh$ we have $E_{0h}\ll E_{0e}$, and we can expand $\mu_{e_{\text{max}}}$ in $\frac{E_{0h}}{E_{0e}}$:
\begin{align}
\mu_{e_{\text{max}}}\approx-\frac{1}{2}E_{0e}+\frac{1}{2\kappa^2}\frac{E_{0h}}{E_{0e}}\left(2 E_D+E_{0e}\right)\geq\mu_{e_{\text{min}}}.\label{Boundary expansion}
\end{align}
Equation (\ref{Boundary expansion}) shows that $\mu_{e_{\text{max}}}>\mu_{e_{\text{min}}}$ holds up to arbitrary small values of $\lh$. As a consequence, whatever are the coupling constants, the transition from $s_{h1h2}^{++}$ to $s_{h1h2}^{+-}$ always occurs through the $s+is$ state at $T=0$. However, the width of the $s+is$ region is sensitive to the values of the coupling constants.

Let us now compare the effect Eq.(\ref{Eq:NumberEquation}) has on LCP and UCP. For the true doping variable $E_{F_e}$, we need to take into account the chemical potential renormalization coming from Eq.~(\ref{Eq:NumberEquation}). In the physically relevant limit $E_{0h}\ll E_{0e}$, we have

\begin{align}
\mu_{e_{min}}-\frac{E_{0h}}{2}<\EFe\leq \enspace\mu_{e_{\text{max}}}&-\frac{E_{0h}}{2}\nonumber\\
+&\frac{1}{2\kappa^2}\frac{E_{0h}}{E_{0e}}\left(2 E_D+E_{0e}\right).\label{Boundaries+correction}
\end{align}

The observation that the correction to the upper compared to the lower boundary  is much more pronounced can be explained by Eq.(\ref{Boundaries+correction}). Using the values taken in Fig. \ref{Fig:FinalPhasediagram}, we see that the second term in the upper boundary shift clearly dominates over $-E_{0h}/2$ in that case.

\subsubsection{ Effect of hole band offset on the {$s+is$} state\label{sec_3}}
Now we consider the case $E_{h_1h_2}=E_{F_{h1}}-E_{F_{h2}} \neq 0$. It is seen from (\ref{Eq:NumberEquation+}) that the corrections to $E_{F_e}$ due to $E_{h_1h_2}$ are simply:
\[
\delta E_{F_e} = 2\delta\mu_e=\frac{\delta|\DDe|^2}{E_{0e}}.
\]

$LCP$: We have $\phi_2=\phi_1+\pi$ and it immediately follows from (\ref{Eq:reim2}) that $\phi_1$ is either $0$ or $\pi$ and from (\ref{Eq:reim3}) that $|\DDe| =\frac{1}{2 \kappa} (|\DDho|-|\DDht|)\approx \frac{ E_{h_1h_2} E_{0h}}{2 \kappa \DDho}$. Note that in the case $E_{0h}\ll\DDho$ (BCS limit for hole bands) this expression is valid even for $E_{h_1h_2}\sim\DDho$. We obtain then
\[
\delta E_{F_e}^{LCP}\approx
\frac{1}{4\kappa^2}\frac{ E_{h_1h_2}^2 E_{0h}}{E_{0e}(2 E_D +E_{0h}+E_{0e})}.
\]

We remark that this result remains valid even for moderate values of $E_{h_1h_2}$, if $E_{0h}\ll E_{0e},\;E_D$.

$UCP$: We have $\phi_2=\phi_1$ and it immediately follows from (\ref{Eq:reim2}) that $\phi_1$ is either $0$ or $\pi$ and from (\ref{Eq:reim3}) that $|\DDe| =\frac{1}{2\kappa} (|\DDho|+|\DDht|)$. The result is
\begin{gather*}
\delta E_{F_e}^{UCP}\approx
-\frac{1}{\kappa^2}\frac{E_{h_1h_2}E_{0h}}{E_{0e}}.
\end{gather*}
This correction $E_{h_1h_2}$ is not suppressed by $E_D$, but comparing with (\ref{Boundaries+correction}) we see that it is still much smaller than the effect of the chemical potential renormalization.

Overall, the effect of the hole band offset at $T=0$ is to shrink the $s+is$ region; however, for $E_D\gg E_{0e},E_{0h}$ this effect is insignificant in comparison to the one introduced by the chemical potential renormalization.
\subsubsection{Critical doping at $T=T_c$\label{sec_3a}}
At $T=T_c$ we can linearize the self-consistency equations (\ref{Eq:reim1}) and (\ref{Eq:reim4}) and obtain (assuming $|\mu_{e,h1,h2}|\gg T_c$):
\begin{gather*}
\ln\left(1.13^2\frac{\Lambda\mu_{h_1}}{T_c^2}\right)=\ln\left(1.13^2\frac{\Lambda\mu_{h_2}}{T_c^2}\right)\nn
=\frac{2}{\lh-\vt_h},
\\
L_e=\ln\left(\frac{\Lambda}{|\mu_e|}\right)=\left(\frac{\lhe^2}{\lh}-\frac{\vt_e}{2}\right)^{-1},\nn
\end{gather*}
with the neglected terms being exponentially small $\sim\exp\{-|\mu_{e,h1,h2}|/T_c\}$. It follows then that $s+is$ state persists up to $T_c$ only if $\mu_{h_1}=\mu_{h_2}$. Moreover, it is confined to a single point:
\begin{align}
\mu_e^{\text{min}}=\mu_e^{\text{max}}:=-\frac{E_{0e}}{2}.
\end{align}
$T_c$ is equal to $0.8 \sqrt{E_{0h}\mu_h}$ and one can see that for the parameters used in Fig.\ref{Fig:FinalPhasediagram} $T_c$ is indeed smaller than both $\mu_h$ and $\mu_e$. From (\ref{Eq:NumberEquation}) taken at $T_c$ we find that:
\begin{align}
E_{F_e}=\mu_e+O\left(T_c \text{e}^{\frac{-|\mu_{e}|}{T_c}}\right)
\end{align}
and thus
\begin{align}
E^{\text{min}}_{F_e}=E^{\text{max}}_{F_e}:=-\frac{E_{0e}}{2}+
O\left(T_c \text{e}^{\frac{-|\mu_{\alpha}|}{T_c}}\right)
\label{Eq:T=TcCp}.
\end{align}
We see that the $s+is$ region shrinks to a single point at $T_c$ for equal hole bands. The doping level for this point is close to $-E_{0e}/2$ with only exponentially small corrections. Thus for $E_{0h}\ll E_{0e}$ the $s+is$ point at $T_c$ and LCP at zero temperature should be close to each other, as is the case in Fig. \ref{Fig:FinalPhasediagram}. For unequal hole bands we find that $s+is$ state onset is below $T_c$ for all dopings. Such a separation between the $s+is$ state onset and $T_c$ is actually observed in a recent experiment \cite{Grinenko.2017}).

\section{STM-signatures of \boldmath{$s+is$}-State\label{sec_4}}
Recently it has been proposed\cite{Hirschfeld.2015} that the sign structure of the order parameter in a multiband system can be extracted from the Fourier transform of the local density of states (QPI pattern near an impurity in the superconducting state). The QPI intensity integrated over the wave vectors corresponding to scattering between two bands has been shown to have a dependence on energy very different for $s^{+-}$ and $s^{++}$ scenarios, leading to a strong enhancement of the integrated response for the $s^{+-}$ but not the $s^{++}$ case. This method has been recently successfully applied to confirm the sign-changing nature of the order parameter in FeSe\cite{Sprau.2016}, where the superconducting gaps are also extremely anisotropic.
More recently, the method has been also applied to (Li$_{1-x}$Fe$_x$)OHFe$_{1-y}$Zn$_y$Se with only electron Fermi surface pockets\cite{Du2017}. For the 122 doped systems, such an experiment has not been yet performed; however, one would expect that such a test is feasible in these compounds as well, given the availability of the high-quality STM data\cite{Wang2012,Shan2012}.

Here we show that the QPI patterns contain information on the phase difference between the order parameters of different bands for the case when it is not $0$ or $\pi$ and even if one of the bands is incipient. Based on the results, we provide several methods for detection of the $s+is$ state. As the discussion here has qualitative character, we present the results obtained in the Born approximation assuming weak impurity potential at $T=0$ and ignore anisotropy. Here we consider nonmagnetic and Andreev impurities only, as the magnetic ones do not contribute to the density of states in the Born approximation. Let us first concentrate on the scattering between the two hole bands in the $s+is$ state. The calculations are similar to those in Ref. \cite{Hirschfeld.2015} and we present the details in Appendix B. It is important to notice that for a single wavevector ${\bf q}$ corresponding to interband scattering, there are contributions from both $h1\to h2$ and $h2\to h1$ scattering. First we consider an impurity of the charge type with the potential given by $t_3\tau_3$, with $\tau_i$ being matrices in Nambu space. One obtains:
\begin{equation}
\begin{gathered}
\delta\rho^{h1h2}_{ch}(\omega) =
2 t_3\pi \rho_{h1}\rho_{h2} \text{sgn}(\omega)
\\
\times{\rm Im}
\frac{2 \omega^2 - 2|\Delta_{h1}| |\Delta_2|\cos(\phi_1-\phi_2)}
{\sqrt{\omega^2-|\Delta_{h1}|^2+i\delta}\sqrt{\omega^2-|\Delta_{h2}|^2+i\delta}}.
\end{gathered}
\label{QPI_h1h2_ch}
\end{equation}

\begin{figure}[h]
	\includegraphics[width=8cm,trim=0cm 0cm 0cm 0cm, clip=true]{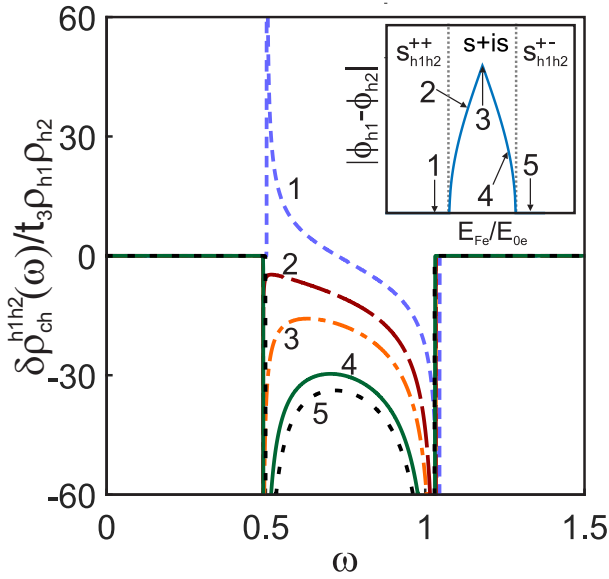}
	\caption{
Interband density of states $\delta\rho^{h1h2}_{ch}(\omega)/t_3 \rho_{h1}\rho_{h2}$ [(\ref{QPI_h1h2_ch}) with $\delta=10^{-6}$] for scattering between the hole bands. Inset shows $\phi_1-\phi_2$ as a function of $E_F/E_{0e}$ for $T=0$ obtained from (\ref{GapEquations})and (\ref{Eq:NumberEquation}) using the same parameters as in Fig.\ref{Fig:FinalPhasediagram} except for $E_{h1h2}=90$ in arbitrary units. Arrows mark the dopings for which $\delta\rho$ curves are presented.
}
\label{Rho_inter}
\end{figure}

In Fig.\ref{Rho_inter} we present the evolution of $\rho^{h1h2}_{ch}(\omega)$ through the $s+is$ state for $\omega>0$. The doping dependence of the order parameters has been obtained from Eqs. (\ref{GapEquations}) and (\ref{Eq:NumberEquation}). The parameters of the model have been taken to yield $|\Delta_{h2}/\Delta_{h1}|\approx0.5$ motivated by the results of ARPES measurements for $\text{K}\text{Fe}_2\text{As}_2$\cite{Okazaki.2012}. One can see that the result evolves continuously from $s^{++}$-like before the $s+is$ region to $s^{+-}$-like after it. Obtaining the value of $\phi_1-\phi_2$ requires then, in principle, fitting the whole curve, which can be rather complicated provided that the exact impurity potential is not known. On the other hand, the edges of the region where the contribution is nonzero (corresponding to $\omega=|\Delta_{h1}|$, $\omega=|\Delta_{h2}|$) are practically unchanged throughout the $s+is$ state, while the form of the curve in between changes dramatically, meaning that QPI pattern is quite sensitive to the value of $\phi_1-\phi_2$. Moreover, as $\phi_1-\phi_2$ depends on temperature (see Fig.\ref{Fig:FinalPhasediagram}), abrupt changes in the QPI pattern with temperature could be also considered as a signature of the $s+is$ state.

Local suppression of superconductivity by an impurity\cite{Hettler.1999} or an individual vortex in a disordered vortex lattice\cite{Pereg.2008} constitutes an Andreev scatterer. The impurity potential in this case is given by $t_A\tau_1$ for the case when order parameters can be taken real. In the $s+is$ state this is not so, so we use the general form $t_A(\alpha \tau_1+\beta\tau_2)$ with real $\alpha,\;\beta$ and fix their values so that the answer is gauge invariant and goes over to $t_A\tau_1$ in the $s_{h1h2}^{++}$ limit. The result is then:
\begin{equation}
\delta \rho^{h1h2}_{A}(\omega) =
4 t_A \pi \rho_1\rho_2
{\rm Im}
\frac{|\omega|(|\Delta_1|+|\Delta_2|)\cos[(\varphi_1-\varphi_2)/2]}
{\sqrt{\omega^2-|\Delta_1|^2+i\delta}\sqrt{\omega^2-|\Delta_2|^2+i\delta}}.
\label{QPI_h1h2_A}
\end{equation}
An interesting feature of this result is that the response vanishes in the $s_{h1h2}^{+-}$ case but is finite throughout the $s+is$ state. This leads to a sufficient qualitative criterion for the $s+is$ state detection: Observation of an $s^{+-}$-like pattern near a charge impurity together with a nonzero $\delta \rho^{\text{inter}}_{A}(\omega)$ near an Andreev scatterer at the same doping suggests the presence of the $s+is$ state.

Let us now move onto $e-h$ scattering. As the electron bands in the $s+is$ state are likely to be incipient or have $|\mu_e|\sim|\Delta_e|$, the BCS-like expressions for the momentum-integrated Green's function are no longer valid. However, in 2D the momentum integration can be performed exactly (assuming a large energy cutoff scale $\Lambda$). For a charge impurity, only the component odd in frequency contains the phase information. The result is
\begin{widetext}
\begin{equation}
\begin{gathered}
\rho^{\text{eh}}_{ch}(\omega)^{odd} =
-2 t_3 \rho_e\rho_h \text{sgn}(\omega)
{\rm Im}
\left\{
\left(
i\pi+
\log\left[\frac{-\mu_e+\sqrt{\omega^2-|\Delta_e|^2+i\delta}}{\mu_e+\sqrt{\omega^2-|\Delta_e|^2+i\delta}}\right]
\right)
\frac{i(2 \omega^2 - |\Delta_h| |\Delta_e|\cos(\phi_h-\phi_e) )}
{2\sqrt{\omega^2-|\Delta_h|^2+i\delta}\sqrt{\omega^2-|\Delta_e|^2+i\delta}}
\right\}
\\
+\frac{t_3 \rho_1\rho_2 \text{sgn}(\omega)}{\pi}
{\rm Im}\left\{
\log\left[\frac{\Lambda^2}{\mu_h^2-\omega^2+|\Delta_h|^2-i\delta}\right]
\log\left[\frac{\Lambda^2}{\mu_e^2-\omega^2+|\Delta_e|^2-i\delta}\right]
\right\}.
\end{gathered}
\label{QPI_eh_ch}
\end{equation}
\end{widetext}

\begin{figure}[h]
\includegraphics[width=8.5cm,trim=0cm 0cm 0cm 0cm, clip=true]{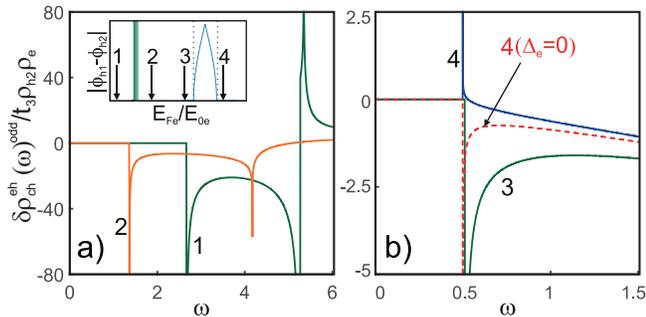}
\caption{
Doping evolution of the interband density of states $\delta\rho^{h2e}_{ch}(\omega)/t_3 \rho_{h2}\rho_{e}$ [(\ref{QPI_eh_ch}) with $\delta=10^{-6}$] for scattering between the electron and the smaller hole (h2) bands.
(a) Dopings close to the Lifshitz transition for the electron band. (b) Dopings around the $s+is$ region. Red dashed line is for the same doping as line $4$ but with $\Delta_e=0$. Inset in pabel (a) shows $\phi_1-\phi_2$ as a function of $E_F/E_{0e}$ obtained from (\ref{GapEquations}) and (\ref{Eq:NumberEquation}) using the same parameters as in Fig.\ref{Rho_inter}. Arrows mark the dopings for which $\delta\rho$ curves are presented}
\label{Rho_eh}
\end{figure}
The results for different doping levels are presented in Fig.\ref{Rho_eh} for the scattering between the electron and the smaller hole (h2) bands. One can see that the square-root singularity for $\mu_e>0$ at $\omega=|\Delta_e|$ ceases to exist in the incipient case $\mu_e<0$. However, a weaker logarithmic singularity is present then at $\omega=\sqrt{|\Delta_e|^2+\mu_e^2}$. The curve evolves throughout $s+is$ state in a similar manner to h1-h2 scattering. The results for h1-e scattering are comparable, except that at dopings beyond the $s+is$ region the phase difference between the order parameters at h1 and e is equal to $\pi$ instead of $0$. This suggests a complementary way to measure $\phi_{h2}$ and $\phi_{h1}$ by extracting $\phi_{h1}-\phi_e$ and $\phi_{h2}-\phi_e$ from the e-h interband density of states.

One could ask if a presence of a superconducting gap at the incipient electron band is at all distinguishable in the result. In Fig.\ref{Rho_eh}(b) we compare the patterns for h2-e scattering at the same doping level with finite $\Delta_e$ obtained from self-consistency equations (solid blue line) and with $\Delta_e=0$. The qualitative difference is due to the order parameter on h2 having the same sign as the one on electron band leading to a sign-changing $s^{++}$-like pattern. Taking $\Delta_e=0$ however leads to an answer not crossing zero and thus the singularity at $\omega=\Delta_{h2}$ changes sign. Thus STM-QPI can be also a useful tool for detecting superconducting gaps at incipient bands, which is a rather nontrivial task in, e.g., ARPES experiments.

\section{Conclusion\label{concl}}

We have studied superconducting pairing in a four-band model as a function of hole doping, where the electron bands can become incipient after a Lifshitz transition. The interactions have been chosen to study the interplay between two superconducting channels of $s$-wave symmetry: With a sign change between the hole and electron pockets ($s_{eh}^{+-}$) [phases at the two hole (electron) pockets being the same], dominant close to optimal doping, and with a sign change between the hole pockets ($s_{h1h2}^{+-}$), leading in the overdoped region.

The phase diagram of the model has been analyzed in the mean-field approximation including BCS-BEC effects\cite{Leggett.1980, Chubukov.2016}. We have found that the crossover from $s_{eh}^{+-}$ to $s_{h1h2}^{+-}$ at low temperatures always occurs via an intermediate $s+is$ state, characterized by a phase difference between the order parameters at the two hole pockets not equal to a multiple of $\pi$. We have shown that the $s+is$ state {\it always} extends into the region where electron bands are incipient.
Additionally, heat capacity anomalies at the transition to the $s+is$ state have been found to be suppressed at temperatures below $T_c$. These findings are in line with the recent experimental data suggesting the presence of $s+is$ state in overdoped $\text{Ba}_{1-x}\text{K}_x\text{Fe}_2\text{As}_2$ at $x=0.73$\cite{Grinenko.2017}, close to the doping where the Lifshitz transition is expected to occur\cite{Sato.2009,Hodovanets.2014}.

A large broadening of the $s+is$-region has been found due to the chemical potential renormalization in the superconducting state (BCS-BEC corrections), previously unaccounted for. This is encouraging for future experiments on $\text{Ba}_{1-x}\text{K}_x\text{Fe}_2\text{As}_2$, as the doping is hard to control precisely due to clustering of K atoms in the lattice \cite{Yeoh.2011,Song.2013}. An energy offset between the $\Gamma$-centered hole pockets, on the contrary, narrows the $s+is$-region, and makes it disappear around $T_c$. The narrowing effect has been found, however, to be strongly suppressed for sufficiently large enough hole pockets. These observations give additional support to the presence of the $s+is$ state in the hole-overdoped pnictides.

Finally, we have studied the interband density of states in the $s+is$ state and to proposed several complementary approaches to detect this state in future STM-QPI experiments. Our results also suggest a general method to detect superconducting gaps on incipient bands.
\begin{acknowledgments}
We acknowledge the discussions with A.V. Chubukov, and P.J. Hirschfeld.
J.B. and I.E. were supported by the joint DFG-ANR Project (ER 463/8-1). K.B.E. acknowledges the financial support of the Ministry of Education and Science of the Russian Federation in the framework of Increase Competitiveness Program of
NUST “MISiS” (Nr. K2-2014-015).\newline
J.B. and P.A.V. contributed equally to this work.
\end{acknowledgments}

\appendix

\section{\label{app} STABILITY OF THE \boldmath{$s+is$} SOLUTION}

To show that the obtained solutions of  Eqs.(\ref{GapEquations}) and (\ref{Eq:NumberEquation}) are thermodynamically stable, one needs to show that the second variation matrix of the action is positive definite. An expression for $S$ can be obtained by decoupling the interaction (\ref{Hint}) in the path integral. A problem arises then because the conventional Hubbard-Stratonovich decoupling yields a diverging integral in the case of repulsive interaction, and the matrix in (\ref{Hint}) has one positive (repulsive) eigenvalue (let us consider for simplicity the case $U_e=0$, where the $\Delta_{e_1}=-\Delta_{e_2}$ channel does not arise). This has been noted previously in Ref. \cite{Maiti.2013} and a method to take the repulsive channel into account has been proposed for a two-band case with interband interaction in Ref. \cite{Fanfarillo.2009}. We use a different procedure, similar to the one used in Ref. \cite{Marciani.2013}, easily generalizable for the multi-band case and arbitrary interactions. First of all, one can bring the interaction Hamiltonian to the form $\sum_i \lambda_i \hat{e}_i^\dagger \hat{e}_i$, where  $\hat{e}_i$ is an eigenvector of matrix $\hat{U}$ composed of operators $\hat{b}_i$. The negative eigenvalues are then decoupled as usual, while for positive ones we use the following relation:
\begin{gather*}
e^{-\lambda b^* b } = \frac{1}{\widetilde{Z}} \int D \Delta^* D\Delta
e^
{
-\frac{|\Delta|^2}{\lambda}- i b^* \Delta - i \Delta^* b
},
\\
\widetilde{Z}=\int D \Delta^* D\Delta
e^
{
-(\Delta^* +i b^* \lambda )\frac{1}{\lambda}(\Delta + i \lambda b)
},
\end{gather*}
where $\int D \Delta^* D\Delta\equiv \int D [{\rm Re}\Delta] D [{\rm Im}\Delta]$. $\widetilde{Z}$ can be shown to be independent of $b$ by shifting the integration contour into the complex plane (${\rm Re}\Delta\to{\rm Re}\Delta-i\lambda {\rm Re}b$, ${\rm Im}\Delta\to{\rm Im}\Delta-i\lambda {\rm Im}b$):
\[
\widetilde{Z}=\int D \Delta^* D\Delta e^{-\frac{|\Delta|^2}{\lambda}}.
\]
Consequently, the decoupling for the full $3\times3$ interaction matrix takes the form:
\begin{widetext}
\begin{equation}
\begin{gathered}
e^{-\frac{1}{\hbar}\int_0^{\hbar\beta} d \tau b^\dagger(\tau) U b(\tau)} = \frac{1}{Z} \int D\tilde{\Delta} D\tilde{\Delta}^*
\exp
\left\{
\frac{1}{\hbar}\int_0^{\hbar\beta} d \tau
\left[
-
    b^\dagger(\tau)
    R^T
    \left(\begin{array}{ccc}
    1&  & \\
    &1&\\
     &  & i
    \end{array} \right)
    \tilde{\Delta}
\right.
\right.
\\
\left.
\left.
-
    \tilde{\Delta}^\dagger
    \left(\begin{array}{ccc}
    1&  & \\
    &1&\\
     &  & i
    \end{array} \right)
    R
    b(\tau)
+
    \tilde{\Delta}^\dagger
    \left(\begin{array}{ccc}
    1/\lambda_1&  & \\
    &1/\lambda_2&\\
     &  & -1/\lambda_3
    \end{array} \right)
    \tilde{\Delta}
\right]
\right\},
\\
Z=\int D\tilde{\Delta} D\tilde{\Delta}^*
\exp\left\{\frac{1}{T}
    \tilde{\Delta}^\dagger
    \left(\begin{array}{ccc}
    1/\lambda_1&  & \\
    &1/\lambda_2&\\
     &  & -1/\lambda_3
    \end{array} \right)
    \tilde{\Delta}
\right\},
\end{gathered}
\label{Eq:HS}
\end{equation}
\end{widetext}
where $\lambda_1,\;\lambda_2<0$ and $\lambda_3>0$ are eigenvalues of the matrix $U$ and $R$ is an orthogonal matrix that satisfies:
\[
R U R^T=
\left(\begin{array}{ccc}
\lambda_1&  & \\
&\lambda_2 &\\
 &  & \lambda_3
\end{array} \right).
\]
In principle, one can next integrate the action over the fermionic fields to obtain an action depending only on $\tilde{\Delta}$ and $\tilde{\Delta}^*$. The convergence of the resulting integral is solely determined by the last term in the exponential in the right-hand side. of Eq.\ref{Eq:HS} and one can evidently see that the described procedure yields a converging integral. The resulting saddle-point equations are rather cumbersome, but their solution is actually related to the one of Eq.(\ref{GapEquations}):
\begin{gather*}
\tilde{\Delta}_0=
\left(\begin{array}{ccc}
1&  & \\
&1 &\\
 &  & -i
\end{array} \right)
R
\Delta_0,
\\
\tilde{\Delta}^*_0=
\left(\begin{array}{ccc}
1&  & \\
&1 &\\
 &  & -i
\end{array} \right)
R
\Delta_0^*,
\end{gather*}
where $\Delta_0=(\Delta_{h1},\Delta_{h2},\Delta_e)^T$. It is evident that $\tilde{\Delta}^*_0\neq(\tilde{\Delta}_0)^*$, meaning that the saddle point of the action is situated in the complex continuation of the integration range. However, as the integral over $\tilde{\Delta},\tilde{\Delta}^*$ converges at $\infty$, one can shift the integration contour such that it includes the point $(\tilde{\Delta}_0,\tilde{\Delta}_0^*)$ without altering the result.
 The value of the action at the saddle point is most easily expressed through the quantities $\Delta^\alpha_0$ and is given by
\begin{equation}
S_0 =-\Delta_0^\dagger U^{-1} \Delta_0
-T \sum_{\omega_n,{\bf k},\alpha} \ln \left[ \omega_n^2+(\xi^\alpha_{{\bf k}})^2+|\Delta^\alpha_0|^2\right]
,
\label{app_gc_pot}
\end{equation}
where $\alpha\in\{h_1,h_2,e_1,e_2\}$ and $\omega_n=(2n+1)\pi T$.

We shall evaluate the second variation matrix of the action with respect to real and imaginary parts of the HS fields $\tilde{\Delta}^r$, $\tilde{\Delta}^{im}$ defined through $\tilde{\Delta} = \tilde{\Delta}^r+i \tilde{\Delta}^{im};\;\tilde{\Delta}^* = \tilde{\Delta}^r-i\tilde{\Delta}^{im}$. Once again, it is easier to evaluate the derivatives with respect to $\Delta_i^a$ and use the transformation
\[
\left.\frac{\partial^2 S}{\partial \tilde{\Delta}^a_i \partial \tilde{\Delta}^b_j}\right|_{\tilde{\Delta}=\tilde{\Delta}_0}=
\sum_{kl}
\frac{\partial \Delta^a_k }{\partial \tilde{\Delta}^a_i}
\left.\frac{\partial^2 S}{\partial \Delta^a_k \partial \Delta^b_l}\right|_{\Delta=\Delta_0}
\frac{\partial \Delta^b_l }{\partial \tilde{\Delta}^b_j},
\]
where $a,b=\{r,\;im\}$. For the considered model, this yields
\begin{widetext}
\begin{equation}
\begin{gathered}
\left.\frac{\partial^2 S}{\partial \tilde{\Delta}^a_i \partial \tilde{\Delta}^b_j}\right|_0=
\left(\begin{array}{ccc}
1&  & \\
&1 &\\
 &  & i
\end{array} \right)
R
\left.\frac{\partial^2 S}{\partial \Delta^a_i \partial \Delta^b_j}\right|_0
R^T
\left(\begin{array}{ccc}
1&  & \\
&1 &\\
 &  & i
\end{array} \right),
\\
\left.\frac{\partial^2 S}{\partial \Delta^r_i \partial \Delta^r_j}\right|_0=-2U^{-1}
-N_0
\left(\begin{array}{ccc}
 L_{h1}+\frac{(\Delta_{h1}^r)^2 L_{h_1}^{\prime}}{|\Delta_{h1}|}&  & \\
& L_{h2}+\frac{(\Delta_{h2}^r)^2 L_{h_2}^{\prime}}{|\Delta_{h2}|} &\\
 &  & 2 L_{e}+2\frac{(\Delta_{e}^r)^2 L_{e}^{\prime}}{|\Delta_e|}
\end{array} \right),
\\
\left.\frac{\partial^2 S}{\partial \Delta^r_i \partial \Delta^{im}_j}\right|_0=
-N_0
\left(\begin{array}{ccc}
 \frac{\Delta_{h1}^r \Delta_{h1}^{im} L_{h_1}^{\prime}}{|\Delta_{h1}|}&  & \\
& \frac{\Delta_{h2}^r \Delta_{h2}^{im} L_{h_2}^{\prime}}{|\Delta_{h2}|} &\\
 &  & 2\frac{\Delta_{e}^r \Delta_{e}^{im} L_{e}^{\prime}}{|\Delta_e|}
\end{array} \right),
\\
\left.\frac{\partial^2 S}{\partial \Delta^{im}_i \partial \Delta^{im}_j}\right|_0=-2U^{-1}
-N_0
\left(\begin{array}{ccc}
 L_{h1}+\frac{(\Delta_{h1}^{im})^2 L_{h_1}^{\prime}}{|\Delta_{h1}|}&  & \\
& L_{h2}+\frac{(\Delta_{h2}^{im})^2 L_{h_2}^{\prime}}{|\Delta_{h2}|} &\\
 &  & 2 L_{e}+2\frac{(\Delta_{e}^{im})^2 L_{e}^{\prime}}{|\Delta_e|}
\end{array} \right)
\end{gathered}
\end{equation}

\begin{figure}
	\includegraphics[width=0.8\linewidth]{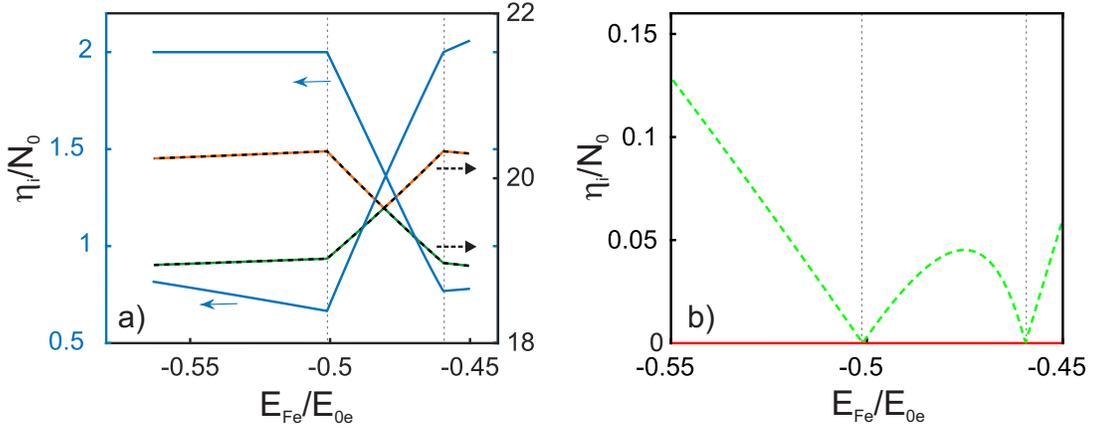}
	\caption{Panels (a) and (b): Eigenvalues of the second variation matrix of the action $\frac{\partial^2 S}{\partial \Delta^a_i \partial \Delta^b_j}$ at $T=0$. Vertical dashed lines mark the boundaries of the $s+is$ state.}
	\label{app_eigen}
\end{figure}
	
\end{widetext}
where $L_{\alpha}^{\prime}=\partial L_{\alpha}/\partial|\Delta_{\alpha}|$ and $|_0$ means value taken at the extremum. In Figs. \ref{app_eigen}(a) and \ref{app_eigen}(b) we present the eigenvalues $\eta_i$ of the resulting $6\times6$ matrix.

All of the eigenvalues are real. One of the eigenvalues [red line in \ref{app_eigen}(b)] is always zero, corresponding to the degeneracy with respect to the overall phase of the superconducting order parameter. The second smallest eigenvalue [green dashed line in Fig.\ref{app_eigen}(b)] becomes zero only at the boundaries of the $s+is$ state corresponding to a soft Leggett mode. All other eigenvalues are clearly positive with the largest ones (black dashed lines) in Fig.\ref{app_eigen} being due  to the repulsive $s_{++++}$ channel. Additionally, while the eigenvectors are complex, they can be normalized, and the corresponding Jacobian is real and has an absolute value of $1$.

Overall, apart from the overall phase mode, present in all superconductors and Leggett mode, softening at the boundaries of the $s+is$ state, we have shown that the second variation matrix of the action is positive definite and the resulting Gaussian integrals converge, proving that the solutions discussed in the main text correspond to a minimum of the free energy due to $\left(\frac{d^2 F}{d \Delta^a_i d\Delta^b_j}\right)_{T,N} = \left(\frac{\partial^2 \Omega}{\partial \Delta^a_i \partial\Delta^b_j}\right)_{T,\mu=\mu(T)}$.

\begin{widetext}

\section{LOCAL DENSITY OF STATES NEAR AN IMPURITY FOR  $\mathbf{s+is}$ STATE AND ROLE OF INCIPIENT BANDS}

We consider the spatially resolved density of states for a multiband superconducting system with a single impurity at $T=0$. Generally one can write the coherent part of Green's function as [we do not take the effects of interaction such as quasiparticle residue ($Z$) renormalization and finite quasiparticle lifetime into account]
\[
G({\bf r},{\bf r}',\omega)=
-i\int d(t-t') e^{i\omega(t-t')}\langle T \Psi({\bf r},t)\Psi^\dagger({\bf r}',t')\rangle
=
\sum_{n}
\frac{\psi_n({\bf r})\psi_n^*({\bf r}')}
{\omega-\epsilon_n+i\delta \text{sgn}(\epsilon_n)},
\]
where $\psi_n({\bf r})$ are eigenfunctions of the Hamiltonian including the impurity potential and the spin indices are suppressed. The density of states is given by
\begin{equation}
\rho({\bf r},\omega) = \sum_{n} |\psi_n({\bf r})|^2\delta(\omega-\epsilon_n)= -\frac{\text{sgn}(\omega)}{\pi} {\rm Im} [G({\bf r},{\bf r},\omega)].
\label{rho}
\end{equation}

One can relate the Fourier transform of $\rho({\bf r},\omega)$ to the Green's functions in momentum space $G({\bf k},{\bf k}',\omega)$:
\begin{equation}
\begin{gathered}
\rho({\bf q},\omega) = \int d{\bf r} e^{i{\bf q r}}\rho({\bf r},\omega) =
-\frac{\text{sgn}(\omega)}{\pi} \int d{\bf r} e^{i{\bf q r}} {\rm Im}
\left[\int \frac{d{\bf k}d{\bf k}'}{(2 \pi)^{2d}} e^{i({\bf k}-{\bf k}'){\bf r}} G({\bf k},{\bf k}',\omega)\right]=
\\
-\frac{\text{sgn}(\omega)}{\pi}
\int \frac{d{\bf k}}{(2 \pi)^{d}}
{\rm Im} [G({\bf k},{\bf k}+{\bf q},\omega)],
\end{gathered}
\label{rho_q_sym}
\end{equation}
where we have assumed inversion symmetric scattering $G({\bf k},{\bf k}+{\bf q},\omega) = G({\bf k},{\bf k}-{\bf q},\omega)$.

\subsection{Green's functions for SC state}
Introducing Gor'kov-Nambu spinors $
\hat{\Psi}=
\left(\begin{array}{c}
\Psi_{{\bf k},\uparrow} \\
\Psi^\dagger_{-{\bf k},\omega,\downarrow}
\end{array} \right)$ we have
\begin{equation}
\begin{gathered}
\hat{G}^0({\bf k},\omega) =
-i\int d(t-t')d({\bf r-r}') e^{i\omega(t-t')-i{\bf k(r-r')}}\langle T \hat{\Psi}({\bf r},t)\hat{\Psi}^\dagger({\bf r}',t')\rangle
\\
=\left(
\begin{array}{cc}
\omega+\xi_{\bf k} & \Delta \\
\Delta^* & \omega-\xi_{\bf k}
\end{array}
\right)\frac{1}{(\omega-\sqrt{\xi_{\bf k}^2+|\Delta|^2}+i\delta)(\omega+\sqrt{\xi_{\bf k}^2+|\Delta|^2}-i\delta)}.
\end{gathered}
\label{G0k}
\end{equation}
Another useful quantity is the momentum-integrated Green's function $\sum_{\bf k} G^0({\bf k},\omega)$. Assuming BCS limit ($\mu\approx E_F\gg\Delta$), we linearize the spectrum near Fermi surface. One obtains
\begin{equation}
\begin{gathered}
\sum_{\bf k} \hat{G}^0({\bf k},\omega)=
\rho_0\int_{-\infty}^{\infty} d \xi
\frac{
	\left(
	\begin{array}{cc}
	\omega\pm\xi & \Delta \\
	\Delta^* & \omega\mp\xi
	\end{array}
	\right)
}
{\omega^2-\xi^2-|\Delta|^2+i\delta}
=
-i \pi \rho_0
\frac{
	\left(
	\begin{array}{cc}
	\omega & \Delta \\
	\Delta^* & \omega
	\end{array}
	\right)
}
{\sqrt{\omega^2-|\Delta|^2+i\delta}}
,
\end{gathered}
\label{G0}
\end{equation}
where $\pm$ is for electron-hole bands. Note that the $i\delta$ in the denominator determines the sign of the imaginary part of the square root for $\omega<\Delta$. Let us use the expressions (\ref{rho_q_sym}) to calculate the density of states for a uniform superconducting system:
\begin{gather*}
\rho_\uparrow^0(0,\omega) =\rho_\downarrow^0(0,\omega)=
-\frac{\text{sgn}(\omega)}{\pi}
{\rm Im} [\hat{G}^0(\omega)]_{11}=
\rho_0{\rm Im} \left[\frac{i|\omega|}{\sqrt{\omega^2-|\Delta|^2+i\delta}}\right]=
\begin{cases}
\frac{|\omega|}{\sqrt{\omega^2-|\Delta|^2}} & \mbox{if } |\omega|>|\Delta|\\
0 & \mbox{if } |\omega|<|\Delta|
\end{cases}.
\end{gather*}

\subsection{Incipient band and e-h assymetry}
We evaluate now the momentum-integrated Green's function for a quadratic band in 2D exactly (assuming only $\Lambda\gg\Delta,\omega,|\mu|$) to account for the smallness of the Fermi energy:
\begin{gather*}
\sum_{\bf k} \hat{G}^0({\bf k},\omega)
=
- \rho_0
\left(
\log\left[\frac{\Lambda}{-\mu-\sqrt{\omega^2-|\Delta|^2+i\delta}}\right]
-\log\left[\frac{\Lambda}{-\mu+\sqrt{\omega^2-|\Delta|^2+i\delta}}\right]
\right)
\frac{
	\left(
	\begin{array}{cc}
	\omega & \Delta \\
	\Delta^* & \omega
	\end{array}
	\right)
}
{2\sqrt{\omega^2-|\Delta|^2+i\delta}}
\\
\mp \frac{\rho_0}{2}
\log\left[\frac{\Lambda^2}{\mu^2-\omega^2+|\Delta|^2-i\delta}\right]
\tau_3.
\end{gather*}
Here one should take care with the sum of logarithms of complex argument. Before joining the logarithms, one needs to ensure that the difference between the phases of the two arguments lies in the interval $(-\pi,\pi)$. This is not so for large $\mu>0$ because then the phase difference is actually $2 \pi$. One can work around this with a following trick:
\begin{gather*}
\log\left[\frac{\Lambda}{-\mu-\sqrt{\omega^2-|\Delta|^2+i\delta}}\right]
-\log\left[\frac{\Lambda}{-\mu+\sqrt{\omega^2-|\Delta|^2+i\delta}}\right]=
\\
=
i\pi+
\log\left[\frac{\Lambda}{\mu+\sqrt{\omega^2-|\Delta|^2+i\delta}}\right]
-\log\left[\frac{\Lambda}{-\mu+\sqrt{\omega^2-|\Delta|^2+i\delta}}\right]
=
i\pi+
\log\left[\frac{-\mu+\sqrt{\omega^2-|\Delta|^2+i\delta}}{\mu+\sqrt{\omega^2-|\Delta|^2+i\delta}}\right].
\end{gather*}
One can see that in this case the correct answer $2 i \pi$ is recovered for $\mu\to +\infty$. However, for an incipient band $\mu<0$ such a trick is not needed, as the phase difference between the arguments of the initial logarithms does not become larger than $\pi$. One obtains the result
\begin{equation}
\begin{gathered}
\sum_{\bf k} \hat{G}^0({\bf k},\omega)_{\mu<0}=
- \rho_0
\log\left[\frac{|\mu|+\sqrt{\omega^2-|\Delta|^2+i\delta}}{|\mu|-\sqrt{\omega^2-|\Delta|^2+i\delta}}\right]
\frac{
	\left(
	\begin{array}{cc}
	\omega & \Delta \\
	\Delta^* & \omega
	\end{array}
	\right)
}
{2\sqrt{\omega^2-|\Delta|^2+i\delta}}
\mp \frac{\rho_0}{2}
\log\left[\frac{\Lambda^2}{\mu^2-\omega^2+|\Delta|^2-i\delta}\right]
\tau_3.
\end{gathered}
\label{G0_incip}
\end{equation}
The square root singularity at $\omega = |\Delta|$ is replaced here by a logarithmic one at $\omega=\sqrt{|\Delta|^2+\mu^2}$. Such a singularity also appears in the $\mu>0$ case, but as we consider the non incipient bands to have rather large $\mu$ we shall simply use BCS answer for those.

\subsection{{\bf q}-integrated $\mathbf{\delta \rho}$ and band space}
The correction to the Green's function due to the impurity potential takes the form:
\begin{equation}
\delta \hat{G} ({\bf k},{\bf k}+{\bf q},\omega) =
\hat{G}^0({\bf k},\omega) \hat{T}({\bf k},{\bf k}+{\bf q},\omega) \hat{G}^0({\bf k}+{\bf q},\omega),
\label{dG}
\end{equation}
where $\hat{T}({\bf k},{\bf k}+{\bf q},\omega)$ satisfies
\begin{equation}
\hat{T}({\bf k},{\bf k}+{\bf q},\omega) =
\hat{V}({\bf q})+
\sum_{{\bf k}_1}
\hat{V}({\bf k}-{\bf k}_1)
\hat{G}^0({\bf k}_1,\omega)
\hat{T}({\bf k}_1,{\bf k}+{\bf q},\omega),
\label{T_gen}
\end{equation}
$\hat{V}({\bf q})$ being the Fourier component of the impurity potential. The Green's functions $\hat{G}^0({\bf k},\omega)$ are concentrated in $k$ space near the Fermi surface at values $|\xi_k| < \sqrt{\omega^2+|\Delta|^2}$ (\ref{G0}) or $|\xi_k|<\Gamma$, where $\Gamma$ is the inverse scattering time due to self-energy corrections. One can use this fact to discriminate between inter- and intraband scattering. Using (\ref{rho_q_sym}) one has
\begin{gather*}
\delta \rho_\uparrow({\bf q},\omega) =
-\frac{\text{sgn}(\omega)}{\pi}
\int \frac{d{\bf k}}{(2 \pi)^{d}}
{\rm Im} [
\hat{G}^0({\bf k},\omega) \hat{T}({\bf k},{\bf k}+{\bf q},\omega) \hat{G}^0({\bf k}+{\bf q},\omega)
]_{11}.
\end{gather*}
Note that in the case when impurity is spinless $\delta \rho_\uparrow({\bf q},\omega)=\delta \rho_\downarrow({\bf q},\omega)$, which is easy to see by taking Gor'kov-Nambu spinors with a different choice of spins, which leads to exactly the same Green's functions.

\begin{figure}[h!]
	\begin{center}
		\includegraphics{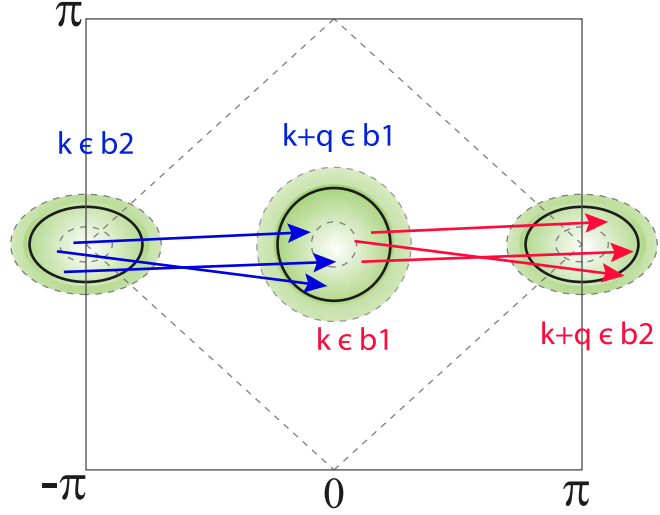}
		\caption{An example for ${\bf q}$-integration for a centrosymmetric two-band system. Green regions represent the extents of Green's function in momentum space,}
	\end{center}
	\label{fig1}
\end{figure}

Let us consider the situation when ${\bf q}$ corresponds to the distance between two bands on the Fermi surface. If the interband scattering wave vector is larger than the momentum extension of the Green's functions within each band and ${\bf k}$ corresponds to the band $b1$, then one can take $\hat{G}^0({\bf k},\omega)\to \hat{G}^0_{b1}({\bf k},\omega)$ and $\hat{G}^0({\bf k}+{\bf q},\omega)\to \hat{G}^0_{b2}({\bf k}+{\bf q},\omega)$. Consequently, integrating the result over ${\bf q}$ such that ${\bf k}+{\bf q}$ is in $b2$, one has
\begin{gather*}
\sum_{{\bf q} \approx{\bf q}_{inter}}\delta \rho_\uparrow({\bf q},\omega) \approx
-\frac{\text{sgn}(\omega)}{\pi}
{\rm Im} [
\hat{G}^0_{b1}(\omega) \hat{T}({\bf k}_1,{\bf k}_2,\omega) \hat{G}_{b2}^0(\omega)
+\hat{G}^0_{b2}(\omega) \hat{T}({\bf k}_2,{\bf k}_1,\omega) \hat{G}_{b1}^0(\omega)
]_{11},
\end{gather*}
where ${\bf k}_1$ and ${\bf k}_2$ are approximate positions of the two bands in momentuum space. Note that because the integration over ${\bf k}$ is performed over all the Brillouin zone and because same set of ${\bf q}$'s gives both $b1\to b2$ and $b2 \to b1$ scattering for a centrosymmetric system one has two contributions for a single set of ${\bf q}$'s (see also Fig.\ref{fig1}). In the same way, one can treat the case when ${\bf q}$ correspond to intra-band scattering. Then, assuming that the dependence of $\hat{V}({\bf q})$ and $\hat{T}({\bf k},{\bf k}+{\bf q},\omega)$ can be neglected for variations of ${\bf q}$ of the order of the Fermi pocket size, one can perform the integral in (\ref{T_gen}) to obtain
\begin{equation}
\underline{T}(\omega) = [1-\underline{V}\underline{G}^0(\omega)]^{-1}\underline{V},
\label{T_band}
\end{equation}
where all the quantities are now matrices also in band space with $\underline{V}^{\mu\nu} = \hat{V}({\bf q}_{\mu\nu})$ and $\underline{G}^0(\omega)_{\mu\nu} = \delta_{\mu\nu} \hat{G}^0_{\mu}(\omega)$. Let us now present the final expression for centrosymmetric $\delta \rho$ for interband scattering for impurities not acting on spin:
\begin{equation}
\begin{gathered}
\delta\rho^{\text{inter}}_{\mu\nu}(\omega) \approx
-\frac{2\text{sgn}(\omega)}{\pi}
{\rm Im} [
\hat{G}^0_{\mu}(\omega) \underline{T}(\omega)_{\mu\nu} \hat{G}_{\nu}^0(\omega)
+\hat{G}^0_{\nu}(\omega) \underline{T}(\omega)_{\mu\nu} \hat{G}_{\mu}^0(\omega)
]_{11}=
\\
-\frac{2\text{sgn}(\omega)}{\pi}
{\rm Im} \; {\rm Tr}\left[\frac{\tau_0+\tau_3}{2}
\left(
\hat{G}^0_{\mu}(\omega) \{\underline{T}(\omega)\}_{\mu\nu} \hat{G}_{\nu}^0(\omega)
+\hat{G}^0_{\nu}(\omega)  \{\underline{T}(\omega)\}_{\mu\nu} \hat{G}_{\mu}^0(\omega)
\right)
\right],
\end{gathered}
\label{drho_band}
\end{equation}
where trace is taken over the Gor'kov-Nambu indices and summation over band indices $\mu,\;\nu$ is not implied.

\subsection{Born approximation}
To consider the qualitative features in the interband DOS, we consider the lowest-order approximation where $\{\underline{T}(\omega)\}_{\mu\nu}=\hat{V}({\bf q}_{\mu\nu})$. We consider impurities with different structure in Nambu space corresponding to different types of scatterers.

\subsubsection{Charge impurity}
Here we consider the impurity acting on the charge density
$\sum_{{\bf q},\sigma} V({\bf q}) \Psi_{{\bf k},\sigma}^{\dagger} \Psi_{{\bf k},\sigma}\to t_3 \tau_3$.
One has
\begin{gather*}
\rho^{\text{inter}}(\omega)= -\frac{2 t_3\text{sgn}(\omega)}{\pi}
{\rm Im} \; {\rm Tr}\left[\frac{\tau_0+\tau_3}{2}
\left(
\hat{G}^0_{1}(\omega) \tau_3 \hat{G}_{2}^0(\omega)
+\hat{G}^0_{2}(\omega) \tau_3  \hat{G}_{1}^0(\omega)
\right)
\right].
\end{gather*}
Evaluating the traces yields:
\begin{equation}
\rho^{\text{inter}}_{ch}(\omega) =
2 t_3\pi \rho_1\rho_2 \text{sgn}(\omega)
{\rm Im}
\frac{2 \omega^2 - \Delta_1 \Delta_2^*- \Delta_2 \Delta_1^*}
{\sqrt{\omega^2-|\Delta_1|^2+i\delta}\sqrt{\omega^2-|\Delta_2|^2+i\delta}}
\end{equation}

Let us also consider scattering between an incipient electron band and a hole band. We also include the e-h asymmetry effects for both but take $\mu_h\gg\Delta,\omega$. One obtains using the Green's functions (\ref{G0_incip}),
\begin{gather*}
\rho^{\text{eh}}_{ch}(\omega) =
-2 t_3 \rho_e\rho_h \text{sgn}(\omega)
{\rm Im}
\left\{
\log\left[\frac{|\mu_e|+\sqrt{\omega^2-|\Delta_e|^2+i\delta}}{|\mu_e|-\sqrt{\omega^2-|\Delta_e|^2+i\delta}}\right]
\frac{i(2 \omega^2 - \Delta_h \Delta_e^*- \Delta_e \Delta_h^*)}
{2\sqrt{\omega^2-|\Delta_h|^2+i\delta}\sqrt{\omega^2-|\Delta_e|^2+i\delta}}
\right\}
\\
+\frac{t_3 \rho_1\rho_2 \text{sgn}(\omega)}{\pi}
{\rm Im}\left\{
\log\left[\frac{\Lambda^2}{\mu_h^2-\omega^2+|\Delta_h|^2-i\delta}\right]
\log\left[\frac{\Lambda^2}{\mu_e^2-\omega^2+|\Delta_e|^2-i\delta}\right]
\right\}
\\
-
2 t_3 \rho_1\rho_2
{\rm Im}\left\{
\frac{i|\omega|\log\left[\frac{\Lambda^2}{\mu_e^2-\omega^2+|\Delta_e|^2-i\delta}\right]}
{\sqrt{\omega^2-|\Delta_h|^2+i\delta}}
-
\log\left[\frac{|\mu_e|+\sqrt{\omega^2-|\Delta_e|^2+i\delta}}{|\mu_e|-\sqrt{\omega^2-|\Delta_e|^2+i\delta}}\right]
\frac{|\omega| \log\left[\frac{\Lambda^2}{\mu_h^2-\omega^2+|\Delta_h|^2-i\delta}\right]}
{2\sqrt{\omega^2-|\Delta_e|^2+i\delta}}
\right\}.
\end{gather*}
The last term is even in $\omega$ and can be thus omitted by considering $\rho^{odd}=[\rho(\omega)-\rho(-\omega)]/2$. It contains only information on the absolute value of the order parameters. The second term remains in the odd-frequency part, but is zero for $\omega<\sqrt{|\Delta_e|^2+\mu_e^2}$. It should affect the curve shape for $\omega>\sqrt{|\Delta_e|^2+\mu_e^2}$ though. The answer for the odd part is then:
\begin{equation}
\begin{gathered}
\rho^{\text{eh}}_{ch}(\omega)^{odd} =
-2 t_3 \rho_e\rho_h \text{sgn}(\omega)
{\rm Im}
\left\{
\log\left[\frac{|\mu_e|+\sqrt{\omega^2-|\Delta_e|^2+i\delta}}{|\mu_e|-\sqrt{\omega^2-|\Delta_e|^2+i\delta}}\right]
\frac{i(2 \omega^2 - \Delta_h \Delta_e^*- \Delta_e \Delta_h^*)}
{2\sqrt{\omega^2-|\Delta_h|^2+i\delta}\sqrt{\omega^2-|\Delta_e|^2+i\delta}}
\right\}+
\\
+\frac{t_3 \rho_1\rho_2 \text{sgn}(\omega)}{\pi}
{\rm Im}\left\{
\log\left[\frac{\Lambda^2}{\mu_h^2-\omega^2+|\Delta_h|^2-i\delta}\right]
\log\left[\frac{\Lambda^2}{\mu_e^2-\omega^2+|\Delta_e|^2-i\delta}\right]
\right\}.
\end{gathered}
\end{equation}

\subsubsection{Andreev impurity}
Andreev impurity is $\sim t_1 \tau_1$ if order parameters on both bands can be taken real. In $s+is$ state this is not so, so to ensure gauge invariance we perform the calculation for
$\alpha \tau_1 + \beta \tau_2 =  \left(
\begin{array}{cc}
0 & Z^* \\
Z & 0
\end{array}
\right)$, where $Z=\alpha+i\beta$ and $\alpha,\;\beta$ are real. Evaluating the traces, one obtains
\begin{gather*}
\rho^{\text{inter}}_A(\omega) =
2 \pi \rho_1\rho_2 |\omega|
\frac{Z\Delta_1+Z^*\Delta_2^*+Z\Delta_2+Z^*\Delta_1^*}
{\sqrt{\omega^2-|\Delta_1|^2+i\delta}\sqrt{\omega^2-|\Delta_2|^2+i\delta}}.
\end{gather*}
To ensure gauge invariance, the answer should be invariant with respect to $\Delta_1\to\Delta_1e^{i\varphi},\;\Delta_2\to\Delta_2e^{i\varphi}$. Taking $Z=t_A \exp\{ -(\varphi_1+\varphi_2)/2\}$ leads then to a gauge-invariant answer with the correct limit $Z\to1$ for $s^{++}$ state. The resulting expression is

\begin{equation}
\rho^{\text{inter}}_{A}(\omega) =
4 t_A \pi \rho_1\rho_2
{\rm Im}
\frac{|\omega|(|\Delta_1|+|\Delta_2|)\cos[(\varphi_1-\varphi_2)/2]}
{\sqrt{\omega^2-|\Delta_1|^2+i\delta}\sqrt{\omega^2-|\Delta_2|^2+i\delta}}
\end{equation}

\subsubsection{Spin impurity}
To consider a spin impurity $\hat{V}\sim {\bf m \cdot \sigma}$, we introduce the four-component Balian-Werthammer spinors:
\[
\hat{\Psi}=
\left(\begin{array}{c}
\Psi_{{\bf k},\uparrow} \\
\Psi_{{\bf k},\downarrow} \\
\Psi^\dagger_{-{\bf k},\omega,\downarrow}\\
-\Psi^\dagger_{-{\bf k},\omega,\uparrow}\\
\end{array} \right).
\]
The Greens's functions are now $4\times4$ matrices; however, they are trivial in the spin space. The expression for $\delta \rho$ is quite similar:
\[
\rho^{\text{inter}}_s(\omega)= -\frac{t_s\text{sgn}(\omega)}{\pi}
{\rm Im} \; {\rm Tr}\left[\frac{\tau_0+\tau_3}{2}
\left(
\hat{G}^0_{1}(\omega){\bf m \cdot \sigma} \hat{G}_{2}^0(\omega)
+\hat{G}^0_{2}(\omega) {\bf m \cdot \sigma}  \hat{G}_{1}^0(\omega)
\right)
\right],
\]
only instead of $2$ in front we evaluate an additional trace over spin indices. It is clear now, that the answer for full DOS correction is $0$; however, one can also calculate the spin-resolved one:
\begin{gather*}
\rho^{\text{inter}}_{\uparrow(\downarrow)}(\omega)_s= -\frac{t_s\text{sgn}(\omega)}{\pi}
{\rm Im} \; {\rm Tr}\left[\frac{\sigma_0\pm\sigma_3}{2}\frac{\tau_0+\tau_3}{2}
\left(
\hat{G}^0_{1}(\omega){\bf m \cdot \sigma} \hat{G}_{2}^0(\omega)
+\hat{G}^0_{2}(\omega) {\bf m \cdot \sigma}  \hat{G}_{1}^0(\omega)
\right)
\right]
\\
=\mp m_z\frac{t_s\text{sgn}(\omega)}{\pi}
{\rm Im} \; {\rm Tr}\left[\frac{\tau_0+\tau_3}{2}
\left(
\hat{G}^0_{1}(\omega) \hat{G}_{2}^0(\omega)
+\hat{G}^0_{2}(\omega) \hat{G}_{1}^0(\omega)
\right)
\right]
\\
=\pm m_z t_s \pi \rho_1\rho_2 \text{sgn}(\omega)
{\rm Im}
\frac{2 \omega^2 + \Delta_1 \Delta_2^*+ \Delta_2 \Delta_1^*}
{\sqrt{\omega^2-|\Delta_1|^2+i\delta}\sqrt{\omega^2-|\Delta_2|^2+i\delta}},
\end{gather*}
where only $m_z$ has entered the expression due to the choice of the quantization axis.

\end{widetext}

\bibliography{Bibliography}

\begin{thebibliography}{80}%
\makeatletter
\providecommand \@ifxundefined [1]{%
 \@ifx{#1\undefined}
}%
\providecommand \@ifnum [1]{%
 \ifnum #1\expandafter \@firstoftwo
 \else \expandafter \@secondoftwo
 \fi
}%
\providecommand \@ifx [1]{%
 \ifx #1\expandafter \@firstoftwo
 \else \expandafter \@secondoftwo
 \fi
}%
\providecommand \natexlab [1]{#1}%
\providecommand \enquote  [1]{``#1''}%
\providecommand \bibnamefont  [1]{#1}%
\providecommand \bibfnamefont [1]{#1}%
\providecommand \citenamefont [1]{#1}%
\providecommand \href@noop [0]{\@secondoftwo}%
\providecommand \href [0]{\begingroup \@sanitize@url \@href}%
\providecommand \@href[1]{\@@startlink{#1}\@@href}%
\providecommand \@@href[1]{\endgroup#1\@@endlink}%
\providecommand \@sanitize@url [0]{\catcode `\\12\catcode `\$12\catcode
  `\&12\catcode `\#12\catcode `\^12\catcode `\_12\catcode `\%12\relax}%
\providecommand \@@startlink[1]{}%
\providecommand \@@endlink[0]{}%
\providecommand \url  [0]{\begingroup\@sanitize@url \@url }%
\providecommand \@url [1]{\endgroup\@href {#1}{\urlprefix }}%
\providecommand \urlprefix  [0]{URL }%
\providecommand \Eprint [0]{\href }%
\providecommand \doibase [0]{http://dx.doi.org/}%
\providecommand \selectlanguage [0]{\@gobble}%
\providecommand \bibinfo  [0]{\@secondoftwo}%
\providecommand \bibfield  [0]{\@secondoftwo}%
\providecommand \translation [1]{[#1]}%
\providecommand \BibitemOpen [0]{}%
\providecommand \bibitemStop [0]{}%
\providecommand \bibitemNoStop [0]{.\EOS\space}%
\providecommand \EOS [0]{\spacefactor3000\relax}%
\providecommand \BibitemShut  [1]{\csname bibitem#1\endcsname}%
\let\auto@bib@innerbib\@empty
\bibitem [{\citenamefont {Kamihara}\ \emph {et~al.}(2008)\citenamefont
  {Kamihara}, \citenamefont {Watanabe}, \citenamefont {Hirano},\ and\
  \citenamefont {Hosono}}]{Kamihara.2008}%
  \BibitemOpen
  \bibfield  {author} {\bibinfo {author} {\bibfnamefont {Y.}~\bibnamefont
  {Kamihara}}, \bibinfo {author} {\bibfnamefont {T.}~\bibnamefont {Watanabe}},
  \bibinfo {author} {\bibfnamefont {M.}~\bibnamefont {Hirano}}, \ and\ \bibinfo
  {author} {\bibfnamefont {H.}~\bibnamefont {Hosono}},\ }\href {\doibase
  10.1021/ja800073m} {\bibfield  {journal} {\bibinfo  {journal} {J. Am. Chem.
  Soc.}\ }\textbf {\bibinfo {volume} {130}},\ \bibinfo {pages} {3296} (\bibinfo
  {year} {2008})}\BibitemShut {NoStop}%
\bibitem [{\citenamefont {Johnston}(2010)}]{Johnston.2010}%
  \BibitemOpen
  \bibfield  {author} {\bibinfo {author} {\bibfnamefont {D.~C.}\ \bibnamefont
  {Johnston}},\ }\href {\doibase 10.1080/00018732.2010.513480} {\bibfield
  {journal} {\bibinfo  {journal} {Adv. Phys.}\ }\textbf {\bibinfo {volume}
  {59}},\ \bibinfo {pages} {803} (\bibinfo {year} {2010})}\BibitemShut
  {NoStop}%
\bibitem [{\citenamefont {Paglione}\ and\ \citenamefont
  {Greene}(2010)}]{Paglione.2010}%
  \BibitemOpen
  \bibfield  {author} {\bibinfo {author} {\bibfnamefont {J.}~\bibnamefont
  {Paglione}}\ and\ \bibinfo {author} {\bibfnamefont {R.~L.}\ \bibnamefont
  {Greene}},\ }\href {\doibase 10.1038/nphys1759} {\bibfield  {journal}
  {\bibinfo  {journal} {Nat. Phys.}\ }\textbf {\bibinfo {volume} {6}},\
  \bibinfo {pages} {645} (\bibinfo {year} {2010})}\BibitemShut {NoStop}%
\bibitem [{\citenamefont {Hirschfeld}\ \emph {et~al.}(2011)\citenamefont
  {Hirschfeld}, \citenamefont {Korshunov},\ and\ \citenamefont
  {Mazin}}]{Hirschfeld.2011}%
  \BibitemOpen
  \bibfield  {author} {\bibinfo {author} {\bibfnamefont {P.~J.}\ \bibnamefont
  {Hirschfeld}}, \bibinfo {author} {\bibfnamefont {M.~M.}\ \bibnamefont
  {Korshunov}}, \ and\ \bibinfo {author} {\bibfnamefont {I.~I.}\ \bibnamefont
  {Mazin}},\ }\href@noop {} {\bibfield  {journal} {\bibinfo  {journal} {Rep.
  Prog. Phys.}\ }\textbf {\bibinfo {volume} {74}},\ \bibinfo {pages} {124508}
  (\bibinfo {year} {2011})}\BibitemShut {NoStop}%
\bibitem [{\citenamefont {Hosono}\ and\ \citenamefont
  {Kuroki}(2015)}]{Hosono.2015}%
  \BibitemOpen
  \bibfield  {author} {\bibinfo {author} {\bibfnamefont {H.}~\bibnamefont
  {Hosono}}\ and\ \bibinfo {author} {\bibfnamefont {K.}~\bibnamefont
  {Kuroki}},\ }\href {\doibase http://dx.doi.org/10.1016/j.physc.2015.02.020}
  {\bibfield  {journal} {\bibinfo  {journal} {Phys. C}\ }\textbf {\bibinfo
  {volume} {514}},\ \bibinfo {pages} {399 } (\bibinfo {year}
  {2015})}\BibitemShut {NoStop}%
\bibitem [{\citenamefont {Miao}\ \emph {et~al.}(2015)\citenamefont {Miao},
  \citenamefont {Qian}, \citenamefont {Shi}, \citenamefont {Richard},
  \citenamefont {Kim}, \citenamefont {Hoesch}, \citenamefont {Xing},
  \citenamefont {Wang}, \citenamefont {Jin}, \citenamefont {Hu},\ and\
  \citenamefont {Ding}}]{Miao.2015}%
  \BibitemOpen
  \bibfield  {author} {\bibinfo {author} {\bibfnamefont {H.}~\bibnamefont
  {Miao}}, \bibinfo {author} {\bibfnamefont {T.}~\bibnamefont {Qian}}, \bibinfo
  {author} {\bibfnamefont {X.}~\bibnamefont {Shi}}, \bibinfo {author}
  {\bibfnamefont {P.}~\bibnamefont {Richard}}, \bibinfo {author} {\bibfnamefont
  {T.~K.}\ \bibnamefont {Kim}}, \bibinfo {author} {\bibfnamefont
  {M.}~\bibnamefont {Hoesch}}, \bibinfo {author} {\bibfnamefont {L.~Y.}\
  \bibnamefont {Xing}}, \bibinfo {author} {\bibfnamefont {X.-C.}\ \bibnamefont
  {Wang}}, \bibinfo {author} {\bibfnamefont {C.-Q.}\ \bibnamefont {Jin}},
  \bibinfo {author} {\bibfnamefont {J.-P.}\ \bibnamefont {Hu}}, \ and\ \bibinfo
  {author} {\bibfnamefont {H.}~\bibnamefont {Ding}},\ }\href {\doibase
  10.1038/ncomms7056} {\bibfield  {journal} {\bibinfo  {journal} {Nat.
  Commun.}\ }\textbf {\bibinfo {volume} {6}},\ \bibinfo {pages} {6056}
  (\bibinfo {year} {2015})}\BibitemShut {NoStop}%
\bibitem [{\citenamefont {Okazaki}\ \emph {et~al.}(2014)\citenamefont
  {Okazaki}, \citenamefont {Ito}, \citenamefont {Ota}, \citenamefont
  {Y.~Shimojima}, \citenamefont {Kiss}, \citenamefont {Watanabe}, \citenamefont
  {Chen}, \citenamefont {Niitaka}, \citenamefont {Hanaguri}, \citenamefont
  {Takagi}, \citenamefont {Chainani},\ and\ \citenamefont
  {Shin}}]{Okazaki.2014}%
  \BibitemOpen
  \bibfield  {author} {\bibinfo {author} {\bibfnamefont {K.}~\bibnamefont
  {Okazaki}}, \bibinfo {author} {\bibfnamefont {Y.}~\bibnamefont {Ito}},
  \bibinfo {author} {\bibfnamefont {Y.~K.}\ \bibnamefont {Ota}}, \bibinfo
  {author} {\bibfnamefont {T.}~\bibnamefont {Y.~Shimojima}}, \bibinfo {author}
  {\bibfnamefont {T.}~\bibnamefont {Kiss}}, \bibinfo {author} {\bibfnamefont
  {S.}~\bibnamefont {Watanabe}}, \bibinfo {author} {\bibfnamefont {C.-T.}\
  \bibnamefont {Chen}}, \bibinfo {author} {\bibfnamefont {S.}~\bibnamefont
  {Niitaka}}, \bibinfo {author} {\bibfnamefont {T.}~\bibnamefont {Hanaguri}},
  \bibinfo {author} {\bibfnamefont {H.}~\bibnamefont {Takagi}}, \bibinfo
  {author} {\bibfnamefont {A.}~\bibnamefont {Chainani}}, \ and\ \bibinfo
  {author} {\bibfnamefont {S.}~\bibnamefont {Shin}},\ }\href
  {https://www.nature.com/articles/srep04109} {\bibfield  {journal} {\bibinfo
  {journal} {Sci. Rep.}\ }\textbf {\bibinfo {volume} {4}},\ \bibinfo {pages}
  {4109} (\bibinfo {year} {2014})}\BibitemShut {NoStop}%
\bibitem [{\citenamefont {Charnukha}\ \emph {et~al.}(2015)\citenamefont
  {Charnukha}, \citenamefont {Thirupathaiah}, \citenamefont {Zabolotnyy},
  \citenamefont {B{\"u}chner}, \citenamefont {Zhigadlo}, \citenamefont
  {Batlogg}, \citenamefont {Yaresko},\ and\ \citenamefont
  {Borisenko}}]{Charnukha.2015}%
  \BibitemOpen
  \bibfield  {author} {\bibinfo {author} {\bibfnamefont {A.}~\bibnamefont
  {Charnukha}}, \bibinfo {author} {\bibfnamefont {S.}~\bibnamefont
  {Thirupathaiah}}, \bibinfo {author} {\bibfnamefont {V.~B.}\ \bibnamefont
  {Zabolotnyy}}, \bibinfo {author} {\bibfnamefont {B.}~\bibnamefont
  {B{\"u}chner}}, \bibinfo {author} {\bibfnamefont {N.~D.}\ \bibnamefont
  {Zhigadlo}}, \bibinfo {author} {\bibfnamefont {B.}~\bibnamefont {Batlogg}},
  \bibinfo {author} {\bibfnamefont {A.~N.}\ \bibnamefont {Yaresko}}, \ and\
  \bibinfo {author} {\bibfnamefont {S.~V.}\ \bibnamefont {Borisenko}},\
  }\href@noop {} {\bibfield  {journal} {\bibinfo  {journal} {Sci. Rep.}\
  }\textbf {\bibinfo {volume} {5}},\ \bibinfo {pages} {10392} (\bibinfo {year}
  {2015})}\BibitemShut {NoStop}%
\bibitem [{\citenamefont {Terashima}\ \emph {et~al.}(2014)\citenamefont
  {Terashima}, \citenamefont {Kikugawa}, \citenamefont {Kiswandhi},
  \citenamefont {Choi}, \citenamefont {Brooks}, \citenamefont {Kasahara},
  \citenamefont {Watashige}, \citenamefont {Ikeda}, \citenamefont {Shibauchi},
  \citenamefont {Matsuda}, \citenamefont {Wolf}, \citenamefont {B\"ohmer},
  \citenamefont {Hardy}, \citenamefont {Meingast}, \citenamefont {L\"ohneysen},
  \citenamefont {Suzuki}, \citenamefont {Arita},\ and\ \citenamefont
  {Uji}}]{Terashima.2014}%
  \BibitemOpen
  \bibfield  {author} {\bibinfo {author} {\bibfnamefont {T.}~\bibnamefont
  {Terashima}}, \bibinfo {author} {\bibfnamefont {N.}~\bibnamefont {Kikugawa}},
  \bibinfo {author} {\bibfnamefont {A.}~\bibnamefont {Kiswandhi}}, \bibinfo
  {author} {\bibfnamefont {E.-S.}\ \bibnamefont {Choi}}, \bibinfo {author}
  {\bibfnamefont {J.~S.}\ \bibnamefont {Brooks}}, \bibinfo {author}
  {\bibfnamefont {S.}~\bibnamefont {Kasahara}}, \bibinfo {author}
  {\bibfnamefont {T.}~\bibnamefont {Watashige}}, \bibinfo {author}
  {\bibfnamefont {H.}~\bibnamefont {Ikeda}}, \bibinfo {author} {\bibfnamefont
  {T.}~\bibnamefont {Shibauchi}}, \bibinfo {author} {\bibfnamefont
  {Y.}~\bibnamefont {Matsuda}}, \bibinfo {author} {\bibfnamefont
  {T.}~\bibnamefont {Wolf}}, \bibinfo {author} {\bibfnamefont {A.~E.}\
  \bibnamefont {B\"ohmer}}, \bibinfo {author} {\bibfnamefont {F.}~\bibnamefont
  {Hardy}}, \bibinfo {author} {\bibfnamefont {C.}~\bibnamefont {Meingast}},
  \bibinfo {author} {\bibfnamefont {H.~v.}\ \bibnamefont {L\"ohneysen}},
  \bibinfo {author} {\bibfnamefont {M.-T.}\ \bibnamefont {Suzuki}}, \bibinfo
  {author} {\bibfnamefont {R.}~\bibnamefont {Arita}}, \ and\ \bibinfo {author}
  {\bibfnamefont {S.}~\bibnamefont {Uji}},\ }\href {\doibase
  10.1103/PhysRevB.90.144517} {\bibfield  {journal} {\bibinfo  {journal} {Phys.
  Rev. B}\ }\textbf {\bibinfo {volume} {90}},\ \bibinfo {pages} {144517}
  (\bibinfo {year} {2014})}\BibitemShut {NoStop}%
\bibitem [{\citenamefont {Watson}\ \emph {et~al.}(2015)\citenamefont {Watson},
  \citenamefont {Kim}, \citenamefont {Haghighirad}, \citenamefont {Davies},
  \citenamefont {McCollam}, \citenamefont {Narayanan}, \citenamefont {Blake},
  \citenamefont {Chen}, \citenamefont {Ghannadzadeh}, \citenamefont
  {Schofield}, \citenamefont {Hoesch}, \citenamefont {Meingast}, \citenamefont
  {Wolf},\ and\ \citenamefont {Coldea}}]{Watson.2015}%
  \BibitemOpen
  \bibfield  {author} {\bibinfo {author} {\bibfnamefont {M.~D.}\ \bibnamefont
  {Watson}}, \bibinfo {author} {\bibfnamefont {T.~K.}\ \bibnamefont {Kim}},
  \bibinfo {author} {\bibfnamefont {A.~A.}\ \bibnamefont {Haghighirad}},
  \bibinfo {author} {\bibfnamefont {N.~R.}\ \bibnamefont {Davies}}, \bibinfo
  {author} {\bibfnamefont {A.}~\bibnamefont {McCollam}}, \bibinfo {author}
  {\bibfnamefont {A.}~\bibnamefont {Narayanan}}, \bibinfo {author}
  {\bibfnamefont {S.~F.}\ \bibnamefont {Blake}}, \bibinfo {author}
  {\bibfnamefont {Y.~L.}\ \bibnamefont {Chen}}, \bibinfo {author}
  {\bibfnamefont {S.}~\bibnamefont {Ghannadzadeh}}, \bibinfo {author}
  {\bibfnamefont {A.~J.}\ \bibnamefont {Schofield}}, \bibinfo {author}
  {\bibfnamefont {M.}~\bibnamefont {Hoesch}}, \bibinfo {author} {\bibfnamefont
  {C.}~\bibnamefont {Meingast}}, \bibinfo {author} {\bibfnamefont
  {T.}~\bibnamefont {Wolf}}, \ and\ \bibinfo {author} {\bibfnamefont {A.~I.}\
  \bibnamefont {Coldea}},\ }\href {\doibase 10.1103/PhysRevB.91.155106}
  {\bibfield  {journal} {\bibinfo  {journal} {Phys. Rev. B}\ }\textbf {\bibinfo
  {volume} {91}},\ \bibinfo {pages} {155106} (\bibinfo {year}
  {2015})}\BibitemShut {NoStop}%
\bibitem [{\citenamefont {Kasahara}\ \emph {et~al.}(2014)\citenamefont
  {Kasahara}, \citenamefont {Watashige}, \citenamefont {Hanaguri},
  \citenamefont {Kohsaka}, \citenamefont {Yamashita}, \citenamefont
  {Shimoyama}, \citenamefont {Mizukami}, \citenamefont {Endo}, \citenamefont
  {Ikeda}, \citenamefont {Aoyama}, \citenamefont {Terashima}, \citenamefont
  {Uji}, \citenamefont {Wolf}, \citenamefont {Löhneysen}, \citenamefont
  {Shibauchi},\ and\ \citenamefont {Matsuda}}]{Kasahara.2014}%
  \BibitemOpen
  \bibfield  {author} {\bibinfo {author} {\bibfnamefont {S.}~\bibnamefont
  {Kasahara}}, \bibinfo {author} {\bibfnamefont {T.}~\bibnamefont {Watashige}},
  \bibinfo {author} {\bibfnamefont {T.}~\bibnamefont {Hanaguri}}, \bibinfo
  {author} {\bibfnamefont {Y.}~\bibnamefont {Kohsaka}}, \bibinfo {author}
  {\bibfnamefont {T.}~\bibnamefont {Yamashita}}, \bibinfo {author}
  {\bibfnamefont {Y.}~\bibnamefont {Shimoyama}}, \bibinfo {author}
  {\bibfnamefont {Y.}~\bibnamefont {Mizukami}}, \bibinfo {author}
  {\bibfnamefont {R.}~\bibnamefont {Endo}}, \bibinfo {author} {\bibfnamefont
  {H.}~\bibnamefont {Ikeda}}, \bibinfo {author} {\bibfnamefont
  {K.}~\bibnamefont {Aoyama}}, \bibinfo {author} {\bibfnamefont
  {T.}~\bibnamefont {Terashima}}, \bibinfo {author} {\bibfnamefont
  {S.}~\bibnamefont {Uji}}, \bibinfo {author} {\bibfnamefont {T.}~\bibnamefont
  {Wolf}}, \bibinfo {author} {\bibfnamefont {H.~v.}\ \bibnamefont
  {Löhneysen}}, \bibinfo {author} {\bibfnamefont {T.}~\bibnamefont
  {Shibauchi}}, \ and\ \bibinfo {author} {\bibfnamefont {Y.}~\bibnamefont
  {Matsuda}},\ }\href {\doibase 10.1073/pnas.1413477111} {\bibfield  {journal}
  {\bibinfo  {journal} {Proc. Nat. Acad. Sci. USA}\ }\textbf {\bibinfo {volume}
  {111}},\ \bibinfo {pages} {16309} (\bibinfo {year} {2014})}\BibitemShut
  {NoStop}%
\bibitem [{\citenamefont {Lubashevsky}\ \emph {et~al.}(2012)\citenamefont
  {Lubashevsky}, \citenamefont {Lahoud}, \citenamefont {Chashka}, \citenamefont
  {Podolsky},\ and\ \citenamefont {Kanigel}}]{Lubashevsky.2012}%
  \BibitemOpen
  \bibfield  {author} {\bibinfo {author} {\bibfnamefont {Y.}~\bibnamefont
  {Lubashevsky}}, \bibinfo {author} {\bibfnamefont {E.}~\bibnamefont {Lahoud}},
  \bibinfo {author} {\bibfnamefont {K.}~\bibnamefont {Chashka}}, \bibinfo
  {author} {\bibfnamefont {D.}~\bibnamefont {Podolsky}}, \ and\ \bibinfo
  {author} {\bibfnamefont {A.}~\bibnamefont {Kanigel}},\ }\href {\doibase
  10.1038/nphys2216} {\bibfield  {journal} {\bibinfo  {journal} {Nat. Phys.}\
  }\textbf {\bibinfo {volume} {8}},\ \bibinfo {pages} {309} (\bibinfo {year}
  {2012})}\BibitemShut {NoStop}%
\bibitem [{\citenamefont {Rinott}\ \emph {et~al.}(2017)\citenamefont {Rinott},
  \citenamefont {Chashka}, \citenamefont {Ribak}, \citenamefont {Rienks},
  \citenamefont {Taleb-Ibrahimi}, \citenamefont {Le~Fevre}, \citenamefont
  {Bertran}, \citenamefont {Randeria},\ and\ \citenamefont
  {Kanigel}}]{Rinott.2017}%
  \BibitemOpen
  \bibfield  {author} {\bibinfo {author} {\bibfnamefont {S.}~\bibnamefont
  {Rinott}}, \bibinfo {author} {\bibfnamefont {K.~B.}\ \bibnamefont {Chashka}},
  \bibinfo {author} {\bibfnamefont {A.}~\bibnamefont {Ribak}}, \bibinfo
  {author} {\bibfnamefont {E.~D.~L.}\ \bibnamefont {Rienks}}, \bibinfo {author}
  {\bibfnamefont {A.}~\bibnamefont {Taleb-Ibrahimi}}, \bibinfo {author}
  {\bibfnamefont {P.}~\bibnamefont {Le~Fevre}}, \bibinfo {author}
  {\bibfnamefont {F.}~\bibnamefont {Bertran}}, \bibinfo {author} {\bibfnamefont
  {M.}~\bibnamefont {Randeria}}, \ and\ \bibinfo {author} {\bibfnamefont
  {A.}~\bibnamefont {Kanigel}},\ }\href
  {http://advances.sciencemag.org/content/3/4/e1602372} {\bibfield  {journal}
  {\bibinfo  {journal} {Sci. Adv.}\ }\textbf {\bibinfo {volume} {3}} (\bibinfo
  {year} {2017})}\BibitemShut {NoStop}%
\bibitem [{\citenamefont {Kasahara}\ \emph {et~al.}(2016)\citenamefont
  {Kasahara}, \citenamefont {Yamashita}, \citenamefont {Shi}, \citenamefont
  {Kobayashi}, \citenamefont {Shimoyama}, \citenamefont {Watashige},
  \citenamefont {Ishida}, \citenamefont {Terashima}, \citenamefont {Wolf},
  \citenamefont {Hardy}, \citenamefont {Meingast}, \citenamefont {Löhneysen},
  \citenamefont {Levchenko}, \citenamefont {T.},\ and\ \citenamefont
  {Matsuda}}]{Kasahara.2016}%
  \BibitemOpen
  \bibfield  {author} {\bibinfo {author} {\bibfnamefont {S.}~\bibnamefont
  {Kasahara}}, \bibinfo {author} {\bibfnamefont {T.}~\bibnamefont {Yamashita}},
  \bibinfo {author} {\bibfnamefont {A.}~\bibnamefont {Shi}}, \bibinfo {author}
  {\bibfnamefont {R.}~\bibnamefont {Kobayashi}}, \bibinfo {author}
  {\bibfnamefont {Y.}~\bibnamefont {Shimoyama}}, \bibinfo {author}
  {\bibfnamefont {T.}~\bibnamefont {Watashige}}, \bibinfo {author}
  {\bibfnamefont {K.}~\bibnamefont {Ishida}}, \bibinfo {author} {\bibfnamefont
  {T.}~\bibnamefont {Terashima}}, \bibinfo {author} {\bibfnamefont
  {T.}~\bibnamefont {Wolf}}, \bibinfo {author} {\bibfnamefont {F.}~\bibnamefont
  {Hardy}}, \bibinfo {author} {\bibfnamefont {C.}~\bibnamefont {Meingast}},
  \bibinfo {author} {\bibfnamefont {H.~v.}\ \bibnamefont {Löhneysen}},
  \bibinfo {author} {\bibfnamefont {A.}~\bibnamefont {Levchenko}}, \bibinfo
  {author} {\bibfnamefont {S.}~\bibnamefont {T.}}, \ and\ \bibinfo {author}
  {\bibfnamefont {Y.}~\bibnamefont {Matsuda}},\ }\href {\doibase
  10.1038/ncomms12843} {\bibfield  {journal} {\bibinfo  {journal} {Nat.
  Commun.}\ }\textbf {\bibinfo {volume} {7}},\ \bibinfo {pages} {12843}
  (\bibinfo {year} {2016})}\BibitemShut {NoStop}%
\bibitem [{\citenamefont {Naidyuk}\ \emph {et~al.}(2016)\citenamefont
  {Naidyuk}, \citenamefont {Fuchs}, \citenamefont {Chareev},\ and\
  \citenamefont {Vasiliev}}]{Naidyuk.2016}%
  \BibitemOpen
  \bibfield  {author} {\bibinfo {author} {\bibfnamefont {Y.~G.}\ \bibnamefont
  {Naidyuk}}, \bibinfo {author} {\bibfnamefont {G.}~\bibnamefont {Fuchs}},
  \bibinfo {author} {\bibfnamefont {D.~A.}\ \bibnamefont {Chareev}}, \ and\
  \bibinfo {author} {\bibfnamefont {A.~N.}\ \bibnamefont {Vasiliev}},\ }\href
  {\doibase 10.1103/PhysRevB.93.144515} {\bibfield  {journal} {\bibinfo
  {journal} {Phys. Rev. B}\ }\textbf {\bibinfo {volume} {93}},\ \bibinfo
  {pages} {144515} (\bibinfo {year} {2016})}\BibitemShut {NoStop}%
\bibitem [{\citenamefont {Sinchenko}\ \emph {et~al.}(2017)\citenamefont
  {Sinchenko}, \citenamefont {Grigoriev}, \citenamefont {Orlov}, \citenamefont
  {Frolov}, \citenamefont {Shakin}, \citenamefont {Chareev}, \citenamefont
  {Volkova},\ and\ \citenamefont {Vasiliev}}]{Sinchenko.2016}%
  \BibitemOpen
  \bibfield  {author} {\bibinfo {author} {\bibfnamefont {A.~A.}\ \bibnamefont
  {Sinchenko}}, \bibinfo {author} {\bibfnamefont {P.~D.}\ \bibnamefont
  {Grigoriev}}, \bibinfo {author} {\bibfnamefont {A.~P.}\ \bibnamefont
  {Orlov}}, \bibinfo {author} {\bibfnamefont {A.~V.}\ \bibnamefont {Frolov}},
  \bibinfo {author} {\bibfnamefont {A.}~\bibnamefont {Shakin}}, \bibinfo
  {author} {\bibfnamefont {D.~A.}\ \bibnamefont {Chareev}}, \bibinfo {author}
  {\bibfnamefont {O.~S.}\ \bibnamefont {Volkova}}, \ and\ \bibinfo {author}
  {\bibfnamefont {A.~N.}\ \bibnamefont {Vasiliev}},\ }\href {\doibase
  10.1103/PhysRevB.95.165120} {\bibfield  {journal} {\bibinfo  {journal} {Phys.
  Rev. B}\ }\textbf {\bibinfo {volume} {95}},\ \bibinfo {pages} {165120}
  (\bibinfo {year} {2017})}\BibitemShut {NoStop}%
\bibitem [{\citenamefont {Randeria}\ and\ \citenamefont
  {Taylor}(2014)}]{Randeria.2014}%
  \BibitemOpen
  \bibfield  {author} {\bibinfo {author} {\bibfnamefont {M.}~\bibnamefont
  {Randeria}}\ and\ \bibinfo {author} {\bibfnamefont {E.}~\bibnamefont
  {Taylor}},\ }\href {\doibase 10.1146/annurev-conmatphys-031113-133829}
  {\bibfield  {journal} {\bibinfo  {journal} {Annu. Rev. Condens. Matter
  Phys.}\ }\textbf {\bibinfo {volume} {5}},\ \bibinfo {pages} {209} (\bibinfo
  {year} {2014})}\BibitemShut {NoStop}%
\bibitem [{\citenamefont {Regal}\ \emph {et~al.}(2004)\citenamefont {Regal},
  \citenamefont {Greiner},\ and\ \citenamefont {Jin}}]{Regal.2004}%
  \BibitemOpen
  \bibfield  {author} {\bibinfo {author} {\bibfnamefont {C.~A.}\ \bibnamefont
  {Regal}}, \bibinfo {author} {\bibfnamefont {M.}~\bibnamefont {Greiner}}, \
  and\ \bibinfo {author} {\bibfnamefont {D.~S.}\ \bibnamefont {Jin}},\ }\href
  {\doibase 10.1103/PhysRevLett.92.040403} {\bibfield  {journal} {\bibinfo
  {journal} {Phys. Rev. Lett.}\ }\textbf {\bibinfo {volume} {92}},\ \bibinfo
  {pages} {040403} (\bibinfo {year} {2004})}\BibitemShut {NoStop}%
\bibitem [{\citenamefont {Bartenstein}\ \emph {et~al.}(2004)\citenamefont
  {Bartenstein}, \citenamefont {Altmeyer}, \citenamefont {Riedl}, \citenamefont
  {Jochim}, \citenamefont {Chin}, \citenamefont {Denschlag},\ and\
  \citenamefont {Grimm}}]{Bartenstein.2004}%
  \BibitemOpen
  \bibfield  {author} {\bibinfo {author} {\bibfnamefont {M.}~\bibnamefont
  {Bartenstein}}, \bibinfo {author} {\bibfnamefont {A.}~\bibnamefont
  {Altmeyer}}, \bibinfo {author} {\bibfnamefont {S.}~\bibnamefont {Riedl}},
  \bibinfo {author} {\bibfnamefont {S.}~\bibnamefont {Jochim}}, \bibinfo
  {author} {\bibfnamefont {C.}~\bibnamefont {Chin}}, \bibinfo {author}
  {\bibfnamefont {J.~H.}\ \bibnamefont {Denschlag}}, \ and\ \bibinfo {author}
  {\bibfnamefont {R.}~\bibnamefont {Grimm}},\ }\href {\doibase
  10.1103/PhysRevLett.92.203201} {\bibfield  {journal} {\bibinfo  {journal}
  {Phys. Rev. Lett.}\ }\textbf {\bibinfo {volume} {92}},\ \bibinfo {pages}
  {203201} (\bibinfo {year} {2004})}\BibitemShut {NoStop}%
\bibitem [{\citenamefont {Ketterle}\ and\ \citenamefont
  {Zwierlein}(2008)}]{Ketterle.2008}%
  \BibitemOpen
  \bibfield  {author} {\bibinfo {author} {\bibfnamefont {W.}~\bibnamefont
  {Ketterle}}\ and\ \bibinfo {author} {\bibfnamefont {M.~W.}\ \bibnamefont
  {Zwierlein}},\ }\href {\doibase 10.1393/ncr/i2008-10033-1} {\bibfield
  {journal} {\bibinfo  {journal} {Riv. Nuovo Cimento}\ }\textbf {\bibinfo
  {volume} {31}},\ \bibinfo {pages} {247} (\bibinfo {year} {2008})}\BibitemShut
  {NoStop}%
\bibitem [{\citenamefont {Zwerger}(2012)}]{Zwerger.2012}%
  \BibitemOpen
  \bibfield  {author} {\bibinfo {author} {\bibfnamefont {W.}~\bibnamefont
  {Zwerger}},\ }\href@noop {} {\emph {\bibinfo {title} {The BCS-BEC Crossover
  and the Unitary Fermi Gas}}}\ (\bibinfo  {publisher} {Springer-Verlag},\
  \bibinfo {address} {Heidelberg, Germany},\ \bibinfo {year}
  {2012})\BibitemShut {NoStop}%
\bibitem [{\citenamefont {Sommer}\ \emph {et~al.}(2012)\citenamefont {Sommer},
  \citenamefont {Cheuk}, \citenamefont {Ku}, \citenamefont {Bakr},\ and\
  \citenamefont {Zwierlein}}]{Sommer.2012}%
  \BibitemOpen
  \bibfield  {author} {\bibinfo {author} {\bibfnamefont {A.~T.}\ \bibnamefont
  {Sommer}}, \bibinfo {author} {\bibfnamefont {L.~W.}\ \bibnamefont {Cheuk}},
  \bibinfo {author} {\bibfnamefont {M.~J.~H.}\ \bibnamefont {Ku}}, \bibinfo
  {author} {\bibfnamefont {W.~S.}\ \bibnamefont {Bakr}}, \ and\ \bibinfo
  {author} {\bibfnamefont {M.~W.}\ \bibnamefont {Zwierlein}},\ }\href {\doibase
  10.1103/PhysRevLett.108.045302} {\bibfield  {journal} {\bibinfo  {journal}
  {Phys. Rev. Lett.}\ }\textbf {\bibinfo {volume} {108}},\ \bibinfo {pages}
  {045302} (\bibinfo {year} {2012})}\BibitemShut {NoStop}%
\bibitem [{\citenamefont {Makhalov}\ \emph {et~al.}(2014)\citenamefont
  {Makhalov}, \citenamefont {Martiyanov},\ and\ \citenamefont
  {Turlapov}}]{Makhalov.2014}%
  \BibitemOpen
  \bibfield  {author} {\bibinfo {author} {\bibfnamefont {V.}~\bibnamefont
  {Makhalov}}, \bibinfo {author} {\bibfnamefont {K.}~\bibnamefont
  {Martiyanov}}, \ and\ \bibinfo {author} {\bibfnamefont {A.}~\bibnamefont
  {Turlapov}},\ }\href {\doibase 10.1103/PhysRevLett.112.045301} {\bibfield
  {journal} {\bibinfo  {journal} {Phys. Rev. Lett.}\ }\textbf {\bibinfo
  {volume} {112}},\ \bibinfo {pages} {045301} (\bibinfo {year}
  {2014})}\BibitemShut {NoStop}%
\bibitem [{\citenamefont {Ries}\ \emph {et~al.}(2015)\citenamefont {Ries},
  \citenamefont {Wenz}, \citenamefont {Z\"urn}, \citenamefont {Bayha},
  \citenamefont {Boettcher}, \citenamefont {Kedar}, \citenamefont {Murthy},
  \citenamefont {Neidig}, \citenamefont {Lompe},\ and\ \citenamefont
  {Jochim}}]{Ries.2015}%
  \BibitemOpen
  \bibfield  {author} {\bibinfo {author} {\bibfnamefont {M.~G.}\ \bibnamefont
  {Ries}}, \bibinfo {author} {\bibfnamefont {A.~N.}\ \bibnamefont {Wenz}},
  \bibinfo {author} {\bibfnamefont {G.}~\bibnamefont {Z\"urn}}, \bibinfo
  {author} {\bibfnamefont {L.}~\bibnamefont {Bayha}}, \bibinfo {author}
  {\bibfnamefont {I.}~\bibnamefont {Boettcher}}, \bibinfo {author}
  {\bibfnamefont {D.}~\bibnamefont {Kedar}}, \bibinfo {author} {\bibfnamefont
  {P.~A.}\ \bibnamefont {Murthy}}, \bibinfo {author} {\bibfnamefont
  {M.}~\bibnamefont {Neidig}}, \bibinfo {author} {\bibfnamefont
  {T.}~\bibnamefont {Lompe}}, \ and\ \bibinfo {author} {\bibfnamefont
  {S.}~\bibnamefont {Jochim}},\ }\href {\doibase
  10.1103/PhysRevLett.114.230401} {\bibfield  {journal} {\bibinfo  {journal}
  {Phys. Rev. Lett.}\ }\textbf {\bibinfo {volume} {114}},\ \bibinfo {pages}
  {230401} (\bibinfo {year} {2015})}\BibitemShut {NoStop}%
\bibitem [{\citenamefont {Murthy}\ \emph {et~al.}(2015)\citenamefont {Murthy},
  \citenamefont {Boettcher}, \citenamefont {Bayha}, \citenamefont {Holzmann},
  \citenamefont {Kedar}, \citenamefont {Neidig}, \citenamefont {Ries},
  \citenamefont {Wenz}, \citenamefont {Z\"urn},\ and\ \citenamefont
  {Jochim}}]{Murthy.2015}%
  \BibitemOpen
  \bibfield  {author} {\bibinfo {author} {\bibfnamefont {P.~A.}\ \bibnamefont
  {Murthy}}, \bibinfo {author} {\bibfnamefont {I.}~\bibnamefont {Boettcher}},
  \bibinfo {author} {\bibfnamefont {L.}~\bibnamefont {Bayha}}, \bibinfo
  {author} {\bibfnamefont {M.}~\bibnamefont {Holzmann}}, \bibinfo {author}
  {\bibfnamefont {D.}~\bibnamefont {Kedar}}, \bibinfo {author} {\bibfnamefont
  {M.}~\bibnamefont {Neidig}}, \bibinfo {author} {\bibfnamefont {M.~G.}\
  \bibnamefont {Ries}}, \bibinfo {author} {\bibfnamefont {A.~N.}\ \bibnamefont
  {Wenz}}, \bibinfo {author} {\bibfnamefont {G.}~\bibnamefont {Z\"urn}}, \ and\
  \bibinfo {author} {\bibfnamefont {S.}~\bibnamefont {Jochim}},\ }\href
  {\doibase 10.1103/PhysRevLett.115.010401} {\bibfield  {journal} {\bibinfo
  {journal} {Phys. Rev. Lett.}\ }\textbf {\bibinfo {volume} {115}},\ \bibinfo
  {pages} {010401} (\bibinfo {year} {2015})}\BibitemShut {NoStop}%
\bibitem [{\citenamefont {Murthy}\ \emph {et~al.}()\citenamefont {Murthy},
  \citenamefont {Neidig}, \citenamefont {Klemt}, \citenamefont {Bayha},
  \citenamefont {Boettcher}, \citenamefont {Enss}, \citenamefont {Holten},
  \citenamefont {Zürn}, \citenamefont {Preiss},\ and\ \citenamefont
  {Jochim}}]{Murthy.2017}%
  \BibitemOpen
  \bibfield  {author} {\bibinfo {author} {\bibfnamefont {P.~A.}\ \bibnamefont
  {Murthy}}, \bibinfo {author} {\bibfnamefont {M.}~\bibnamefont {Neidig}},
  \bibinfo {author} {\bibfnamefont {R.}~\bibnamefont {Klemt}}, \bibinfo
  {author} {\bibfnamefont {L.}~\bibnamefont {Bayha}}, \bibinfo {author}
  {\bibfnamefont {I.}~\bibnamefont {Boettcher}}, \bibinfo {author}
  {\bibfnamefont {T.}~\bibnamefont {Enss}}, \bibinfo {author} {\bibfnamefont
  {M.}~\bibnamefont {Holten}}, \bibinfo {author} {\bibfnamefont
  {G.}~\bibnamefont {Zürn}}, \bibinfo {author} {\bibfnamefont {P.~M.}\
  \bibnamefont {Preiss}}, \ and\ \bibinfo {author} {\bibfnamefont
  {S.}~\bibnamefont {Jochim}},\ }\href@noop {} {\ }\Eprint
  {http://arxiv.org/abs/1705.10577} {arXiv:1705.10577} \BibitemShut {NoStop}%
\bibitem [{\citenamefont {Guidini}\ and\ \citenamefont
  {Perali}(2014)}]{Guidini.2014}%
  \BibitemOpen
  \bibfield  {author} {\bibinfo {author} {\bibfnamefont {A.}~\bibnamefont
  {Guidini}}\ and\ \bibinfo {author} {\bibfnamefont {A.}~\bibnamefont
  {Perali}},\ }\href {http://stacks.iop.org/0953-2048/27/i=12/a=124002}
  {\bibfield  {journal} {\bibinfo  {journal} {Supercond. Sci. Technol.}\
  }\textbf {\bibinfo {volume} {27}},\ \bibinfo {pages} {124002} (\bibinfo
  {year} {2014})}\BibitemShut {NoStop}%
\bibitem [{\citenamefont {Chubukov}\ \emph {et~al.}(2016)\citenamefont
  {Chubukov}, \citenamefont {Eremin},\ and\ \citenamefont
  {Efremov}}]{Chubukov.2016}%
  \BibitemOpen
  \bibfield  {author} {\bibinfo {author} {\bibfnamefont {A.~V.}\ \bibnamefont
  {Chubukov}}, \bibinfo {author} {\bibfnamefont {I.}~\bibnamefont {Eremin}}, \
  and\ \bibinfo {author} {\bibfnamefont {D.~V.}\ \bibnamefont {Efremov}},\
  }\href {\doibase 10.1103/PhysRevB.93.174516} {\bibfield  {journal} {\bibinfo
  {journal} {Phys. Rev. B}\ }\textbf {\bibinfo {volume} {93}},\ \bibinfo
  {pages} {174516} (\bibinfo {year} {2016})}\BibitemShut {NoStop}%
\bibitem [{\citenamefont {Lifshitz}(1960)}]{Lifshitz.1960}%
  \BibitemOpen
  \bibfield  {author} {\bibinfo {author} {\bibfnamefont {I.~M.}\ \bibnamefont
  {Lifshitz}},\ }\href@noop {} {\bibfield  {journal} {\bibinfo  {journal} {Zh.
  Eksp. Teor. Fiz.}\ }\textbf {\bibinfo {volume} {38}},\ \bibinfo {pages}
  {1569} (\bibinfo {year} {1960})}\BibitemShut {NoStop}%
\bibitem [{\citenamefont {Ptok}\ \emph {et~al.}(2017)\citenamefont {Ptok},
  \citenamefont {Kapcia}, \citenamefont {Cichy}, \citenamefont {Oleś},\ and\
  \citenamefont {Piekarz}}]{Ptok.2017}%
  \BibitemOpen
  \bibfield  {author} {\bibinfo {author} {\bibfnamefont {A.}~\bibnamefont
  {Ptok}}, \bibinfo {author} {\bibfnamefont {K.~J.}\ \bibnamefont {Kapcia}},
  \bibinfo {author} {\bibfnamefont {A.}~\bibnamefont {Cichy}}, \bibinfo
  {author} {\bibfnamefont {A.~M.}\ \bibnamefont {Oleś}}, \ and\ \bibinfo
  {author} {\bibfnamefont {P.}~\bibnamefont {Piekarz}},\ }\href
  {http://doi.org/10.1038/srep41979} {\bibfield  {journal} {\bibinfo  {journal}
  {Sci. Rep.}\ }\textbf {\bibinfo {volume} {7}},\ \bibinfo {pages} {41979}
  (\bibinfo {year} {2017})}\BibitemShut {NoStop}%
\bibitem [{\citenamefont {Bianconi}\ \emph {et~al.}(2001)\citenamefont
  {Bianconi}, \citenamefont {Castro}, \citenamefont {Agrestini}, \citenamefont
  {Campi}, \citenamefont {Saini}, \citenamefont {Saccone}, \citenamefont
  {Negri},\ and\ \citenamefont {Giovannini}}]{Bianconi.2001}%
  \BibitemOpen
  \bibfield  {author} {\bibinfo {author} {\bibfnamefont {A.}~\bibnamefont
  {Bianconi}}, \bibinfo {author} {\bibfnamefont {D.~D.}\ \bibnamefont
  {Castro}}, \bibinfo {author} {\bibfnamefont {S.}~\bibnamefont {Agrestini}},
  \bibinfo {author} {\bibfnamefont {G.}~\bibnamefont {Campi}}, \bibinfo
  {author} {\bibfnamefont {N.~L.}\ \bibnamefont {Saini}}, \bibinfo {author}
  {\bibfnamefont {A.}~\bibnamefont {Saccone}}, \bibinfo {author} {\bibfnamefont
  {S.~D.}\ \bibnamefont {Negri}}, \ and\ \bibinfo {author} {\bibfnamefont
  {M.}~\bibnamefont {Giovannini}},\ }\href
  {http://stacks.iop.org/0953-8984/13/i=33/a=318} {\bibfield  {journal}
  {\bibinfo  {journal} {J. Phys. Condens. Matter}\ }\textbf {\bibinfo {volume}
  {13}},\ \bibinfo {pages} {7383} (\bibinfo {year} {2001})}\BibitemShut
  {NoStop}%
\bibitem [{\citenamefont {Perali}\ \emph {et~al.}(1996)\citenamefont {Perali},
  \citenamefont {Bianconi}, \citenamefont {Lanzara},\ and\ \citenamefont
  {Saini}}]{Perali.1996}%
  \BibitemOpen
  \bibfield  {author} {\bibinfo {author} {\bibfnamefont {A.}~\bibnamefont
  {Perali}}, \bibinfo {author} {\bibfnamefont {A.}~\bibnamefont {Bianconi}},
  \bibinfo {author} {\bibfnamefont {A.}~\bibnamefont {Lanzara}}, \ and\
  \bibinfo {author} {\bibfnamefont {N.}~\bibnamefont {Saini}},\ }\href
  {\doibase http://dx.doi.org/10.1016/0038-1098(96)00373-0} {\bibfield
  {journal} {\bibinfo  {journal} {Solid State Commun.}\ }\textbf {\bibinfo
  {volume} {100}},\ \bibinfo {pages} {181 } (\bibinfo {year}
  {1996})}\BibitemShut {NoStop}%
\bibitem [{\citenamefont {Valletta}\ \emph {et~al.}(1997)\citenamefont
  {Valletta}, \citenamefont {Bianconi}, \citenamefont {Perali},\ and\
  \citenamefont {Saini}}]{Valetta.1997}%
  \BibitemOpen
  \bibfield  {author} {\bibinfo {author} {\bibfnamefont {A.}~\bibnamefont
  {Valletta}}, \bibinfo {author} {\bibfnamefont {A.}~\bibnamefont {Bianconi}},
  \bibinfo {author} {\bibfnamefont {A.}~\bibnamefont {Perali}}, \ and\ \bibinfo
  {author} {\bibfnamefont {N.}~\bibnamefont {Saini}},\ }\href {\doibase
  10.1007/s002570050513} {\bibfield  {journal} {\bibinfo  {journal} {Z. Phys.}\
  }\textbf {\bibinfo {volume} {104}},\ \bibinfo {pages} {707} (\bibinfo {year}
  {1997})}\BibitemShut {NoStop}%
\bibitem [{\citenamefont {Innocenti}\ \emph {et~al.}(2010)\citenamefont
  {Innocenti}, \citenamefont {Poccia}, \citenamefont {Ricci}, \citenamefont
  {Valletta}, \citenamefont {Caprara}, \citenamefont {Perali},\ and\
  \citenamefont {Bianconi}}]{Innocenti.2010}%
  \BibitemOpen
  \bibfield  {author} {\bibinfo {author} {\bibfnamefont {D.}~\bibnamefont
  {Innocenti}}, \bibinfo {author} {\bibfnamefont {N.}~\bibnamefont {Poccia}},
  \bibinfo {author} {\bibfnamefont {A.}~\bibnamefont {Ricci}}, \bibinfo
  {author} {\bibfnamefont {A.}~\bibnamefont {Valletta}}, \bibinfo {author}
  {\bibfnamefont {S.}~\bibnamefont {Caprara}}, \bibinfo {author} {\bibfnamefont
  {A.}~\bibnamefont {Perali}}, \ and\ \bibinfo {author} {\bibfnamefont
  {A.}~\bibnamefont {Bianconi}},\ }\href {\doibase 10.1103/PhysRevB.82.184528}
  {\bibfield  {journal} {\bibinfo  {journal} {Phys. Rev. B}\ }\textbf {\bibinfo
  {volume} {82}},\ \bibinfo {pages} {184528} (\bibinfo {year}
  {2010})}\BibitemShut {NoStop}%
\bibitem [{\citenamefont {Innocenti}\ \emph {et~al.}(2011)\citenamefont
  {Innocenti}, \citenamefont {Valletta},\ and\ \citenamefont
  {Bianconi}}]{Innocenti2011}%
  \BibitemOpen
  \bibfield  {author} {\bibinfo {author} {\bibfnamefont {D.}~\bibnamefont
  {Innocenti}}, \bibinfo {author} {\bibfnamefont {A.}~\bibnamefont {Valletta}},
  \ and\ \bibinfo {author} {\bibfnamefont {A.}~\bibnamefont {Bianconi}},\
  }\href {\doibase 10.1007/s10948-010-1096-y} {\bibfield  {journal} {\bibinfo
  {journal} {J. Supercond. Nov. Magn.}\ }\textbf {\bibinfo {volume} {24}},\
  \bibinfo {pages} {1137} (\bibinfo {year} {2011})}\BibitemShut {NoStop}%
\bibitem [{\citenamefont {Valentinis}\ \emph {et~al.}(2016)\citenamefont
  {Valentinis}, \citenamefont {van~der Marel},\ and\ \citenamefont
  {Berthod}}]{Valentinis.2016}%
  \BibitemOpen
  \bibfield  {author} {\bibinfo {author} {\bibfnamefont {D.}~\bibnamefont
  {Valentinis}}, \bibinfo {author} {\bibfnamefont {D.}~\bibnamefont {van~der
  Marel}}, \ and\ \bibinfo {author} {\bibfnamefont {C.}~\bibnamefont
  {Berthod}},\ }\href {\doibase 10.1103/PhysRevB.94.024511} {\bibfield
  {journal} {\bibinfo  {journal} {Phys. Rev. B}\ }\textbf {\bibinfo {volume}
  {94}},\ \bibinfo {pages} {024511} (\bibinfo {year} {2016})}\BibitemShut
  {NoStop}%
\bibitem [{\citenamefont {Shi}\ \emph {et~al.}(2017)\citenamefont {Shi},
  \citenamefont {Han}, \citenamefont {Peng}, \citenamefont {Richard},
  \citenamefont {Qian}, \citenamefont {Wu}, \citenamefont {Qiu}, \citenamefont
  {Wang}, \citenamefont {Hu}, \citenamefont {Sun},\ and\ \citenamefont
  {Ding}}]{Shi.2017}%
  \BibitemOpen
  \bibfield  {author} {\bibinfo {author} {\bibfnamefont {X.}~\bibnamefont
  {Shi}}, \bibinfo {author} {\bibfnamefont {Z.-Q.}\ \bibnamefont {Han}},
  \bibinfo {author} {\bibfnamefont {X.-L.}\ \bibnamefont {Peng}}, \bibinfo
  {author} {\bibfnamefont {P.}~\bibnamefont {Richard}}, \bibinfo {author}
  {\bibfnamefont {T.}~\bibnamefont {Qian}}, \bibinfo {author} {\bibfnamefont
  {X.-X.}\ \bibnamefont {Wu}}, \bibinfo {author} {\bibfnamefont {M.-W.}\
  \bibnamefont {Qiu}}, \bibinfo {author} {\bibfnamefont {S.}~\bibnamefont
  {Wang}}, \bibinfo {author} {\bibfnamefont {J.}~\bibnamefont {Hu}}, \bibinfo
  {author} {\bibfnamefont {Y.-J.}\ \bibnamefont {Sun}}, \ and\ \bibinfo
  {author} {\bibfnamefont {H.}~\bibnamefont {Ding}},\ }\href {\doibase
  10.1038/ncomms14988} {\bibfield  {journal} {\bibinfo  {journal} {Nat.
  Commun.}\ }\textbf {\bibinfo {volume} {8}},\ \bibinfo {pages} {14988}
  (\bibinfo {year} {2017})}\BibitemShut {NoStop}%
\bibitem [{\citenamefont {Sato}\ \emph {et~al.}(2009)\citenamefont {Sato},
  \citenamefont {Nakayama}, \citenamefont {Sekiba}, \citenamefont {Richard},
  \citenamefont {Xu}, \citenamefont {Souma}, \citenamefont {Takahashi},
  \citenamefont {Chen}, \citenamefont {Luo}, \citenamefont {Wang},\ and\
  \citenamefont {Ding}}]{Sato.2009}%
  \BibitemOpen
  \bibfield  {author} {\bibinfo {author} {\bibfnamefont {T.}~\bibnamefont
  {Sato}}, \bibinfo {author} {\bibfnamefont {K.}~\bibnamefont {Nakayama}},
  \bibinfo {author} {\bibfnamefont {Y.}~\bibnamefont {Sekiba}}, \bibinfo
  {author} {\bibfnamefont {P.}~\bibnamefont {Richard}}, \bibinfo {author}
  {\bibfnamefont {Y.-M.}\ \bibnamefont {Xu}}, \bibinfo {author} {\bibfnamefont
  {S.}~\bibnamefont {Souma}}, \bibinfo {author} {\bibfnamefont
  {T.}~\bibnamefont {Takahashi}}, \bibinfo {author} {\bibfnamefont {G.~F.}\
  \bibnamefont {Chen}}, \bibinfo {author} {\bibfnamefont {J.~L.}\ \bibnamefont
  {Luo}}, \bibinfo {author} {\bibfnamefont {N.~L.}\ \bibnamefont {Wang}}, \
  and\ \bibinfo {author} {\bibfnamefont {H.}~\bibnamefont {Ding}},\ }\href
  {\doibase 10.1103/PhysRevLett.103.047002} {\bibfield  {journal} {\bibinfo
  {journal} {Phys. Rev. Lett.}\ }\textbf {\bibinfo {volume} {103}},\ \bibinfo
  {pages} {047002} (\bibinfo {year} {2009})}\BibitemShut {NoStop}%
\bibitem [{\citenamefont {Xu}\ \emph {et~al.}(2013)\citenamefont {Xu},
  \citenamefont {Richard}, \citenamefont {Shi}, \citenamefont {van Roekeghem},
  \citenamefont {Qian}, \citenamefont {Razzoli}, \citenamefont {Rienks},
  \citenamefont {Chen}, \citenamefont {Ieki}, \citenamefont {Nakayama},
  \citenamefont {Sato}, \citenamefont {Takahashi}, \citenamefont {Shi},\ and\
  \citenamefont {Ding}}]{Xu.2013}%
  \BibitemOpen
  \bibfield  {author} {\bibinfo {author} {\bibfnamefont {N.}~\bibnamefont
  {Xu}}, \bibinfo {author} {\bibfnamefont {P.}~\bibnamefont {Richard}},
  \bibinfo {author} {\bibfnamefont {X.}~\bibnamefont {Shi}}, \bibinfo {author}
  {\bibfnamefont {A.}~\bibnamefont {van Roekeghem}}, \bibinfo {author}
  {\bibfnamefont {T.}~\bibnamefont {Qian}}, \bibinfo {author} {\bibfnamefont
  {E.}~\bibnamefont {Razzoli}}, \bibinfo {author} {\bibfnamefont
  {E.}~\bibnamefont {Rienks}}, \bibinfo {author} {\bibfnamefont {G.-F.}\
  \bibnamefont {Chen}}, \bibinfo {author} {\bibfnamefont {E.}~\bibnamefont
  {Ieki}}, \bibinfo {author} {\bibfnamefont {K.}~\bibnamefont {Nakayama}},
  \bibinfo {author} {\bibfnamefont {T.}~\bibnamefont {Sato}}, \bibinfo {author}
  {\bibfnamefont {T.}~\bibnamefont {Takahashi}}, \bibinfo {author}
  {\bibfnamefont {M.}~\bibnamefont {Shi}}, \ and\ \bibinfo {author}
  {\bibfnamefont {H.}~\bibnamefont {Ding}},\ }\href {\doibase
  10.1103/PhysRevB.88.220508} {\bibfield  {journal} {\bibinfo  {journal} {Phys.
  Rev. B}\ }\textbf {\bibinfo {volume} {88}},\ \bibinfo {pages} {220508}
  (\bibinfo {year} {2013})}\BibitemShut {NoStop}%
\bibitem [{\citenamefont {Hodovanets}\ \emph {et~al.}(2014)\citenamefont
  {Hodovanets}, \citenamefont {Liu}, \citenamefont {Jesche}, \citenamefont
  {Ran}, \citenamefont {Mun}, \citenamefont {Lograsso}, \citenamefont
  {Bud'ko},\ and\ \citenamefont {Canfield}}]{Hodovanets.2014}%
  \BibitemOpen
  \bibfield  {author} {\bibinfo {author} {\bibfnamefont {H.}~\bibnamefont
  {Hodovanets}}, \bibinfo {author} {\bibfnamefont {Y.}~\bibnamefont {Liu}},
  \bibinfo {author} {\bibfnamefont {A.}~\bibnamefont {Jesche}}, \bibinfo
  {author} {\bibfnamefont {S.}~\bibnamefont {Ran}}, \bibinfo {author}
  {\bibfnamefont {E.~D.}\ \bibnamefont {Mun}}, \bibinfo {author} {\bibfnamefont
  {T.~A.}\ \bibnamefont {Lograsso}}, \bibinfo {author} {\bibfnamefont {S.~L.}\
  \bibnamefont {Bud'ko}}, \ and\ \bibinfo {author} {\bibfnamefont {P.~C.}\
  \bibnamefont {Canfield}},\ }\href {\doibase 10.1103/PhysRevB.89.224517}
  {\bibfield  {journal} {\bibinfo  {journal} {Phys. Rev. B}\ }\textbf {\bibinfo
  {volume} {89}},\ \bibinfo {pages} {224517} (\bibinfo {year}
  {2014})}\BibitemShut {NoStop}%
\bibitem [{\citenamefont {Nakayama}\ \emph {et~al.}(2011)\citenamefont
  {Nakayama}, \citenamefont {Sato}, \citenamefont {Richard}, \citenamefont
  {Xu}, \citenamefont {Kawahara}, \citenamefont {Umezawa}, \citenamefont
  {Qian}, \citenamefont {Neupane}, \citenamefont {Chen}, \citenamefont {Ding},\
  and\ \citenamefont {Takahashi}}]{Nakayama.2011}%
  \BibitemOpen
  \bibfield  {author} {\bibinfo {author} {\bibfnamefont {K.}~\bibnamefont
  {Nakayama}}, \bibinfo {author} {\bibfnamefont {T.}~\bibnamefont {Sato}},
  \bibinfo {author} {\bibfnamefont {P.}~\bibnamefont {Richard}}, \bibinfo
  {author} {\bibfnamefont {Y.-M.}\ \bibnamefont {Xu}}, \bibinfo {author}
  {\bibfnamefont {T.}~\bibnamefont {Kawahara}}, \bibinfo {author}
  {\bibfnamefont {K.}~\bibnamefont {Umezawa}}, \bibinfo {author} {\bibfnamefont
  {T.}~\bibnamefont {Qian}}, \bibinfo {author} {\bibfnamefont {M.}~\bibnamefont
  {Neupane}}, \bibinfo {author} {\bibfnamefont {G.~F.}\ \bibnamefont {Chen}},
  \bibinfo {author} {\bibfnamefont {H.}~\bibnamefont {Ding}}, \ and\ \bibinfo
  {author} {\bibfnamefont {T.}~\bibnamefont {Takahashi}},\ }\href {\doibase
  10.1103/PhysRevB.83.020501} {\bibfield  {journal} {\bibinfo  {journal} {Phys.
  Rev. B}\ }\textbf {\bibinfo {volume} {83}},\ \bibinfo {pages} {020501}
  (\bibinfo {year} {2011})}\BibitemShut {NoStop}%
\bibitem [{\citenamefont {Ding}\ \emph {et~al.}(2008)\citenamefont {Ding},
  \citenamefont {Richard}, \citenamefont {Nakayama}, \citenamefont {Sugawara},
  \citenamefont {Arakane}, \citenamefont {Sekiba}, \citenamefont {Takayama},
  \citenamefont {Souma}, \citenamefont {Sato}, \citenamefont {Takahashi},
  \citenamefont {Wang}, \citenamefont {Dai}, \citenamefont {Fang},
  \citenamefont {Chen}, \citenamefont {Luo},\ and\ \citenamefont
  {Wang}}]{Ding.2008}%
  \BibitemOpen
  \bibfield  {author} {\bibinfo {author} {\bibfnamefont {H.}~\bibnamefont
  {Ding}}, \bibinfo {author} {\bibfnamefont {P.}~\bibnamefont {Richard}},
  \bibinfo {author} {\bibfnamefont {K.}~\bibnamefont {Nakayama}}, \bibinfo
  {author} {\bibfnamefont {K.}~\bibnamefont {Sugawara}}, \bibinfo {author}
  {\bibfnamefont {T.}~\bibnamefont {Arakane}}, \bibinfo {author} {\bibfnamefont
  {Y.}~\bibnamefont {Sekiba}}, \bibinfo {author} {\bibfnamefont
  {A.}~\bibnamefont {Takayama}}, \bibinfo {author} {\bibfnamefont
  {S.}~\bibnamefont {Souma}}, \bibinfo {author} {\bibfnamefont
  {T.}~\bibnamefont {Sato}}, \bibinfo {author} {\bibfnamefont {T.}~\bibnamefont
  {Takahashi}}, \bibinfo {author} {\bibfnamefont {Z.}~\bibnamefont {Wang}},
  \bibinfo {author} {\bibfnamefont {X.}~\bibnamefont {Dai}}, \bibinfo {author}
  {\bibfnamefont {Z.}~\bibnamefont {Fang}}, \bibinfo {author} {\bibfnamefont
  {G.~F.}\ \bibnamefont {Chen}}, \bibinfo {author} {\bibfnamefont {J.~L.}\
  \bibnamefont {Luo}}, \ and\ \bibinfo {author} {\bibfnamefont {N.~L.}\
  \bibnamefont {Wang}},\ }\href {\doibase 10.1209/0295-5075/83/47001}
  {\bibfield  {journal} {\bibinfo  {journal} {EPL.}\ }\textbf {\bibinfo
  {volume} {83}},\ \bibinfo {pages} {47001} (\bibinfo {year}
  {2008})}\BibitemShut {NoStop}%
\bibitem [{\citenamefont {Christianson}\ \emph {et~al.}(2008)\citenamefont
  {Christianson}, \citenamefont {Goremychkin}, \citenamefont {Osborn},
  \citenamefont {Rosenkranz}, \citenamefont {Lumsden}, \citenamefont
  {Malliakas}, \citenamefont {Todorov}, \citenamefont {Claus}, \citenamefont
  {Chung}, \citenamefont {Kanatzidis}, \citenamefont {Bewley},\ and\
  \citenamefont {Guidi}}]{Christianson.2008}%
  \BibitemOpen
  \bibfield  {author} {\bibinfo {author} {\bibfnamefont {A.~D.}\ \bibnamefont
  {Christianson}}, \bibinfo {author} {\bibfnamefont {E.~A.}\ \bibnamefont
  {Goremychkin}}, \bibinfo {author} {\bibfnamefont {R.}~\bibnamefont {Osborn}},
  \bibinfo {author} {\bibfnamefont {S.}~\bibnamefont {Rosenkranz}}, \bibinfo
  {author} {\bibfnamefont {M.~D.}\ \bibnamefont {Lumsden}}, \bibinfo {author}
  {\bibfnamefont {C.~D.}\ \bibnamefont {Malliakas}}, \bibinfo {author}
  {\bibfnamefont {I.~S.}\ \bibnamefont {Todorov}}, \bibinfo {author}
  {\bibfnamefont {H.}~\bibnamefont {Claus}}, \bibinfo {author} {\bibfnamefont
  {D.~Y.}\ \bibnamefont {Chung}}, \bibinfo {author} {\bibfnamefont {M.~G.}\
  \bibnamefont {Kanatzidis}}, \bibinfo {author} {\bibfnamefont {R.~I.}\
  \bibnamefont {Bewley}}, \ and\ \bibinfo {author} {\bibfnamefont
  {T.}~\bibnamefont {Guidi}},\ }\href {\doibase 10.1038/nature07625} {\bibfield
   {journal} {\bibinfo  {journal} {Nature (London)}\ }\textbf {\bibinfo
  {volume} {456}},\ \bibinfo {pages} {930} (\bibinfo {year}
  {2008})}\BibitemShut {NoStop}%
\bibitem [{\citenamefont {Luo}\ \emph {et~al.}(2009)\citenamefont {Luo},
  \citenamefont {Tanatar}, \citenamefont {Reid}, \citenamefont {Shakeripour},
  \citenamefont {Doiron-Leyraud}, \citenamefont {Ni}, \citenamefont {Bud’ko},
  \citenamefont {Canfield}, \citenamefont {Luo}, \citenamefont {Wang},
  \citenamefont {Wen}, \citenamefont {Prozorov},\ and\ \citenamefont
  {Taillefer}}]{Luo.2009}%
  \BibitemOpen
  \bibfield  {author} {\bibinfo {author} {\bibfnamefont {X.~G.}\ \bibnamefont
  {Luo}}, \bibinfo {author} {\bibfnamefont {M.~A.}\ \bibnamefont {Tanatar}},
  \bibinfo {author} {\bibfnamefont {J.-P.}\ \bibnamefont {Reid}}, \bibinfo
  {author} {\bibfnamefont {H.}~\bibnamefont {Shakeripour}}, \bibinfo {author}
  {\bibfnamefont {N.}~\bibnamefont {Doiron-Leyraud}}, \bibinfo {author}
  {\bibfnamefont {N.}~\bibnamefont {Ni}}, \bibinfo {author} {\bibfnamefont
  {S.~L.}\ \bibnamefont {Bud’ko}}, \bibinfo {author} {\bibfnamefont {P.~C.}\
  \bibnamefont {Canfield}}, \bibinfo {author} {\bibfnamefont {H.}~\bibnamefont
  {Luo}}, \bibinfo {author} {\bibfnamefont {Z.}~\bibnamefont {Wang}}, \bibinfo
  {author} {\bibfnamefont {H.-H.}\ \bibnamefont {Wen}}, \bibinfo {author}
  {\bibfnamefont {R.}~\bibnamefont {Prozorov}}, \ and\ \bibinfo {author}
  {\bibfnamefont {L.}~\bibnamefont {Taillefer}},\ }\href {\doibase
  10.1103/PhysRevB.80.140503} {\bibfield  {journal} {\bibinfo  {journal} {Phys.
  Rev. B}\ }\textbf {\bibinfo {volume} {80}},\ \bibinfo {pages} {140503}
  (\bibinfo {year} {2009})}\BibitemShut {NoStop}%
\bibitem [{\citenamefont {Okazaki}\ \emph {et~al.}(2012)\citenamefont
  {Okazaki}, \citenamefont {Ota}, \citenamefont {Kotani}, \citenamefont
  {Malaeb}, \citenamefont {Ishida}, \citenamefont {Shimojima}, \citenamefont
  {Kiss}, \citenamefont {Watanabe}, \citenamefont {Chen}, \citenamefont
  {Kihou}, \citenamefont {Lee}, \citenamefont {Iyo}, \citenamefont {Eisaki},
  \citenamefont {Saito}, \citenamefont {Fukazawa}, \citenamefont {Kohori},
  \citenamefont {Hashimoto}, \citenamefont {Shibauchi}, \citenamefont
  {Matsuda}, \citenamefont {Ikeda}, \citenamefont {Miyahara}, \citenamefont
  {Arita}, \citenamefont {Chainani},\ and\ \citenamefont
  {Shin}}]{Okazaki.2012}%
  \BibitemOpen
  \bibfield  {author} {\bibinfo {author} {\bibfnamefont {K.}~\bibnamefont
  {Okazaki}}, \bibinfo {author} {\bibfnamefont {Y.}~\bibnamefont {Ota}},
  \bibinfo {author} {\bibfnamefont {Y.}~\bibnamefont {Kotani}}, \bibinfo
  {author} {\bibfnamefont {W.}~\bibnamefont {Malaeb}}, \bibinfo {author}
  {\bibfnamefont {Y.}~\bibnamefont {Ishida}}, \bibinfo {author} {\bibfnamefont
  {T.}~\bibnamefont {Shimojima}}, \bibinfo {author} {\bibfnamefont
  {T.}~\bibnamefont {Kiss}}, \bibinfo {author} {\bibfnamefont {S.}~\bibnamefont
  {Watanabe}}, \bibinfo {author} {\bibfnamefont {C.-T.}\ \bibnamefont {Chen}},
  \bibinfo {author} {\bibfnamefont {K.}~\bibnamefont {Kihou}}, \bibinfo
  {author} {\bibfnamefont {C.~H.}\ \bibnamefont {Lee}}, \bibinfo {author}
  {\bibfnamefont {A.}~\bibnamefont {Iyo}}, \bibinfo {author} {\bibfnamefont
  {H.}~\bibnamefont {Eisaki}}, \bibinfo {author} {\bibfnamefont
  {T.}~\bibnamefont {Saito}}, \bibinfo {author} {\bibfnamefont
  {H.}~\bibnamefont {Fukazawa}}, \bibinfo {author} {\bibfnamefont
  {Y.}~\bibnamefont {Kohori}}, \bibinfo {author} {\bibfnamefont
  {K.}~\bibnamefont {Hashimoto}}, \bibinfo {author} {\bibfnamefont
  {T.}~\bibnamefont {Shibauchi}}, \bibinfo {author} {\bibfnamefont
  {Y.}~\bibnamefont {Matsuda}}, \bibinfo {author} {\bibfnamefont
  {H.}~\bibnamefont {Ikeda}}, \bibinfo {author} {\bibfnamefont
  {H.}~\bibnamefont {Miyahara}}, \bibinfo {author} {\bibfnamefont
  {R.}~\bibnamefont {Arita}}, \bibinfo {author} {\bibfnamefont
  {A.}~\bibnamefont {Chainani}}, \ and\ \bibinfo {author} {\bibfnamefont
  {S.}~\bibnamefont {Shin}},\ }\href@noop {} {\bibfield  {journal} {\bibinfo
  {journal} {Science}\ }\textbf {\bibinfo {volume} {337}},\ \bibinfo {pages}
  {1314} (\bibinfo {year} {2012})}\BibitemShut {NoStop}%
\bibitem [{\citenamefont {Watanabe}\ \emph {et~al.}(2014)\citenamefont
  {Watanabe}, \citenamefont {Yamashita}, \citenamefont {Kawamoto},
  \citenamefont {Kurata}, \citenamefont {Mizukami}, \citenamefont {Ohta},
  \citenamefont {Kasahara}, \citenamefont {Yamashita}, \citenamefont {Saito},
  \citenamefont {Fukazawa}, \citenamefont {Kohori}, \citenamefont {Ishida},
  \citenamefont {Kihou}, \citenamefont {Lee}, \citenamefont {Iyo},
  \citenamefont {Eisaki}, \citenamefont {Vorontsov}, \citenamefont
  {Shibauchi},\ and\ \citenamefont {Matsuda}}]{Watanabe.2014}%
  \BibitemOpen
  \bibfield  {author} {\bibinfo {author} {\bibfnamefont {D.}~\bibnamefont
  {Watanabe}}, \bibinfo {author} {\bibfnamefont {T.}~\bibnamefont {Yamashita}},
  \bibinfo {author} {\bibfnamefont {Y.}~\bibnamefont {Kawamoto}}, \bibinfo
  {author} {\bibfnamefont {S.}~\bibnamefont {Kurata}}, \bibinfo {author}
  {\bibfnamefont {Y.}~\bibnamefont {Mizukami}}, \bibinfo {author}
  {\bibfnamefont {T.}~\bibnamefont {Ohta}}, \bibinfo {author} {\bibfnamefont
  {S.}~\bibnamefont {Kasahara}}, \bibinfo {author} {\bibfnamefont
  {M.}~\bibnamefont {Yamashita}}, \bibinfo {author} {\bibfnamefont
  {T.}~\bibnamefont {Saito}}, \bibinfo {author} {\bibfnamefont
  {H.}~\bibnamefont {Fukazawa}}, \bibinfo {author} {\bibfnamefont
  {Y.}~\bibnamefont {Kohori}}, \bibinfo {author} {\bibfnamefont
  {S.}~\bibnamefont {Ishida}}, \bibinfo {author} {\bibfnamefont
  {K.}~\bibnamefont {Kihou}}, \bibinfo {author} {\bibfnamefont {C.~H.}\
  \bibnamefont {Lee}}, \bibinfo {author} {\bibfnamefont {A.}~\bibnamefont
  {Iyo}}, \bibinfo {author} {\bibfnamefont {H.}~\bibnamefont {Eisaki}},
  \bibinfo {author} {\bibfnamefont {A.~B.}\ \bibnamefont {Vorontsov}}, \bibinfo
  {author} {\bibfnamefont {T.}~\bibnamefont {Shibauchi}}, \ and\ \bibinfo
  {author} {\bibfnamefont {Y.}~\bibnamefont {Matsuda}},\ }\href {\doibase
  10.1103/PhysRevB.89.115112} {\bibfield  {journal} {\bibinfo  {journal} {Phys.
  Rev. B}\ }\textbf {\bibinfo {volume} {89}},\ \bibinfo {pages} {115112}
  (\bibinfo {year} {2014})}\BibitemShut {NoStop}%
\bibitem [{\citenamefont {Cho}\ \emph {et~al.}(2016)\citenamefont {Cho},
  \citenamefont {Ko{\'n}czykowski}, \citenamefont {Teknowijoyo}, \citenamefont
  {Tanatar}, \citenamefont {Liu}, \citenamefont {Lograsso}, \citenamefont
  {Straszheim}, \citenamefont {Mishra}, \citenamefont {Maiti}, \citenamefont
  {Hirschfeld},\ and\ \citenamefont {Prozorov}}]{Cho.2016}%
  \BibitemOpen
  \bibfield  {author} {\bibinfo {author} {\bibfnamefont {K.}~\bibnamefont
  {Cho}}, \bibinfo {author} {\bibfnamefont {M.}~\bibnamefont
  {Ko{\'n}czykowski}}, \bibinfo {author} {\bibfnamefont {S.}~\bibnamefont
  {Teknowijoyo}}, \bibinfo {author} {\bibfnamefont {M.~A.}\ \bibnamefont
  {Tanatar}}, \bibinfo {author} {\bibfnamefont {Y.}~\bibnamefont {Liu}},
  \bibinfo {author} {\bibfnamefont {T.~A.}\ \bibnamefont {Lograsso}}, \bibinfo
  {author} {\bibfnamefont {W.~E.}\ \bibnamefont {Straszheim}}, \bibinfo
  {author} {\bibfnamefont {V.}~\bibnamefont {Mishra}}, \bibinfo {author}
  {\bibfnamefont {S.}~\bibnamefont {Maiti}}, \bibinfo {author} {\bibfnamefont
  {P.~J.}\ \bibnamefont {Hirschfeld}}, \ and\ \bibinfo {author} {\bibfnamefont
  {R.}~\bibnamefont {Prozorov}},\ }\href
  {http://advances.sciencemag.org/content/2/9/e1600807} {\bibfield  {journal}
  {\bibinfo  {journal} {Sci. Adv.}\ }\textbf {\bibinfo {volume} {2}},\ \bibinfo
  {pages} {e1600807} (\bibinfo {year} {2016})}\BibitemShut {NoStop}%
\bibitem [{\citenamefont {Tafti}\ \emph {et~al.}(2013)\citenamefont {Tafti},
  \citenamefont {Juneau-Fecteau}, \citenamefont {Delage}, \citenamefont {{René
  de Cotret}}, \citenamefont {Reid}, \citenamefont {Wang}, \citenamefont {Luo},
  \citenamefont {Chen}, \citenamefont {Doiron-Leyraud},\ and\ \citenamefont
  {Taillefer}}]{Tafti.2013}%
  \BibitemOpen
  \bibfield  {author} {\bibinfo {author} {\bibfnamefont {F.~F.}\ \bibnamefont
  {Tafti}}, \bibinfo {author} {\bibfnamefont {A.}~\bibnamefont
  {Juneau-Fecteau}}, \bibinfo {author} {\bibfnamefont {M.-.}\ \bibnamefont
  {Delage}}, \bibinfo {author} {\bibfnamefont {S.}~\bibnamefont {{René de
  Cotret}}}, \bibinfo {author} {\bibfnamefont {J.-P.}\ \bibnamefont {Reid}},
  \bibinfo {author} {\bibfnamefont {A.~F.}\ \bibnamefont {Wang}}, \bibinfo
  {author} {\bibfnamefont {X.-G.}\ \bibnamefont {Luo}}, \bibinfo {author}
  {\bibfnamefont {X.~H.}\ \bibnamefont {Chen}}, \bibinfo {author}
  {\bibfnamefont {N.}~\bibnamefont {Doiron-Leyraud}}, \ and\ \bibinfo {author}
  {\bibfnamefont {L.}~\bibnamefont {Taillefer}},\ }\href {\doibase
  10.1038/nphys2617} {\bibfield  {journal} {\bibinfo  {journal} {Nat. Phys.}\
  }\textbf {\bibinfo {volume} {9}},\ \bibinfo {pages} {349} (\bibinfo {year}
  {2013})}\BibitemShut {NoStop}%
\bibitem [{\citenamefont {Reid}\ \emph {et~al.}(2012)\citenamefont {Reid},
  \citenamefont {Tanatar}, \citenamefont {Juneau-Fecteau}, \citenamefont
  {Gordon}, \citenamefont {deCotret}, \citenamefont {Doiron-Leyraud},
  \citenamefont {Saito}, \citenamefont {Fukazawa}, \citenamefont {Kohori},
  \citenamefont {Kihou}, \citenamefont {Lee}, \citenamefont {Iyo},
  \citenamefont {Eisaki}, \citenamefont {Prozorov},\ and\ \citenamefont
  {Taillefer}}]{Reid.2012}%
  \BibitemOpen
  \bibfield  {author} {\bibinfo {author} {\bibfnamefont {J.-P.}\ \bibnamefont
  {Reid}}, \bibinfo {author} {\bibfnamefont {M.~A.}\ \bibnamefont {Tanatar}},
  \bibinfo {author} {\bibfnamefont {A.}~\bibnamefont {Juneau-Fecteau}},
  \bibinfo {author} {\bibfnamefont {R.~T.}\ \bibnamefont {Gordon}}, \bibinfo
  {author} {\bibfnamefont {S.~R.}\ \bibnamefont {deCotret}}, \bibinfo {author}
  {\bibfnamefont {N.}~\bibnamefont {Doiron-Leyraud}}, \bibinfo {author}
  {\bibfnamefont {T.}~\bibnamefont {Saito}}, \bibinfo {author} {\bibfnamefont
  {H.}~\bibnamefont {Fukazawa}}, \bibinfo {author} {\bibfnamefont
  {Y.}~\bibnamefont {Kohori}}, \bibinfo {author} {\bibfnamefont
  {K.}~\bibnamefont {Kihou}}, \bibinfo {author} {\bibfnamefont {C.~H.}\
  \bibnamefont {Lee}}, \bibinfo {author} {\bibfnamefont {A.}~\bibnamefont
  {Iyo}}, \bibinfo {author} {\bibfnamefont {H.}~\bibnamefont {Eisaki}},
  \bibinfo {author} {\bibfnamefont {R.}~\bibnamefont {Prozorov}}, \ and\
  \bibinfo {author} {\bibfnamefont {L.}~\bibnamefont {Taillefer}},\ }\href
  {\doibase 10.1103/PhysRevLett.109.087001} {\bibfield  {journal} {\bibinfo
  {journal} {Phys. Rev. Lett.}\ }\textbf {\bibinfo {volume} {109}},\ \bibinfo
  {pages} {087001} (\bibinfo {year} {2012})}\BibitemShut {NoStop}%
\bibitem [{\citenamefont {Stanev}\ and\ \citenamefont
  {Te\v{s}anovi\'{c}}(2010)}]{Stanev.2010}%
  \BibitemOpen
  \bibfield  {author} {\bibinfo {author} {\bibfnamefont {V.}~\bibnamefont
  {Stanev}}\ and\ \bibinfo {author} {\bibfnamefont {Z.}~\bibnamefont
  {Te\v{s}anovi\'{c}}},\ }\href {\doibase 10.1103/PhysRevB.81.134522}
  {\bibfield  {journal} {\bibinfo  {journal} {Phys. Rev. B}\ }\textbf {\bibinfo
  {volume} {81}},\ \bibinfo {pages} {134522} (\bibinfo {year}
  {2010})}\BibitemShut {NoStop}%
\bibitem [{\citenamefont {Carlstr\"om}\ \emph {et~al.}(2011)\citenamefont
  {Carlstr\"om}, \citenamefont {Garaud},\ and\ \citenamefont
  {Babaev}}]{Carlstrom.2011}%
  \BibitemOpen
  \bibfield  {author} {\bibinfo {author} {\bibfnamefont {J.}~\bibnamefont
  {Carlstr\"om}}, \bibinfo {author} {\bibfnamefont {J.}~\bibnamefont {Garaud}},
  \ and\ \bibinfo {author} {\bibfnamefont {E.}~\bibnamefont {Babaev}},\ }\href
  {\doibase 10.1103/PhysRevB.84.134518} {\bibfield  {journal} {\bibinfo
  {journal} {Phys. Rev. B}\ }\textbf {\bibinfo {volume} {84}},\ \bibinfo
  {pages} {134518} (\bibinfo {year} {2011})}\BibitemShut {NoStop}%
\bibitem [{\citenamefont {Hu}\ and\ \citenamefont {Wang}(2012)}]{Hu.2012}%
  \BibitemOpen
  \bibfield  {author} {\bibinfo {author} {\bibfnamefont {X.}~\bibnamefont
  {Hu}}\ and\ \bibinfo {author} {\bibfnamefont {Z.}~\bibnamefont {Wang}},\
  }\href {\doibase 10.1103/PhysRevB.85.064516} {\bibfield  {journal} {\bibinfo
  {journal} {Phys. Rev. B}\ }\textbf {\bibinfo {volume} {85}},\ \bibinfo
  {pages} {064516} (\bibinfo {year} {2012})}\BibitemShut {NoStop}%
\bibitem [{\citenamefont {Maiti}\ and\ \citenamefont
  {Chubukov}(2013)}]{Maiti.2013}%
  \BibitemOpen
  \bibfield  {author} {\bibinfo {author} {\bibfnamefont {S.}~\bibnamefont
  {Maiti}}\ and\ \bibinfo {author} {\bibfnamefont {A.~V.}\ \bibnamefont
  {Chubukov}},\ }\href {\doibase 10.1103/PhysRevB.87.144511} {\bibfield
  {journal} {\bibinfo  {journal} {Phys. Rev. B}\ }\textbf {\bibinfo {volume}
  {87}},\ \bibinfo {pages} {144511} (\bibinfo {year} {2013})}\BibitemShut
  {NoStop}%
\bibitem [{\citenamefont {Thomale}\ \emph {et~al.}(2011)\citenamefont
  {Thomale}, \citenamefont {Platt}, \citenamefont {Hanke},\ and\ \citenamefont
  {Bernevig}}]{Thomale.2011}%
  \BibitemOpen
  \bibfield  {author} {\bibinfo {author} {\bibfnamefont {R.}~\bibnamefont
  {Thomale}}, \bibinfo {author} {\bibfnamefont {C.}~\bibnamefont {Platt}},
  \bibinfo {author} {\bibfnamefont {W.}~\bibnamefont {Hanke}}, \ and\ \bibinfo
  {author} {\bibfnamefont {B.~A.}\ \bibnamefont {Bernevig}},\ }\href {\doibase
  10.1103/PhysRevLett.106.187003} {\bibfield  {journal} {\bibinfo  {journal}
  {Phys. Rev. Lett.}\ }\textbf {\bibinfo {volume} {106}},\ \bibinfo {pages}
  {187003} (\bibinfo {year} {2011})}\BibitemShut {NoStop}%
\bibitem [{\citenamefont {Platt}\ \emph {et~al.}(2009)\citenamefont {Platt},
  \citenamefont {Honerkamp},\ and\ \citenamefont {Hanke}}]{Platt.2009}%
  \BibitemOpen
  \bibfield  {author} {\bibinfo {author} {\bibfnamefont {C.}~\bibnamefont
  {Platt}}, \bibinfo {author} {\bibfnamefont {C.}~\bibnamefont {Honerkamp}}, \
  and\ \bibinfo {author} {\bibfnamefont {W.}~\bibnamefont {Hanke}},\ }\href
  {\doibase 10.1088/1367-2630/11/5/055058} {\bibfield  {journal} {\bibinfo
  {journal} {New J. Phys.}\ }\textbf {\bibinfo {volume} {11}},\ \bibinfo
  {pages} {055058} (\bibinfo {year} {2009})}\BibitemShut {NoStop}%
\bibitem [{\citenamefont {Silaev}\ and\ \citenamefont
  {Babaev}(2013)}]{Silaev.2013}%
  \BibitemOpen
  \bibfield  {author} {\bibinfo {author} {\bibfnamefont {M.}~\bibnamefont
  {Silaev}}\ and\ \bibinfo {author} {\bibfnamefont {E.}~\bibnamefont
  {Babaev}},\ }\href {\doibase 10.1103/PhysRevB.88.220504} {\bibfield
  {journal} {\bibinfo  {journal} {Phys. Rev. B}\ }\textbf {\bibinfo {volume}
  {88}},\ \bibinfo {pages} {220504} (\bibinfo {year} {2013})}\BibitemShut
  {NoStop}%
\bibitem [{\citenamefont {Lin}\ and\ \citenamefont {Hu}(2012)}]{Lin.2012}%
  \BibitemOpen
  \bibfield  {author} {\bibinfo {author} {\bibfnamefont {S.-Z.}\ \bibnamefont
  {Lin}}\ and\ \bibinfo {author} {\bibfnamefont {X.}~\bibnamefont {Hu}},\
  }\href {\doibase 10.1103/PhysRevLett.108.177005} {\bibfield  {journal}
  {\bibinfo  {journal} {Phys. Rev. Lett.}\ }\textbf {\bibinfo {volume} {108}},\
  \bibinfo {pages} {177005} (\bibinfo {year} {2012})}\BibitemShut {NoStop}%
\bibitem [{\citenamefont {Maiti}\ \emph {et~al.}(2015)\citenamefont {Maiti},
  \citenamefont {Sigrist},\ and\ \citenamefont {Chubukov}}]{Maiti.2015}%
  \BibitemOpen
  \bibfield  {author} {\bibinfo {author} {\bibfnamefont {S.}~\bibnamefont
  {Maiti}}, \bibinfo {author} {\bibfnamefont {M.}~\bibnamefont {Sigrist}}, \
  and\ \bibinfo {author} {\bibfnamefont {A.}~\bibnamefont {Chubukov}},\ }\href
  {\doibase 10.1103/PhysRevB.91.161102} {\bibfield  {journal} {\bibinfo
  {journal} {Phys. Rev. B}\ }\textbf {\bibinfo {volume} {91}},\ \bibinfo
  {pages} {161102} (\bibinfo {year} {2015})}\BibitemShut {NoStop}%
\bibitem [{\citenamefont {Garaud}\ and\ \citenamefont
  {Babaev}(2014)}]{Garaud.2014}%
  \BibitemOpen
  \bibfield  {author} {\bibinfo {author} {\bibfnamefont {J.}~\bibnamefont
  {Garaud}}\ and\ \bibinfo {author} {\bibfnamefont {E.}~\bibnamefont
  {Babaev}},\ }\href {\doibase 10.1103/PhysRevLett.112.017003} {\bibfield
  {journal} {\bibinfo  {journal} {Phys. Rev. Lett.}\ }\textbf {\bibinfo
  {volume} {112}},\ \bibinfo {pages} {017003} (\bibinfo {year}
  {2014})}\BibitemShut {NoStop}%
\bibitem [{\citenamefont {Mahyari}\ \emph {et~al.}(2014)\citenamefont
  {Mahyari}, \citenamefont {Cannell}, \citenamefont {Gomez}, \citenamefont
  {Tezok}, \citenamefont {Zelati}, \citenamefont {de~Mello}, \citenamefont
  {Yan}, \citenamefont {Mandrus},\ and\ \citenamefont {Sonier}}]{Lotfi.2014}%
  \BibitemOpen
  \bibfield  {author} {\bibinfo {author} {\bibfnamefont {Z.~L.}\ \bibnamefont
  {Mahyari}}, \bibinfo {author} {\bibfnamefont {A.}~\bibnamefont {Cannell}},
  \bibinfo {author} {\bibfnamefont {C.}~\bibnamefont {Gomez}}, \bibinfo
  {author} {\bibfnamefont {S.}~\bibnamefont {Tezok}}, \bibinfo {author}
  {\bibfnamefont {A.}~\bibnamefont {Zelati}}, \bibinfo {author} {\bibfnamefont
  {E.~V.~L.}\ \bibnamefont {de~Mello}}, \bibinfo {author} {\bibfnamefont
  {J.-Q.}\ \bibnamefont {Yan}}, \bibinfo {author} {\bibfnamefont {D.~G.}\
  \bibnamefont {Mandrus}}, \ and\ \bibinfo {author} {\bibfnamefont {J.~E.}\
  \bibnamefont {Sonier}},\ }\href {\doibase 10.1103/PhysRevB.89.020502}
  {\bibfield  {journal} {\bibinfo  {journal} {Phys. Rev. B}\ }\textbf {\bibinfo
  {volume} {89}},\ \bibinfo {pages} {020502} (\bibinfo {year}
  {2014})}\BibitemShut {NoStop}%
\bibitem [{\citenamefont {Grinenko}\ \emph {et~al.}(2017)\citenamefont
  {Grinenko}, \citenamefont {Materne}, \citenamefont {Sarkar}, \citenamefont
  {Luetkens}, \citenamefont {Kihou}, \citenamefont {Lee}, \citenamefont
  {Akhmadaliev}, \citenamefont {Efremov}, \citenamefont {Drechsler},\ and\
  \citenamefont {Klauss}}]{Grinenko.2017}%
  \BibitemOpen
  \bibfield  {author} {\bibinfo {author} {\bibfnamefont {V.}~\bibnamefont
  {Grinenko}}, \bibinfo {author} {\bibfnamefont {P.}~\bibnamefont {Materne}},
  \bibinfo {author} {\bibfnamefont {R.}~\bibnamefont {Sarkar}}, \bibinfo
  {author} {\bibfnamefont {H.}~\bibnamefont {Luetkens}}, \bibinfo {author}
  {\bibfnamefont {K.}~\bibnamefont {Kihou}}, \bibinfo {author} {\bibfnamefont
  {C.~H.}\ \bibnamefont {Lee}}, \bibinfo {author} {\bibfnamefont
  {S.}~\bibnamefont {Akhmadaliev}}, \bibinfo {author} {\bibfnamefont {D.~V.}\
  \bibnamefont {Efremov}}, \bibinfo {author} {\bibfnamefont {S.-L.}\
  \bibnamefont {Drechsler}}, \ and\ \bibinfo {author} {\bibfnamefont {H.-H.}\
  \bibnamefont {Klauss}},\ }\href {\doibase 10.1103/PhysRevB.95.214511}
  {\bibfield  {journal} {\bibinfo  {journal} {Phys. Rev. B}\ }\textbf {\bibinfo
  {volume} {95}},\ \bibinfo {pages} {214511} (\bibinfo {year}
  {2017})}\BibitemShut {NoStop}%
\bibitem [{\citenamefont {Garaud}\ \emph {et~al.}(2016)\citenamefont {Garaud},
  \citenamefont {Silaev},\ and\ \citenamefont {Babaev}}]{Garaud.2016}%
  \BibitemOpen
  \bibfield  {author} {\bibinfo {author} {\bibfnamefont {J.}~\bibnamefont
  {Garaud}}, \bibinfo {author} {\bibfnamefont {M.}~\bibnamefont {Silaev}}, \
  and\ \bibinfo {author} {\bibfnamefont {E.}~\bibnamefont {Babaev}},\ }\href
  {\doibase 10.1103/PhysRevLett.116.097002} {\bibfield  {journal} {\bibinfo
  {journal} {Phys. Rev. Lett.}\ }\textbf {\bibinfo {volume} {116}},\ \bibinfo
  {pages} {097002} (\bibinfo {year} {2016})}\BibitemShut {NoStop}%
\bibitem [{\citenamefont {Yerin}\ \emph {et~al.}(2015)\citenamefont {Yerin},
  \citenamefont {Omelyanchouk},\ and\ \citenamefont {Il’ichev}}]{Yerin.2015}%
  \BibitemOpen
  \bibfield  {author} {\bibinfo {author} {\bibfnamefont {Y.}~\bibnamefont
  {Yerin}}, \bibinfo {author} {\bibfnamefont {A.}~\bibnamefont {Omelyanchouk}},
  \ and\ \bibinfo {author} {\bibfnamefont {E.}~\bibnamefont {Il’ichev}},\
  }\href {\doibase 10.1088/0953-2048/28/9/095006} {\bibfield  {journal}
  {\bibinfo  {journal} {Supercond. Sci. Technol.}\ }\textbf {\bibinfo {volume}
  {28}},\ \bibinfo {pages} {095006} (\bibinfo {year} {2015})}\BibitemShut
  {NoStop}%
\bibitem [{\citenamefont {Hirschfeld}\ \emph {et~al.}(2015)\citenamefont
  {Hirschfeld}, \citenamefont {Altenfeld}, \citenamefont {Eremin},\ and\
  \citenamefont {Mazin}}]{Hirschfeld.2015}%
  \BibitemOpen
  \bibfield  {author} {\bibinfo {author} {\bibfnamefont {P.~J.}\ \bibnamefont
  {Hirschfeld}}, \bibinfo {author} {\bibfnamefont {D.}~\bibnamefont
  {Altenfeld}}, \bibinfo {author} {\bibfnamefont {I.}~\bibnamefont {Eremin}}, \
  and\ \bibinfo {author} {\bibfnamefont {I.~I.}\ \bibnamefont {Mazin}},\ }\href
  {\doibase 10.1103/PhysRevB.92.184513} {\bibfield  {journal} {\bibinfo
  {journal} {Phys. Rev. B}\ }\textbf {\bibinfo {volume} {92}},\ \bibinfo
  {pages} {184513} (\bibinfo {year} {2015})}\BibitemShut {NoStop}%
\bibitem [{\citenamefont {Leggett}(1980)}]{Leggett.1980}%
  \BibitemOpen
  \bibfield  {author} {\bibinfo {author} {\bibfnamefont {A.}~\bibnamefont
  {Leggett}},\ }\href@noop {} {\emph {\bibinfo {title} {\textup{in} Modern
  Trends in the Theory of Condensed Matter}}},\ edited by\ \bibinfo {editor}
  {\bibfnamefont {A.}~\bibnamefont {Pekalski}}\ and\ \bibinfo {editor}
  {\bibfnamefont {R.}~\bibnamefont {Przystawa}}\ (\bibinfo  {publisher}
  {Springer-Verlag},\ \bibinfo {address} {Berlin},\ \bibinfo {year}
  {1980})\BibitemShut {NoStop}%
\bibitem [{\citenamefont {Nozieres}\ and\ \citenamefont
  {Schmitt-Rink}(1985)}]{Nozieres.1985}%
  \BibitemOpen
  \bibfield  {author} {\bibinfo {author} {\bibfnamefont {P.}~\bibnamefont
  {Nozieres}}\ and\ \bibinfo {author} {\bibfnamefont {S.}~\bibnamefont
  {Schmitt-Rink}},\ }\href@noop {} {\bibfield  {journal} {\bibinfo  {journal}
  {J. Low Temp. Phys.}\ }\textbf {\bibinfo {volume} {59}},\ \bibinfo {pages}
  {195} (\bibinfo {year} {1985})}\BibitemShut {NoStop}%
\bibitem [{\citenamefont {Bojesen}\ \emph {et~al.}(2013)\citenamefont
  {Bojesen}, \citenamefont {Babaev},\ and\ \citenamefont
  {Sudb\o{}}}]{Bojesen.2013}%
  \BibitemOpen
  \bibfield  {author} {\bibinfo {author} {\bibfnamefont {T.~A.}\ \bibnamefont
  {Bojesen}}, \bibinfo {author} {\bibfnamefont {E.}~\bibnamefont {Babaev}}, \
  and\ \bibinfo {author} {\bibfnamefont {A.}~\bibnamefont {Sudb\o{}}},\ }\href
  {\doibase 10.1103/PhysRevB.88.220511} {\bibfield  {journal} {\bibinfo
  {journal} {Phys. Rev. B}\ }\textbf {\bibinfo {volume} {88}},\ \bibinfo
  {pages} {220511} (\bibinfo {year} {2013})}\BibitemShut {NoStop}%
\bibitem [{\citenamefont {Bojesen}\ \emph {et~al.}(2014)\citenamefont
  {Bojesen}, \citenamefont {Babaev},\ and\ \citenamefont
  {Sudb\o{}}}]{Bojesen.2014}%
  \BibitemOpen
  \bibfield  {author} {\bibinfo {author} {\bibfnamefont {T.~A.}\ \bibnamefont
  {Bojesen}}, \bibinfo {author} {\bibfnamefont {E.}~\bibnamefont {Babaev}}, \
  and\ \bibinfo {author} {\bibfnamefont {A.}~\bibnamefont {Sudb\o{}}},\ }\href
  {\doibase 10.1103/PhysRevB.89.104509} {\bibfield  {journal} {\bibinfo
  {journal} {Phys. Rev. B}\ }\textbf {\bibinfo {volume} {89}},\ \bibinfo
  {pages} {104509} (\bibinfo {year} {2014})}\BibitemShut {NoStop}%
\bibitem [{\citenamefont {Carlstr\"om}\ and\ \citenamefont
  {Babaev}(2015)}]{Carlstrom.2015}%
  \BibitemOpen
  \bibfield  {author} {\bibinfo {author} {\bibfnamefont {J.}~\bibnamefont
  {Carlstr\"om}}\ and\ \bibinfo {author} {\bibfnamefont {E.}~\bibnamefont
  {Babaev}},\ }\href {\doibase 10.1103/PhysRevB.91.140504} {\bibfield
  {journal} {\bibinfo  {journal} {Phys. Rev. B}\ }\textbf {\bibinfo {volume}
  {91}},\ \bibinfo {pages} {140504} (\bibinfo {year} {2015})}\BibitemShut
  {NoStop}%
\bibitem [{\citenamefont {Gor'kov}\ and\ \citenamefont
  {Melik-Barkhudarov}(1961)}]{Gor'kov.1961}%
  \BibitemOpen
  \bibfield  {author} {\bibinfo {author} {\bibfnamefont {L.~P.}\ \bibnamefont
  {Gor'kov}}\ and\ \bibinfo {author} {\bibfnamefont {T.~K.}\ \bibnamefont
  {Melik-Barkhudarov}},\ }\href@noop {} {\bibfield  {journal} {\bibinfo
  {journal} {Soviet Physics JETP}\ }\textbf {\bibinfo {volume} {13}} (\bibinfo
  {year} {1961})}\BibitemShut {NoStop}%
\bibitem [{\citenamefont {Yeoh}\ \emph {et~al.}(2011)\citenamefont {Yeoh},
  \citenamefont {Gault}, \citenamefont {Cui}, \citenamefont {Zhu},
  \citenamefont {Moody}, \citenamefont {Li}, \citenamefont {Zheng},
  \citenamefont {Li}, \citenamefont {Wang}, \citenamefont {Dou}, \citenamefont
  {Sun}, \citenamefont {Lin},\ and\ \citenamefont {Ringer}}]{Yeoh.2011}%
  \BibitemOpen
  \bibfield  {author} {\bibinfo {author} {\bibfnamefont {W.~K.}\ \bibnamefont
  {Yeoh}}, \bibinfo {author} {\bibfnamefont {B.}~\bibnamefont {Gault}},
  \bibinfo {author} {\bibfnamefont {X.~Y.}\ \bibnamefont {Cui}}, \bibinfo
  {author} {\bibfnamefont {C.}~\bibnamefont {Zhu}}, \bibinfo {author}
  {\bibfnamefont {M.~P.}\ \bibnamefont {Moody}}, \bibinfo {author}
  {\bibfnamefont {L.}~\bibnamefont {Li}}, \bibinfo {author} {\bibfnamefont
  {R.~K.}\ \bibnamefont {Zheng}}, \bibinfo {author} {\bibfnamefont {W.~X.}\
  \bibnamefont {Li}}, \bibinfo {author} {\bibfnamefont {X.~L.}\ \bibnamefont
  {Wang}}, \bibinfo {author} {\bibfnamefont {S.~X.}\ \bibnamefont {Dou}},
  \bibinfo {author} {\bibfnamefont {G.~L.}\ \bibnamefont {Sun}}, \bibinfo
  {author} {\bibfnamefont {C.~T.}\ \bibnamefont {Lin}}, \ and\ \bibinfo
  {author} {\bibfnamefont {S.~P.}\ \bibnamefont {Ringer}},\ }\href {\doibase
  10.1103/PhysRevLett.106.247002} {\bibfield  {journal} {\bibinfo  {journal}
  {Phys. Rev. Lett.}\ }\textbf {\bibinfo {volume} {106}},\ \bibinfo {pages}
  {247002} (\bibinfo {year} {2011})}\BibitemShut {NoStop}%
\bibitem [{\citenamefont {Song}\ \emph {et~al.}(2013)\citenamefont {Song},
  \citenamefont {Yin}, \citenamefont {Zech}, \citenamefont {Williams},
  \citenamefont {Yee}, \citenamefont {Chen}, \citenamefont {Luo}, \citenamefont
  {Wang}, \citenamefont {Hudson},\ and\ \citenamefont {Hoffman}}]{Song.2013}%
  \BibitemOpen
  \bibfield  {author} {\bibinfo {author} {\bibfnamefont {C.-L.}\ \bibnamefont
  {Song}}, \bibinfo {author} {\bibfnamefont {Y.}~\bibnamefont {Yin}}, \bibinfo
  {author} {\bibfnamefont {M.}~\bibnamefont {Zech}}, \bibinfo {author}
  {\bibfnamefont {T.}~\bibnamefont {Williams}}, \bibinfo {author}
  {\bibfnamefont {M.~M.}\ \bibnamefont {Yee}}, \bibinfo {author} {\bibfnamefont
  {G.-F.}\ \bibnamefont {Chen}}, \bibinfo {author} {\bibfnamefont {J.-L.}\
  \bibnamefont {Luo}}, \bibinfo {author} {\bibfnamefont {N.-L.}\ \bibnamefont
  {Wang}}, \bibinfo {author} {\bibfnamefont {E.~W.}\ \bibnamefont {Hudson}}, \
  and\ \bibinfo {author} {\bibfnamefont {J.~E.}\ \bibnamefont {Hoffman}},\
  }\href {\doibase 10.1103/PhysRevB.87.214519} {\bibfield  {journal} {\bibinfo
  {journal} {Phys. Rev. B}\ }\textbf {\bibinfo {volume} {87}},\ \bibinfo
  {pages} {214519} (\bibinfo {year} {2013})}\BibitemShut {NoStop}%
\bibitem [{\citenamefont {Sprau}\ \emph {et~al.}(2017)\citenamefont {Sprau},
  \citenamefont {Kostin}, \citenamefont {Kreisel}, \citenamefont {B{\"o}hmer},
  \citenamefont {Taufour}, \citenamefont {Canfield}, \citenamefont {Mukherjee},
  \citenamefont {Hirschfeld}, \citenamefont {Andersen},\ and\ \citenamefont
  {Davis}}]{Sprau.2016}%
  \BibitemOpen
  \bibfield  {author} {\bibinfo {author} {\bibfnamefont {P.~O.}\ \bibnamefont
  {Sprau}}, \bibinfo {author} {\bibfnamefont {A.}~\bibnamefont {Kostin}},
  \bibinfo {author} {\bibfnamefont {A.}~\bibnamefont {Kreisel}}, \bibinfo
  {author} {\bibfnamefont {A.~E.}\ \bibnamefont {B{\"o}hmer}}, \bibinfo
  {author} {\bibfnamefont {V.}~\bibnamefont {Taufour}}, \bibinfo {author}
  {\bibfnamefont {P.~C.}\ \bibnamefont {Canfield}}, \bibinfo {author}
  {\bibfnamefont {S.}~\bibnamefont {Mukherjee}}, \bibinfo {author}
  {\bibfnamefont {P.~J.}\ \bibnamefont {Hirschfeld}}, \bibinfo {author}
  {\bibfnamefont {B.~M.}\ \bibnamefont {Andersen}}, \ and\ \bibinfo {author}
  {\bibfnamefont {J.~C.~S.}\ \bibnamefont {Davis}},\ }\href {\doibase
  10.1126/science.aal1575} {\bibfield  {journal} {\bibinfo  {journal}
  {Science}\ }\textbf {\bibinfo {volume} {357}},\ \bibinfo {pages} {75}
  (\bibinfo {year} {2017})}\BibitemShut {NoStop}%
\bibitem [{\citenamefont {Du}\ \emph {et~al.}()\citenamefont {Du},
  \citenamefont {Yang}, \citenamefont {Altenfeld}, \citenamefont {Gu},
  \citenamefont {Yang}, \citenamefont {Eremin}, \citenamefont {Hirschfeld},
  \citenamefont {Mazin}, \citenamefont {Lin}, \citenamefont {Zhu},\ and\
  \citenamefont {Wen}}]{Du2017}%
  \BibitemOpen
  \bibfield  {author} {\bibinfo {author} {\bibfnamefont {Z.}~\bibnamefont
  {Du}}, \bibinfo {author} {\bibfnamefont {X.}~\bibnamefont {Yang}}, \bibinfo
  {author} {\bibfnamefont {D.}~\bibnamefont {Altenfeld}}, \bibinfo {author}
  {\bibfnamefont {Q.}~\bibnamefont {Gu}}, \bibinfo {author} {\bibfnamefont
  {H.}~\bibnamefont {Yang}}, \bibinfo {author} {\bibfnamefont {I.}~\bibnamefont
  {Eremin}}, \bibinfo {author} {\bibfnamefont {P.~J.}\ \bibnamefont
  {Hirschfeld}}, \bibinfo {author} {\bibfnamefont {I.~I.}\ \bibnamefont
  {Mazin}}, \bibinfo {author} {\bibfnamefont {H.}~\bibnamefont {Lin}}, \bibinfo
  {author} {\bibfnamefont {X.}~\bibnamefont {Zhu}}, \ and\ \bibinfo {author}
  {\bibfnamefont {H.-H.}\ \bibnamefont {Wen}},\ }\href@noop {} {\ }\Eprint
  {http://arxiv.org/abs/1704.06141 (unpublished)} {arXiv:1704.06141
  (unpublished)} \BibitemShut {NoStop}%
\bibitem [{\citenamefont {Wang}\ \emph {et~al.}(2012)\citenamefont {Wang},
  \citenamefont {Yang}, \citenamefont {Fang}, \citenamefont {Shen},
  \citenamefont {Wang}, \citenamefont {Shan}, \citenamefont {Zhang},
  \citenamefont {Dai},\ and\ \citenamefont {Wen}}]{Wang2012}%
  \BibitemOpen
  \bibfield  {author} {\bibinfo {author} {\bibfnamefont {Z.}~\bibnamefont
  {Wang}}, \bibinfo {author} {\bibfnamefont {H.}~\bibnamefont {Yang}}, \bibinfo
  {author} {\bibfnamefont {D.}~\bibnamefont {Fang}}, \bibinfo {author}
  {\bibfnamefont {B.}~\bibnamefont {Shen}}, \bibinfo {author} {\bibfnamefont
  {Q.-H.}\ \bibnamefont {Wang}}, \bibinfo {author} {\bibfnamefont
  {L.}~\bibnamefont {Shan}}, \bibinfo {author} {\bibfnamefont {C.}~\bibnamefont
  {Zhang}}, \bibinfo {author} {\bibfnamefont {P.}~\bibnamefont {Dai}}, \ and\
  \bibinfo {author} {\bibfnamefont {H.-H.}\ \bibnamefont {Wen}},\ }\href
  {\doibase 10.1038/NPHYS2478} {\bibfield  {journal} {\bibinfo  {journal} {Nat.
  Phys.}\ }\textbf {\bibinfo {volume} {9}},\ \bibinfo {pages} {42} (\bibinfo
  {year} {2012})}\BibitemShut {NoStop}%
\bibitem [{\citenamefont {Shan}\ \emph {et~al.}(2012)\citenamefont {Shan},
  \citenamefont {Gong}, \citenamefont {Wang}, \citenamefont {Shen},
  \citenamefont {Hou}, \citenamefont {Ren}, \citenamefont {Li}, \citenamefont
  {Yang}, \citenamefont {Wen}, \citenamefont {Li},\ and\ \citenamefont
  {Dai}}]{Shan2012}%
  \BibitemOpen
  \bibfield  {author} {\bibinfo {author} {\bibfnamefont {L.}~\bibnamefont
  {Shan}}, \bibinfo {author} {\bibfnamefont {J.}~\bibnamefont {Gong}}, \bibinfo
  {author} {\bibfnamefont {Y.-L.}\ \bibnamefont {Wang}}, \bibinfo {author}
  {\bibfnamefont {B.}~\bibnamefont {Shen}}, \bibinfo {author} {\bibfnamefont
  {X.}~\bibnamefont {Hou}}, \bibinfo {author} {\bibfnamefont {C.}~\bibnamefont
  {Ren}}, \bibinfo {author} {\bibfnamefont {C.}~\bibnamefont {Li}}, \bibinfo
  {author} {\bibfnamefont {H.}~\bibnamefont {Yang}}, \bibinfo {author}
  {\bibfnamefont {H.-H.}\ \bibnamefont {Wen}}, \bibinfo {author} {\bibfnamefont
  {S.}~\bibnamefont {Li}}, \ and\ \bibinfo {author} {\bibfnamefont
  {P.}~\bibnamefont {Dai}},\ }\href {\doibase 10.1103/PhysRevLett.108.227002}
  {\bibfield  {journal} {\bibinfo  {journal} {Phys. Rev. Lett.}\ }\textbf
  {\bibinfo {volume} {108}},\ \bibinfo {pages} {227002} (\bibinfo {year}
  {2012})}\BibitemShut {NoStop}%
\bibitem [{\citenamefont {Hettler}\ and\ \citenamefont
  {Hirschfeld}(1999)}]{Hettler.1999}%
  \BibitemOpen
  \bibfield  {author} {\bibinfo {author} {\bibfnamefont {M.~H.}\ \bibnamefont
  {Hettler}}\ and\ \bibinfo {author} {\bibfnamefont {P.~J.}\ \bibnamefont
  {Hirschfeld}},\ }\href {\doibase 10.1103/PhysRevB.59.9606} {\bibfield
  {journal} {\bibinfo  {journal} {Phys. Rev. B}\ }\textbf {\bibinfo {volume}
  {59}},\ \bibinfo {pages} {9606} (\bibinfo {year} {1999})}\BibitemShut
  {NoStop}%
\bibitem [{\citenamefont {Pereg-Barnea}\ and\ \citenamefont
  {Franz}(2008)}]{Pereg.2008}%
  \BibitemOpen
  \bibfield  {author} {\bibinfo {author} {\bibfnamefont {T.}~\bibnamefont
  {Pereg-Barnea}}\ and\ \bibinfo {author} {\bibfnamefont {M.}~\bibnamefont
  {Franz}},\ }\href {\doibase 10.1103/PhysRevB.78.020509} {\bibfield  {journal}
  {\bibinfo  {journal} {Phys. Rev. B}\ }\textbf {\bibinfo {volume} {78}},\
  \bibinfo {pages} {020509} (\bibinfo {year} {2008})}\BibitemShut {NoStop}%
\bibitem [{\citenamefont {Fanfarillo}\ \emph {et~al.}(2009)\citenamefont
  {Fanfarillo}, \citenamefont {Benfatto}, \citenamefont {Caprara},
  \citenamefont {Castellani},\ and\ \citenamefont {Grilli}}]{Fanfarillo.2009}%
  \BibitemOpen
  \bibfield  {author} {\bibinfo {author} {\bibfnamefont {L.}~\bibnamefont
  {Fanfarillo}}, \bibinfo {author} {\bibfnamefont {L.}~\bibnamefont
  {Benfatto}}, \bibinfo {author} {\bibfnamefont {S.}~\bibnamefont {Caprara}},
  \bibinfo {author} {\bibfnamefont {C.}~\bibnamefont {Castellani}}, \ and\
  \bibinfo {author} {\bibfnamefont {M.}~\bibnamefont {Grilli}},\ }\href
  {\doibase 10.1103/PhysRevB.79.172508} {\bibfield  {journal} {\bibinfo
  {journal} {Phys. Rev. B}\ }\textbf {\bibinfo {volume} {79}},\ \bibinfo
  {pages} {172508} (\bibinfo {year} {2009})}\BibitemShut {NoStop}%
\bibitem [{\citenamefont {Marciani}\ \emph {et~al.}(2013)\citenamefont
  {Marciani}, \citenamefont {Fanfarillo}, \citenamefont {Castellani},\ and\
  \citenamefont {Benfatto}}]{Marciani.2013}%
  \BibitemOpen
  \bibfield  {author} {\bibinfo {author} {\bibfnamefont {M.}~\bibnamefont
  {Marciani}}, \bibinfo {author} {\bibfnamefont {L.}~\bibnamefont
  {Fanfarillo}}, \bibinfo {author} {\bibfnamefont {C.}~\bibnamefont
  {Castellani}}, \ and\ \bibinfo {author} {\bibfnamefont {L.}~\bibnamefont
  {Benfatto}},\ }\href {\doibase 10.1103/PhysRevB.88.214508} {\bibfield
  {journal} {\bibinfo  {journal} {Phys. Rev. B}\ }\textbf {\bibinfo {volume}
  {88}},\ \bibinfo {pages} {214508} (\bibinfo {year} {2013})}\BibitemShut
  {NoStop}%
\end{thebibliography}%

\end{document}